\documentclass[final,11pt]{article}
\pdfoutput=1
\usepackage{MRnotes}

\newcommand{\RNAdS}[1]{Reissner-Nordstr\"om-AdS$_{#1}$}

\newcommand{\RQ}{r_{_{\!Q}}}    
\newcommand{\bRQ}{\mathfrak{C}}   
\newcommand{\sdc}{\mathfrak{S}_{_Q}}  

\newcommand{\Mser}[2]{\varphi_{_{#1}}^{#2}}  
\newcommand{\Mserh}[2]{\hat{\varphi}_{_{#1}}^{#2}} 
\newcommand{\Dfn}[2]{\Delta_{_{#1}}^{#2}} 
\newcommand{\Dfnh}[2]{\hat{\Delta}_{_{#1}}^{#2}} 

\newcommand{\ann}{\mathscr{M}}   
\newcommand{\sen}[1]{\varphi_{_{#1}}} 

\newcommand{\bwt}{\mathfrak{w}}   
\newcommand{\bqt}{\mathfrak{q}}    
\newcommand{\bk}{\vb{k}}  
\newcommand{\bx}{\vb{x}}   

\newcommand{\ctor}{\zeta}  
\newcommand{\ri}{\varrho}   
\newcommand{\rib}{u}    
\newcommand{\Dz}{\mathbb{D}}  

\newcommand{\ScS}{\mathbb{S}} 
\newcommand{\ScST}{\mathbb{S}^{\scriptscriptstyle{\text{T}}} } 
\newcommand{\ai}{\alpha}   

\newcommand{\HH}{\Psi}
\newcommand{\HHT}{\Psi_{_\text{T}}}

\newcommand{\gOrb}{\Upsilon}

\newcommand{\HS}{\Psi_{_\text{S}}}
\newcommand{\PHE}{\Phi_{_\text{E}}}
\newcommand{\PHO}{\Phi_{_\text{O}}}
\newcommand{\PHW}{\Phi_{_\text{W}}}

\newcommand{\PHB}{\Phi_{_\text{B}}}

\newcommand{\EEin}{\mathcal{E}^{\text{\tiny{Ein}}}}   
\newcommand{\EMax}{\mathcal{E}^{\text{\tiny{Max}}}}  
\newcommand{\EEq}{\mathbb{E}}
\newcommand{\EEqT}{\mathbb{E}_{_\text{T}}}

\newcommand{\EEqB}{\mathbb{E}_{_\text{B}}}

\newcommand{\AGR}{\mathscr{A}}

\newcommand{\HA}{\widehat{\Psi}}

\newcommand{\TcftD}{T^{\text{\tiny{CFT}}}}
\newcommand{\TcftU}{T_{\text{\tiny{CFT}}}}

\newcommand{\Lk}{\Lambda_k}
\newcommand{\MW}{\Theta}
\newcommand{\MZ}{\mathscr{Z}}

\newcommand{\Tone}{\mathscr{T}}

\newcommand{\Gatt}{\Gamma_s}
\newcommand{\Ost}{\widetilde{\Gamma}_s(\omega,\bk)}

\newcommand{\KS}{K_s}
\newcommand{\Kc}{K_c}

\newcommand{\Vd}{\mathsf{V}}
\newcommand{\Zd}{\mathsf{Z}} 
\newcommand{\VZd}{\mathfrak{V}_{\mathsf{Z}}}
\newcommand{\VVd}{\mathfrak{V}_{\mathsf{V}}}
\newcommand{\Np}[1]{\mathsf{N}_{_#1}}

\newcommand{\PiZ}{\Pi_{_\Zd}}
\newcommand{\PiV}{\Pi_{_\Vd}}

\newcommand{\OpV}{\breve{\mathcal{O}}_{_\Vd}}
\newcommand{\OpZ}{\breve{\mathcal{O}}_{_\Zd}}
\newcommand{\PoV}{\breve{\mathcal{V}}}
\newcommand{\PoZ}{\breve{\mathcal{Z}}}
\newcommand{\JoV}{\breve{\bm{\xi}}}
\newcommand{\JoZ}{\breve{\bm{\zeta}}}

\newcommand{\QOp}{\breve{\mathcal{Q}} }
\newcommand{\EOp}{\breve{\mathcal{E}} }

\newcommand{\Pbg}{P_0}
\newcommand{\Qbg}{\rho_0}

\newcommand{\In}{\text{\tiny{in}}}     
\newcommand{\Rev}{\text{\tiny{rev}}}  
\newcommand{\SKs}{\text{\tiny{SK}}}   

\newcommand{\nB}{n_{_B}}   
\newcommand{\Gin}[1]{G_{_{#1}}^\In}     
\newcommand{\Kin}[1]{K_{_{#1}}^\In}      
\newcommand{\Grev}[1]{G_{_{#1}}^\Rev}   
\newcommand{\Krev}[1]{K_{_{#1}}^\Rev}    

\newcommand{\MV}{\mathscr{V}}

\newcommand{\Jcft}{J^\text{\tiny{CFT}}}

 \newcommand{\MYd}{\mathscr{Y}}    
 \newcommand{\MY}{\mathscr{Y}}   
\newcommand{\BQT}{\mathfrak{p}}

\newcommand{\skR}{\text{\tiny R}}
\newcommand{\skL}{\text{\tiny L}}

\title{An effective description of  charge diffusion and energy transport in a charged plasma from holography}
\author[a]{Temple He,}
\author[b]{R. Loganayagam,}
\author[a]{Mukund Rangamani,}  
\author[a]{Julio Virrueta}

\affiliation[a]{
	Center for Quantum Mathematics and Physics (QMAP)\\
	Department of Physics \& Astronomy, University of California, Davis, CA 95616 USA}
\affiliation[b]{
	International Centre for Theoretical Sciences (ICTS-TIFR), \\ 
	Tata Institute of Fundamental Research, Shivakote, Hesaraghatta, Bangalore 560089, India.}

\emailAdd{tmhe@ucdavis.edu}
\emailAdd{nayagam@icts.res.in}
\emailAdd{mukund@physics.ucdavis.edu}
\emailAdd{jvirrueta@ucdavis.edu}

\abstract{
We discuss the physics of sound propagation and charge diffusion in a plasma with non-vanishing charge density. Our analysis culminates the program initiated in \cite{He:2021jna} to construct an open effective field theory of low-lying modes of the stress tensor and charge current in such plasmas. We model the plasma holographically as a \RNAdS{d+1} black hole, and study linearized  fluctuations of longitudinally polarized scalar gravitons and photons in this  background. We demonstrate that the perturbations can be decoupled and repackaged into the dynamics of two designer scalars, whose gravitational coupling is modulated by a non-trivial dilatonic factor. The holographic analysis allows us to isolate the phonon mode from the charge diffusion mode, and identify the combination of currents that corresponds to each of them. We use these results to obtain the real-time  Gaussian effective action,  which includes both the retarded response and the associated stochastic (Hawking) fluctuations, accurate to quartic order in gradients. }

\begin{document} 
\maketitle


\section{Introduction}
\label{sec:intro}

The holographic AdS/CFT correspondence provides a useful arena to understand real-time dynamics of thermal plasmas.  Under the duality, the response functions and fluctuations of the plasma map onto the study of perturbations of asymptotically AdS black holes. The response data is encoded in the quasinormal modes \cite{Horowitz:1999jd}, while the thermal fluctuations map onto the Hawking quanta.  Of interest to us are the low-lying,  or long-lived,  quasinormal modes that characterize near-equilibrium hydrodynamic behaviour \cite{Policastro:2002se,Policastro:2002tn}. The non-linear dissipative response of these modes is well understood within the context of the fluid/gravity correspondence \cite{Bhattacharyya:2008jc, Hubeny:2011hd}, but it is only recently that we have begun to systematically understand the associated stochastic fluctuations. 

A unified framework for capturing both the response and the  fluctuations is in terms of an \emph{open effective theory} \cite{Jana:2020vyx}. This description seeks to capture the real-time (Schwinger-Keldysh) dynamics of the thermal plasma. This has been aided by the improved prescriptions for analyzing real-time dynamics in holography \cite{Glorioso:2018mmw}, building on earlier works \cite{Son:2002sd,Herzog:2002pc,Skenderis:2008dg,vanRees:2009rw}. 

One can understand this open effective theory as follows: consider coupling the thermal plasma  to an external probe (measurement device). To obtain the low energy effective field theory, we integrate out the fast modes of the plasma, in a suitable Born-Oppenheimer approximation, and obtain the dynamics valid at long-distances and late-times. Since in this limit, the low-lying modes are those that are associated to conserved currents, we end up with an effective description of the hydrodynamic modes, allowing us thus to view hydrodynamics as an open effective field theory.
 
Motivated by this logic, in a recent series of papers, the preliminary steps for constructing such an open effective field theory have been undertaken. In particular, \cite{Ghosh:2020lel, He:2022jnc} have succeeded in constructing a description of a neutral plasma   (dual to a \SAdS{} black hole) at the Gaussian order. The first of these works focused on momentum diffusion, while the second tackled attenuated  sound propagation.\footnote{
	Non-Gaussian effects, while straightforward to include in this framework \cite{Jana:2020vyx}, are yet to be analyzed in the hydrodynamic context. }
For other works analyzing thermal real-time dynamics from AdS black hole backgrounds see \cite{deBoer:2018qqm,Chakrabarty:2019aeu,Loganayagam:2020eue,Loganayagam:2020iol,Chakrabarty:2020ohe,Bu:2020jfo,Bu:2021clf,Bu:2022esd}.

From a hydrodynamic viewpoint, the intrinsic challenge is to separate out the slow and fast modes of the conserved current (energy-momentum tensor  for a neutral plasma). Moreover, the diffusive dynamics of momentum is qualitatively different from that of energy transport (which produces sound). One way to proceed is to use symmetries: pick a direction of spatial momentum in $\mathbb{R}^{d-1,1}$ and decompose the energy-momentum tensor into polarizations labeled by the transverse $SO(d-2)$ rotation group. Transverse traceless tensors are short-lived, transverse vectors diffuse, and the remaining longitudinal mode transports energy. 

This perspective turns out to be particularly efficacious in the dual gravitational setting. The real-time gravitational saddle is the gravitational Schwinger-Keldysh (grSK) geometry introduced in \cite{Glorioso:2018mmw},  which is a two-sheeted complex spacetime with a thermal monodromy condition around the horizon. As explained in \cite{Jana:2020vyx} this geometry allows for directly computing real-time thermal correlation functions of the dual field theory. Operationally, one starts with the Lorentzian black hole solution and restricts attention to the domain of outer communication to focus on the ingoing modes. This can be achieved by working in ingoing coordinates that are smooth across the future horizon. Once one has obtained the ingoing wavefunction, one can construct a suitable boundary-bulk propagator with sources on a boundary Schwinger-Keldysh contour. This involves an admixture of both ingoing and outgoing modes, but as explained in \cite{Jana:2020vyx}, the latter can be obtained from the ingoing solution by  using a suitable covariance property under the  discrete time-reflection isometry. Crucially, the thermal monodromy picked up by the outgoing mode is important in ensuring that the correlation functions satisfy the thermal KMS condition.

We will here be interested in the correlation functions of stress tensor and charge current of the dual plama.  Since the conserved currents of the field theory map to gauge fields in the bulk geometry, one has to additionally confront the issue of gauge choices and boundary conditions. As argued in \cite{Ghosh:2020lel,He:2022jnc} the oft-used radial gauge choice is not well suited for the analysis. A natural way to proceed is to use the gauge invariant combination of perturbations analyzed in \cite{Kodama:2003jz, Kodama:2003kk}. This description not only exploits the $SO(d-2)$ representation structure, but it also repackages the dynamics of dual gravitational fluctuations into a set of designer scalar fields.

These designer fields are non-minimally coupled scalars, with their gravitational coupling modulated radially in the AdS geometry, and fall into two very natural classes. Short-lived modes, dubbed Markovian, have the dilaton blowing up (as a power law) near the AdS boundary. On the other hand, long-lived non-Markovian modes have a dilatonic coupling that is decaying near the boundary and growing somewhere in the interior of the spacetime.

For Markovian fields, the asymptotically growing dilaton requires one to freeze its source, the  non-normalizable part, at the AdS boundary. Furthermore, we  want to integrate out the dual boundary degrees of freedom to derive the open effective dynamics. Operationally, therefore, Markovian modes have Dirichlet  (standard) boundary conditions imposed on them, both for the purposes of computing the generating function of correlators, and for the purposes of obtaining the open effective field theory. 

Non-Markovian fields, on the other hand, have a decaying dilatonic coupling, which implies that the bulk wavefunctions are dominantly supported near the boundary. Associated to this is the fact that such fields naturally come with boundary terms which require us to freeze the normalizable part. Thus, for purposes of computing the generating function of correlators, we impose Neumann (alternate) boundary conditions on the non-Markovian fields.  

However, in deriving the open effective field theory, we should not integrate out the non-Markovian modes. They are the physical low-energy degrees of freedom which must be retained in the effective action. Therefore, \cite{Ghosh:2020lel} proposed to compute not the generating function of correlators, but the Wilsonian influence functional (WIF), parameterized by the boundary values of the non-Markovian fields. From the dual plasma viewpoint, we are parameterizing the effective dynamics by the expectation value of long-lived modes.  Operationally, this turns out to be quite easy to do: the switch from the generating functional to the WIF is implemented by a Legendre transform. This step has the effect of removing the boundary term that implements the Neumann boundary condition. The upshot is simply that the computation of the WIF for non-Markovian fields is achieved by quantizing them with Dirichlet boundary conditions.

As noted above, the analysis of \cite{Ghosh:2020lel,He:2022jnc} made these ideas quite precise in the context of a neutral fluid. A more challenging proposition is the dynamics of a charged plasma, where we have not only the energy-momentum tensor but also the conserved charge current. There are certain additional novelties in this case: the non-zero density in the background causes mode mixing between the two currents. In the transverse vector sector, we find a mixing between the short-lived propagating current modes and the momentum diffusion modes, while the scalar sector comprises  two long-lived modes: sound and charge diffusion.

In  \cite{He:2021jna} we initiated the analysis of such charged plasmas which are dual to \RNAdS{} black holes.\footnote{ 
	The dynamics of a probe charge current in a neutral plasma was discussed in \cite{Ghosh:2020lel}. This problem was also analyzed before in \cite{Glorioso:2018mmw,deBoer:2018qqm} and revisited in \cite{Bu:2020jfo}. These references work in radial gauge, which is a poor choice for reasons elucidated in \cite{Ghosh:2020lel,He:2021jna}. Physically, while the ingoing solutions are fine in radial gauge, the time-reversed outgoing solutions are singular and lead to certain ambiguities, which were handled by certain ad hoc choices in the aforementioned references.  }
Therein, we analyzed the transverse tensor and vector perturbations, demonstrating that the Markovian charge current can be decoupled from the non-Markovian momentum diffusion at the Gaussian order. The gravitational description helps provide this clean separation and allows one to find linear combination of currents which correspond to the two different modes. 

In the present work, we conclude the analysis of charged plasmas by including scalar perturbations. The main novelty here is that for the first time we encounter mixing between two non-Markovian modes corresponding to energy transport and charge diffusion, respectively. 
From a technical standpoint, scalar perturbations of \RNAdS{} are challenging: there are a-priori ten functions in the metric and gauge field, which need to be distilled into two independent physical degrees of freedom.  Furthermore, as discussed in  \cite{He:2022jnc}, the gravity dual of the sound mode is a designer scalar whose dialtonic coupling is modulated as a function of spatial momentum.\footnote{ One consequence of this behaviour is that the mode is Markovian at low orders in the gradient expansion, with its true non-Markovian character only emerging at quartic order in gradients. } Heuristically, this can be traced to the fact that energy transport results in a physical propagating Goldstone mode, the phonon, whose attenuation only kicks in at higher orders in gradients.\footnote{ Relativistic conformal fluids are compressible, and thus always have a low-lying mode with linear dispersion, which characterizes the physical sound mode in the fluid. This was first discussed in the context of holography in \cite{Policastro:2002tn}. We refer to this as the phonon Goldstone mode as we did in the neutral plasma case \cite{He:2022jnc}. This is natural from a holographic perspective, as originally explained in \cite{Bhattacharyya:2008jc} in the context of the fluid/gravity correspondence. Lest we cause confusion, we should remark that this phonon is produced by compression and rarefaction of energy gradients and is a longitudinal mode. It is not associated to translational symmetry breaking, which would be the case for phonon modes in solids, which have both longitudinal and transverse components. } On the contrary, charge diffusion is expected to behave as a long-lived non-Markovian field all through.  

Despite these complications, we demonstrate that the dynamics of energy transport and charge diffusion can be sensibly decoupled and packaged into two independent designer scalars. The bulk dynamics of these fields is somewhat involved, but surprisingly tractable in the gradient expansion analysis we undertake. To this end, we are guided by the previous analysis in \cite{Ghosh:2020lel,He:2022jnc,He:2021jna}; in fact, we use the function basis constructed in the latter two works to find an efficient presentation of the bulk  perturbation equations.  

Using this data, we derive the Gaussian effective field theory for the scalar sector of a charged plasma, and obtain a convenient parameterization of the conserved currents in terms of the phonon and charge diffusion operators. We match our results with the limited data available in the literature. Curiously, while there is a large body of work analyzing the low-lying quasinormal modes of \AdS{} black holes, much of the literature on linearized hydrodynamics focuses on the vector perturbations of \RNAdS{5} black holes. For the scalar sector there is little data available; our results agree to low orders with data extracted from the fluid/gravity literature for $d=4$, \cite{Banerjee:2008th,Erdmenger:2008rm,Plewa:2012vt}.\footnote{ 
	Note added in v2: We recently became aware of \cite{Abbasi:2020ykq}, who derive the dispersion relations for sound and charge diffusion to quadratic order in momenta in $d=4$. We give expressions for the dispersion relations accurate to quartic order (in fact, a prediction of the sound dispersion to sextic order, cf., \cref{sec:ZWasym}) for arbitrary dimensions $d>3$.} 
As a consequence, the results we describe herein are the first comprehensive analysis of the longitudinal energy transport and charge diffusion mode in a holographic system. This allows us not only the extraction of the dispersion relations, but also gives a clear picture of the physical combinations of the energy momentum tensor and charge current which drive these two independent modes (at quadratic order). We also give a heuristic picture of the result from a hydrodynamic perspective. This discussion will be brief as we hope to explain the connections of our linearized analysis  with the non-linear effective actions obtained using symmetry principles by various authors in the past decade in a separate work.

The outline of the paper is as follows. In \cref{sec:background} we will quickly review the background   \RNAdS{d+1} solution and its grSK uplift, using this to establish our conventions. In \cref{sec:dynamics} we describe how the perturbations of the black hole geometry are packaged into designer fields. Using their dynamics, we extract the Wilsonian influence functional and conserved currents in \cref{sec:ZSK} and discuss some physical implications. We conclude with a brief discussion in \cref{sec:discuss}. 

As much of the technicalities parallel earlier works, we have chosen to keep only the salient details in the main text. Readers interested in the details of our computations are invited to consult the appendices. \cref{sec:dynamicsderive} describes how to distill the dynamics into two designer scalars. This discussion is inspired by the original work of \cite{Kodama:2003kk}, but we have endeavored to clarify some aspects in our presentation. \cref{sec:actionderive}, which is new, explains how to obtain these dynamical equations from the Einstein-Maxwell action, deriving in the process the correct boundary conditions for the designer fields. In \cref{sec:gradexpfns} we explain the gradient expansion solution of the designer fields and use it to construct the physical metric and gauge field  perturbations. This data is then used in \cref{sec:bdyobs} where we outline how to obtain the boundary observables. 

\section{Background}
\label{sec:background}
 
We give a  brief summary of the background geometry, parameterizing the  data of \RNAdS{d+1} geometry, which is a solution to the Einstein-Maxwell theory, in a suitable manner. We adhere to the conventions of \cite{He:2021jna}, which the reader is encouraged to consult for additional details. 

Consider the  Einstein-Maxwell theory in $d+1$ dimensions with a negative cosmological constant, viz., 
\begin{equation}\label{eq:SEMax}
\begin{split}
S_\text{EM} 
&= 
	\frac{1}{16\pi G_N}\, \int d^{d+1} x\, \sqrt{-g} \, 
		\left[ R + d(d-1) - \frac{1}{2} \,F_{AB}\, F^{AB}\right] + 
		  S_\text{bdy} + S_\text{ct} \,,\\
S_\text{bdy}
&=  \frac{1}{8\pi G_N}\, \int d^d x\, \sqrt{-\gamma} \, K		  \,.
\end{split}
\end{equation}
 Here $g_{AB}$ is the bulk metric, $\gamma_{\mu\nu}$ the induced metric on the timelike asymptotic boundary, and $K$ is the extrinsic curvature of the boundary.\footnote{
 	Conventions: uppercase Latin alphabet ($A,B,\cdots$)  indicate bulk spacetime indices,  Greek alphabets ($\mu,\nu, \cdots$)  refer to boundary spacetime indices, and lowercase Latin alphabets ($i,j, \cdots$) are used to refer to the spatial directions along the boundary. \label{fn:conventions}} 
The counterterm action $S_\text{ct}$ is necessary to obtain finite physical answers; it is given in \cref{sec:dynamicsderive}. We have chosen the electromagnetic coupling to simplify some expressions; it will lead to factors of two in boundary conservation laws later. 

The equations of motion following  from \eqref{eq:SEMax} are
\begin{equation}\label{eq:EMeqns}
\begin{split}
\EEin_{AB}&\equiv
R_{AB} - \frac{1}{2}\, R\, g_{AB} - \frac{d(d-1)}{2}\, g_{AB} 
=
	 g^{CD}\, F_{AC} \, F_{BD} - \frac{1}{4}\, g_{AB}\, F_{CD}\, F^{CD} \,, \\
\EMax_B&\equiv
\nabla^A F_{AB} = 0	\,,
\end{split}
\end{equation}
with the \RNAdS{d+1} geometry being a two-parameter family of solutions, parametrized by $r_+$ (outer horizon scale) and $Q$ (a measure of the charge). In ingoing Eddington-Finkelstein coordinates the line element and gauge potential take the form
\begin{equation}\label{eq:RNAdS}
\begin{aligned}
ds^2 =
	2 dv dr - r^2\, f(r)\, dv^2 + r^2\, d\vb{x}^2 \,,  \qquad   \vb{A} =  -a(r)\, dv \,,
\end{aligned}
\end{equation}
with
\begin{equation}\label{eq:RNAdSfns}
f(r) = 1 - (1+Q^2)\,\left(\frac{r_+}{r}\right)^d + Q^2\, \left(\frac{r_+}{r}\right)^{2(d-1)}\,, \qquad
a(r) = \sqrt{\frac{d-1}{d-2}}\, Q\, \frac{r_+^{d-1}}{r^{d-2}}\,.
\end{equation}	
This solution  describes a charged thermal plasma of the dual CFT, with intensive thermodynamic parameters temperature and chemical potential   being\footnote{ 
	We define the effective central charge of the boundary theory as $c_\text{eff} = \frac{\lads^{d-1}}{16\pi G_N}$ and work  in units  where  $\lads =1$. Dimensions of physical quantities can be restored using the latter, eg.,  $T \propto r_+/\lads^2$.\label{fn:lads}} 
\begin{equation}\label{eq:TRN}
\begin{split}
T 
&=  
	\frac{d -(d-2)\, Q^2}{4\pi}\, r_+  \,, \qquad \mu =\sqrt{\frac{d-1}{d-2}} \, Q \, r_+  \,. 
\end{split}
\end{equation}	
The parameter $Q$ lives in a bounded domain 
\begin{equation}\label{eq:Qbound}
0 \leq Q \leq  \sqrt{\frac{d}{d-2}}\,.
\end{equation}	
$Q=0$ is the neutral \SAdS{d+1} solution and the upper limit corresponds to the extremal solution with vanishing temperature.

As described in \cite{He:2021jna}, it will be useful to introduce  a  function parameterized by the  ohmic radius $\sdc$ (which determines the DC conductivity)
\begin{equation}\label{eq:hfndef}
h(r) =  1-\sdc\, \frac{r_+^{d-2}}{r^{d-2}} \,,
\end{equation}	
where 
\begin{equation}\label{eq:RQdef}
\sdc \equiv  \frac{\RQ^{d-2}}{r_+^{d-2}} \,, \qquad \RQ^{d-2} = \frac{d-1}{d}\, \frac{2\,Q^2}{1+Q^2}   \, r_+^{d-2}  \,.
\end{equation}	
 For \SAdS{d+1} $\sdc =0$, while $\sdc=1$ for the extremal \RNAdS{d+1} solution. This parameter characterizes the DC conductivity, $\sigma_\text{dc} = r_+^{d-3} \, (1-\sdc)^2$, and serves as a proxy for the charge. The physical ohmic radius $\RQ$ is sandwiched between the inner and outer horizons (see \cite[Fig 6]{He:2021jna}). We will call $h$ the ohmic function.

Let us collect some useful thermodynamic and transport properties of the unperturbed charged plasma. We have the pressure $P_0$, the charge density $\Qbg$, and the dc conductivity  $ \sigma_\text{dc}$ to be given (up to central charge factors) by 
\begin{subequations}
\begin{equation}\label{eq:bgP}
\Pbg = (1+Q^2)\, r_+^d \,,
\end{equation}	
\begin{equation}\label{eq:rho0def}
\Qbg = (d-2)\, \mu\, r_+^{d-2}\,,
\end{equation}	
\begin{equation}\label{eq:sigmadc}
 \sigma_\text{dc} = r_+^{d-3}\, (1-\sdc)^2 \,.
\end{equation}	
\label{eq:bgPQs}
\end{subequations}
The pressure and charge density can be read off directly from the thermodynamic formulae. The aforementioned expression for the dc conductivity was introduced in \cite{He:2021jna}, where it demonstrated to be equivalent to the earlier result obtained in \cite{Hartnoll:2007ip}. Note that we are stripping off a factor of the  central charge $c_\text{eff}$ from the energy-momentum tensor  and a factor of $2\,c_\text{eff}$ from the charge current in these expressions for convenience.\footnote{
	The relative factor of $2$ between the charge current and the energy-momentum tensor originates from the normalization of the Maxwell term in \eqref{eq:SEMax}, see also \cref{fn:Jouleheat}. \label{fn:J2ceff}
}

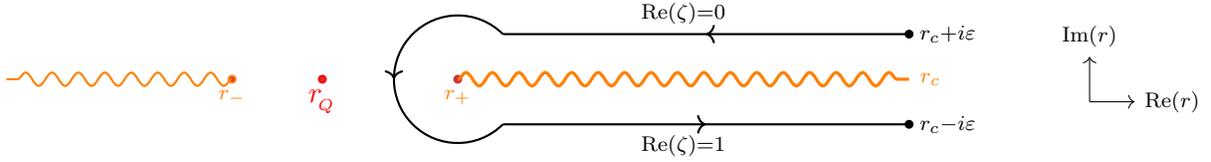
\begin{figure}[h!]
\begin{center}
\begin{tikzpicture}[scale=0.6]
\draw[thick,color=rust,fill=rust] (-5,0) circle (0.45ex);
\draw[thick,color=black,fill=black] (5,1) circle (0.45ex);
\draw[thick,color=black,fill=black] (5,-1) circle (0.45ex);
\draw[very thick,snake it, color=orange] (-5,0) node [below] {$\scriptstyle{r_+}$} -- (5,0) node [right] {$\scriptstyle{r_c}$};
\draw[thick,color=black, ->-] (5,1)  node [right] {$\scriptstyle{r_c+i\varepsilon}$} -- (0,1) node [above] {$\scriptstyle{\Re(\ctor) =0}$} -- (-4,1);
\draw[thick,color=black,->-] (-4,-1) -- (0,-1) node [below] {$\scriptstyle{\Re(\ctor) =1}$} -- (5,-1) node [right] {$\scriptstyle{r_c-i\varepsilon}$};
\draw[thick,color=black,->-] (-4,1) arc (45:315:1.414);
\draw[thin, color=black,  ->] (9,-0.5) -- (9,0.5) node [above] {$\scriptstyle{\Im(r)}$};
\draw[thin, color=black,  ->] (9,-0.5) -- (10,-0.5) node [right] {$\scriptstyle{\Re(r)}$};  
\draw[thick,color=orange,fill=rust] (-10,0) circle (0.45ex);
\draw[thick,color=red,fill=rust] (-8,0) circle (0.45ex) node[below] {$\RQ$};
\draw[thick,snake it, color=orange] (-10,0) node [below] {$\scriptstyle{r_-}$} -- (-15,0) ;
\end{tikzpicture}
\caption{ The complex $r$ plane with the locations of the two regulated boundaries (with cut-off $r_c$), the outer and inner horizons at $r_\pm$ and the ohmic radius $\RQ$ marked. The grSK contour is a codimension-1 surface in this plane (drawn at fixed $v$). As indicated the direction of the contour is counter-clockwise,  and it encircles the branch point at the outer horizon with the cut running out to the boundary. The cut emanating from the inner horizon  and the ohmic radius, are not encountered by the contour.}
\label{fig:mockt}
\end{center}
\end{figure}

Real-time correlation functions of the energy-momentum tensor  and charge currents are  computed by uplifting this solution to an appropriate Schwinger-Keldysh form. This results in the grSK geometry, a particular complexification of the \RNAdS{d+1} geometry, with line element 
\begin{equation}\label{eq:grSKgeom}
ds^2 = -r^2\, f\, dv^2 + i\,\beta r^2\, f\, dv \,d \ctor + r^2 \, d\mathbf{x}^2 \,, \qquad \dv{r}{\ctor} = \frac{i\,\beta}{2}\, r^2f \,.
\end{equation}	
The coordinate $\ctor$ is the mock tortoise coordinate with $\beta = T^{-1}$,  defined on the complex $r$ plane along a contour that encircles the cut emanating from the horizon at $r=r_+$, cf., \cref{fig:mockt}.  Our analysis employs the time-reversal\footnote{
	Time-reversal is a $\mathbb{Z}_2$ involution $v\mapsto i\beta\ctor-v$  which leaves \eqref{eq:grSKgeom} invariant.
} covariant bases introduced in 
\cite{Ghosh:2020lel}. The tangent space basis is $\{\Dz_+,\partial_v, \partial_i\}$, while 
$\{\frac{dr}{r^2 f} , dv - \frac{dr}{r^2 f}, dx^i\}$ provides the dual cotangent space basis. The derivation $\Dz_+$ is defined to be a dressed radial derivative
\begin{equation}\label{eq:Dz}
\Dz_\pm = r^2f \, \pdv{}{r} \pm \, \pdv{}{v} \,, \qquad \Dz_\pm = r^2f \, \pdv{}{r} \mp \, i \, \omega \,, 
\end{equation}	
in the time and frequency domain, respectively.  

\paragraph{Some conventions:}
We adopt a shorthand notation for the measure on the Fourier domain
\begin{equation}
\begin{split}
\int_k \equiv \int\frac{d\omega}{2\pi}\int\frac{d^{d-1}\bk}{(2\pi)^{d-1}} 
\end{split}
\end{equation}
to keep expressions compact. In writing various expressions, we will use the outer horizon radius to scale out dimensions, in particular, defining
\begin{equation}\label{eq:bwq}
\bwt = \frac{\omega}{r_+} \,, \qquad \bqt = \frac{k}{r_+}\,.
\end{equation}	
%

\section{Dynamics: designer sound and charge diffusion fields}
\label{sec:dynamics}

We are interested in analyzing the dynamics of sound and charge diffusion in a charged holographic plasma. On the dual gravitational side, the two modes correspond to perturbations of the \RNAdS{d+1} black hole, specifically, the scalar polarizations of gravitons and photons. We pick a direction for the spatial momentum and decompose the fields into planar harmonics with respect to the transverse $SO(d-2)$ little group. The scalar polarizations are the only modes we shall consider here, tensor and vector modes having previously been analyzed in \cite{He:2021jna}. The metric and gauge field including linearized perturbations, schematically are
\begin{equation}\label{eq:EMpert}
\begin{split}
ds^2 = ds_{(0)}^2 + ds_{(1)}^2 \,, \qquad \vb{A} = - a\, dv + \AGR_A\, dx^A \,, 
\end{split}
\end{equation}
with the subscript `$(0)$' referring to the background solution \eqref{eq:RNAdS}.

A priori there are seven metric components and three gauge potentials. Owing to diffeomorphism and gauge symmetry, not all of these ten functions are physical. By analyzing gauge invariants (which are suitable components of the curvatures) and the transformation under the time-reversal, charge conjugation $\mathbb{Z}_2$ involution, we can zero in onto a useful gauge choice. We employ the \emph{Debye gauge}, one where components that multiply derivatives of the $SO(d-2)$ scalar harmonic are set to zero. This leaves behind six functions, four in the metric and two in the gauge potential. Examining the component of the Einstein equation that transforms as a tensor, we find that one of the metric functions can be algebraically eliminated. The Maxwell constraint equation furthermore implies that the vector potential itself can be parameterized in terms of a single scalar field. All told, we can, with a suitable rescaling of functions by powers of the radial coordinate, write the perturbation ansatz as 
\begin{equation}\label{eq:EOWDeb}
\begin{split}
 ds_{(1)}^2
 &=  
 	\frac{\PHE-r f\,\PHW}{r^{d-3}}\, dv^2 
	 +\frac{2}{r^{d-1}f} \left(\PHO-\PHE +rf\,\PHW\right) dv\,dr
	 + r^2\, \frac{\PHW}{r^{d-2}}\, d\vb{x}^2 \\
& \qquad \qquad 
	-\frac{1}{r^{d+1}f^2}\left[2(\PHO-\PHE) + r f \,(d-1)\, \PHW\right] dr^2  \,,\\
\AGR_A\, dx^A 
&=
	 \frac{1}{r^{d-3}} \left(dv\, \Dz_+ -  dr \,\dv{r}\right) \MV	\,.
\end{split}
\end{equation}	

The above ansatz does not yet account for the momentum constraint equation. This can be solved by introducing a single field $\MW$ to parameterize $\PHE$ and $\PHO$. Consistency of the parameterization for metric fields says that $\MW$ and $\PHW$ are not independent, but can be expressed in terms of a single field $\MZ$. This parameterization takes the form:
\begin{equation}\label{eq:EOWMZ}
\begin{split}
\PHO
&= 
	-i\omega\, \MW\,, \qquad 
\PHE 
=   
	\Dz_+ \MW -2\, a'\,r^2f\, \MV \,,\\ \medskip
\MW 
&= 
	\frac{r}{\Lk}\left[\Dz_+  -\frac{r^2\, f'}{2}\right]   \MZ + \frac{2\,(d-1)\, r^3f\, a'}{\Lk}\, \MV \,, \\
\PHW 
&= 
	   \frac{1}{ \Lk}\left[r\,\Dz_+ + \frac{k^2}{d-1} \right]\, \MZ+ \frac{2\,(d-1)\, r^3f\, a'}{\Lk}\, 
	   \MV\,.
\end{split}
\end{equation}	

The admixture of the function $\MV$, which encodes the gauge potential in the above parameterization, originates from the coupling between the gravitational and gauge degrees of freedom. As such one can show that $\MV$ and  $\MZ$ obey a pair of coupled second order equations, which in turn imply all of the Einstein-Maxwell equations of motion. 

One can, however, do better.   It is possible to decouple the dynamics with a final change of variables. We introduce two functions $\Vd$ and $\Zd$, which are gravity duals of  the charge diffusion mode and the phonon mode, respectively, as
\begin{equation}\label{eq:MZVdiagonal}
\begin{split}
\MZ &= \frac{\Lk}{\bRQ\, h} \, \Vd + h\, \Zd \,, \\
\MV 
&=  
	\left(\frac{d-2}{\bRQ}\, a - r_+\, \frac{\BQT^2+2}{2}\right) \frac{\Vd}{2\, h} +\left( (d-2) a + \frac{r_+}{2} \, \bRQ\, \BQT^2\right) \frac{h}{2\, \Lk} \, \Zd \,.
\end{split}
\end{equation}
We have introduced in the above a modulation function 
\begin{equation}\label{eq:Lkdef}
\Lk =	k^2 + \frac{1}{2}(d-1)r^3 f' \,, 
\end{equation}	
and a  deformed momentum parameter\footnote{
	It is important to realize that $\BQT$ is defined with no approximations and satisfies 
	$\BQT^2 \left(2+\BQT^2\right) = \frac{2\,d\,\nu_s}{\bRQ^2}\, \bqt^2$. When we solve 
	the equations \eqref{eq:ZVDiagEom} we will expand $\BQT$ in powers of $\bqt^2$. However, while computing boundary data, especially the on-shell action and the boundary sources we will treat them exactly. There will be instances in the computation of the spatial part of the stress tensor where we guess that certain terms resum into $\BQT$. These we shall highlight; therefore, unless otherwise noted all such factors can be traced back to the diagonalization of the equations above. \label{fn:BQTprops}
} 
\begin{equation}\label{eq:diagDefs}
\begin{split}
\BQT^2 
=  \sqrt{1+2\,d\,\nu_s\, \frac{\bqt^2}{\bRQ^2}}-1\,, \qquad 
\bRQ 
= (d-2)\frac{\mu}{r_{+}\sdc}\,, \qquad 
\nu_s
= \frac{2(d-2)}{d(d-1)}\,. 
\end{split}
\end{equation}
The function $\Lk$ is the charged analog of the sound modulation function that was encountered in the analysis of the neutral plasma \cite{He:2022jnc}. The deformed momentum $\BQT$ arises from the decoupling of the metric and gauge field perturbations. We note here a  useful identity
\begin{equation}\label{eq:bRQPQreln} 
\frac{\Qbg}{\Pbg} = \frac{d\, (d-2)}{2\, \bRQ\, r_+} \,, 
\end{equation}	
which we will employ in the sequel to simplify various expressions. 

These two fields obey the following decoupled equations:
\begin{equation}\label{eq:ZVDiagEom}
\begin{split}
&
	r^{d-3}\,h^2\, \Dz_+\left(\frac{1}{r^{d-3}\,h^2}\, \Dz_+\Vd\right) + \left( \omega^2 -  k^2 f   +\VVd\right)  \Vd = 0 \,,\\ 
& 
	\frac{r^{d-3}\,\Lk^2}{h^2}\, \Dz_+\left(\frac{h^2}{r^{d-3}\,\Lk^2}\, \Dz_+\Zd\right) + \left( \omega^2 - \left(1-\frac{(d-2)}{2}\,\frac{2+\BQT^2}{1+\BQT^2}\,\frac{r^3 f'}{h\,\Lk}\right) k^2f + \VZd \right) \Zd =0 \,.
\end{split}
\end{equation}
The functions $\VVd$ and $\VZd$  appearing in the effective potentials are somewhat involved; they can be found in \eqref{eq:VdPot} and \eqref{eq:ZdPot}, respectively. These terms contribute only at $\order{k^2}$, i.e., they do not affect solutions in a boundary gradient expansion which solely depend on temporal variations. What is crucial for our discussion is that the equations are manifestly time-reversal invariant. In particular, this implies that it suffices to solve them with ingoing boundary conditions for $\Zd,\Vd$. The full grSK solution is then obtained by using time-reversal properties as elucidated in \cite{Jana:2020vyx}. The equations were originally obtained in \cite{Kodama:2003kk}; we give a clean presentation outlining various intermediate terms in \cref{sec:dynamicsderive}.

The fields $\Vd$ and $\Zd$ are generalizations of the designer scalars introduced in \cite{Ghosh:2020lel, He:2021jna, He:2022jnc}. They are both non-Markovian, with index $\ann = 3-d$, and their bulk dynamics is modulated by the dilatons:\footnote{
	The Markovianity index was  defined in \cite{Ghosh:2020lel} by the large $r$  behaviour of the dilaton, specifically, $\lim_{r\to \infty} e^{\chi} \to r^{\ann - d +1}$. In particular, that of a minimally coupled scalar is $\ann =d-1$. }
\begin{equation}\label{eq:dilZV}
\begin{split}
e^{\chi_{\Vd}} &	= \frac{1}{r^{2(d-2)}h^2}  \,, \\
e^{\chi_{\Zd}} &= \frac{h^2}{r^{2(d-2)}\Lk^2} \,. 
\end{split}
\end{equation}
The non-Markovianity of $\Vd$ is explicit (since $h\to 1$ asymptotically), while that for $\Zd$ is only valid for $k\neq0$.\footnote{ 
	Spatially homogeneous modes in the background can be understood by examining large diffeomorphisms and gauge transformations in the background \eqref{eq:RNAdS}, as explained in  \cite{He:2022jnc}. We expect the dynamics of gravitational zero modes be Markovian, similar to the zero momentum behaviour of a minimally coupled scalar field. }

The field $\Vd$ is the gravitational dual of the charge diffusion mode; in fact, taking $Q\to 0$ we recover the equation for a probe Maxwell field analyzed in \cite{Ghosh:2020lel}  (see their equation (8.14) with $\ann =3-d$). In that case the description corresponds to charge diffusion in a neutral plasma, whereas here we are interested in the dynamics of charge diffusion in a background with non-trivial charge density. This results in the effective dilaton having further charge modulation through the ohmic function $h$. This behaviour is analogous to what we encountered in our analysis of charge propagation; the transverse vector photons in the \RNAdS{d+1} background analyzed in \cite{He:2021jna} obey Markovian dynamics with a similar modulation by $h$. The designer field in that case was denoted $\MY_\ai$ and  had Markovianity index $\ann =d-3$. We will be able to use its solutions, with suitable analytical continuations, to write down the solutions for $\Vd$. 

The field $\Zd$ is the  gravitational dual of the sound mode of the plasma. Setting $Q\to 0$ we recover the equation for the analogous field analyzed in \cite{He:2022jnc} (denoted in that work as $\MZ$). The modulation function $\Lk$ has similar origins -- in fact, \eqref{eq:Lkdef} is identical when expressed in terms of the emblackening function $f$. 

The main novelty here is that charge diffusion interacts with sound propagation. With a judicious choice of variables they can be decoupled, as exemplified by the above equations. We will 
interpret these in terms of the conserved current components in due course. 

Plugging in the above ansatz and field redefinitions into the Einstein-Maxwell action and the associated Gibbons-Hawking boundary term, one can obtain a (decoupled) action for the fields $\Vd$ and $\Zd$.  While our field redefinitions involve higher derivatives (eg., $\PHE\sim \Dz_+^2 \Zd$), the action is second order in the designer fields and can be shown to be
\begin{equation}\label{eq:ZVDBulkmain}
\begin{split}
S[\Vd,\Zd]
&=
	- \int dr\, \int_k \, \frac{\sqrt{-g}}{8\, r^{2(d-1)}\, f} 
	\left[S_{_\Vd}^{\text{\tiny{bulk}}} + S_{_\Zd}^{\text{\tiny{bulk}}}\right] 	 + S_\text{bdy}[\Vd,\Zd] \,,\\
S_{_\Vd}^{\text{\tiny{bulk}}}
&=	\frac{8\,\Np{\Vd}(\BQT)}{h^2}\, k^2\, 
		 	\bigg[\left(\Dz_+\Vd\right)^2- \left( \omega^2 -  k^2 f   +\VVd\right)\Vd^2\bigg] \,, \\
 S_{_\Zd}^{\text{\tiny{bulk}}}
&= 
	 d\,\nu_s\, \Np{\Zd}(\BQT) \frac{k^2 \,h^2}{\Lk^2}\left[\left(\Dz_+\Zd\right)^2  - \left( \omega^2 - \left(1-\frac{(d-2)}{2}\,\frac{2+\BQT^2}{1+\BQT^2}\,\frac{r^3 f'}{h\,\Lk}\right) k^2f + \VZd\right)\Zd^2 \right] ,
\end{split}
\end{equation}
with\footnote{
	The normalization of $\Np{\Zd}(\BQT) $ has been chosen to ensure that it limits to $k^2$ as $Q \to 0$.}
\begin{equation}\label{eq:NCdef}
\begin{split}
\Np{\Vd}(\BQT) 
&= 
	\frac{r_+^2}{8} \left(1+\BQT^2\right) \left(2+\BQT^2\right) , \\
\Np{\Zd}(\BQT)
&= 
	 \frac{\bRQ^2\,r_+^2}{d\,\nu_s}\, \BQT^2 \left(1+\BQT^2\right)   \,.
\end{split}
\end{equation}
The complications of the field redefinitions are confined to the boundary term $S_\text{bdy}[\Vd,\Zd]$, which is an involved functional of $\{\Vd, \Zd, \Dz_+ \Vd, \Dz_+ \Zd, \Dz_+^2 \Vd, \Dz_+^2\Zd\}$. The structure, however, closely parallels the discussion of the designer field dynamics for the sound propagation in a neutral plasma described in \cite{He:2022jnc}.  As we demonstrate in \cref{sec:actionderive} the boundary term after various simplifications can be shown to be
\begin{equation}\label{eq:SbdyVZmain}
\begin{split}
S_{\text{bdy}}\left[\Vd, \Zd\right] 
&=
	\int_k \Bigg\{ 
			\frac{(d-6)\, (d-1)\,\Pbg}{8\, (d-2)} \left(\frac{\Dz_+\MW}{r^{d-1}}\right)^2  
  			-2 \,k^2\, \Np{\Vd}(\BQT)\, \Vd \, \PiV
  			-\frac{2\, \Np{\Zd}(\BQT)}{\Lk} \; \Zd\,  \PiZ + \cdots
  		\Bigg\} .
\end{split}
\end{equation}
The ellipses above indicate subleading terms which have a vanishing limit at the asymptotic boundary (as $r\to \infty$).  We have also expressed the final answer in terms of the background plasma pressure  $\Pbg$ defined in \eqref{eq:bgP}. 

The conjugate momenta for the  fields $\Vd$ and $\Zd$ are defined as
\begin{equation}\label{eq:PiD}
\PiV = -\frac{1}{r^{d-3}}\Dz_+\Vd \, ,
\qquad 
\PiZ = - \frac{d\,\nu_s}{8}\frac{1}{r^{d-3}}\Dz_+\Zd\,.
\end{equation}
These have finite limit asymptotically, unlike the fields which need to be renormalized. It is the  presence of the $\PiV\, \Vd$ and $\PiZ\, \Zd$ terms that dictates that these designer fields obey Neumann boundary conditions asymptotically at large $r$, viz., $\delta \PiV = \delta \PiZ =0$ in order for the variational principle to give the equations of motion \eqref{eq:ZVDiagEom}.  
We emphasize that this statement applies to the computation of correlation function (or generating functional thereof) using the standard AdS/CFT dictionary. 

We will however, be interested in computing a different object: a Wilsonian influence functional (WIF), in the grSK geometry, for which we will  impose different boundary conditions. This functional is a Legendre transform of the generating functional parameterized by field values or operators,  as opposed to sources in the generating functional. The Legendre transform effectively amounts to removing the $\PiV\, \Vd$ and $\PiZ\, \Zd$ boundary terms, and changes the boundary conditions to Dirichlet for these decoupled non-Markovian fields.

\section{Sound and charge diffusion in the grSK geometry}
\label{sec:ZSK}

With the dynamics for the Einstein-Maxwell system distilled into that of the designer scalars $\Vd$ and $\Zd$, we are now in a position to analyze the solutions on the grSK geometry. We wish to obtain the full solution for the two fields parameterized by their boundary values, i.e., by the expectation values of the dual boundary operator, on the two boundaries of the grSK geometry. Thankfully, the decoupled dynamics obeys time-reversal invariant equations of motion \eqref{eq:ZVDiagEom}. Therefore, as explained in  \cite{Jana:2020vyx},  it suffices for us to solve the equations with ingoing boundary conditions; we can then use time-reversal invariance to obtain the full grSK solution. 

With this in mind, we characterize the solution in terms of the ingoing boundary to bulk Green's functions for the two fields, $\Gin{\Vd}$ and $\Gin{\Zd}$. We parameterize these in a boundary gradient expansion as:
\begin{equation}\label{eq:Ginexp}
\Gin{\Vd,\Zd}(r,\bwt, \bqt)  = 
	\exp\left(\sum_{n,m=1}^\infty \,  (-i)^m \,\bwt^m\, \bqt^{2n}\, \Mser{\Vd,\Zd}{m,2n}(r) \right) .
\end{equation}	
We note that we have normalized the ingoing Green's functions to be unity at the boundary and are demanding regularity at the outer horizon.   The functions $\Mser{\Vd,\Zd}{m,2n}(r)$ can be obtained by solving the equations of motion recursively. In fact, the natural way to parameterize them is using the basis of functions discussed in the context of tensor and vector perturbations in \cite{He:2021jna}.\footnote{
	The solutions we are constructing are linearized versions (in amplitude) of  the fluid/gravity solutions \cite{Banerjee:2008th,Erdmenger:2008rm,Plewa:2012vt} (all of who studied a four dimensional plasma). The operative point here is that those solutions are  characterized  using $SO(d-1)$ representations, which is natural if one wants to parameterize the currents by intensive thermodynamic quantities (temperature and chemical potential) and hydrodynamic velocities. We, however, employ $SO(d-2)$ representations, to disambiguate different physical modes of the plasma. Nevertheless, the solutions in different sectors we study must have similar origins to be assembled into complete $SO(d-1)$ irreps, which can then be checked explicitly.} 
We demonstrate this explicitly in \cref{sec:gradexpfns} and construct the solutions to quartic order in the boundary gradients.  For now, it suffices that there exists a clean function basis to work in, and that one can extract the physical information such as asymptotics therefrom.

\subsection{The grSK solution for the designer fields}
\label{sec:ZVgrSK}

As explained above, we want to parameterize the grSK solution by the boundary values of the 
fields themselves, i.e., the boundary data are the expectation values of the dual operators, viz.,
\begin{equation}\label{eq:ZVvevs}
 \expval{\left(\OpV(\omega, \bk)\right)_{\skL/\skR} }
 	= \PoV_{\skL/\skR}(\omega, \bk)\,,
 \qquad
\expval{\left(\OpZ(\omega, \bk)\right)_{\skL/\skR} }
	= \PoZ_{\skL/\skR}(\omega, \bk)\,, 
\end{equation}
where
\begin{equation}\label{eq:ZVvevdefs}
\begin{split}
\PoV_{\skL/\skR}(\omega, \bk) &= \lim_{r\rightarrow \infty\pm i0} \left[\Vd + \text{counterterms}\right] ,\\
\PoZ_{\skL/\skR}(\omega, \bk) &= \lim_{r\rightarrow \infty \pm i0} \left[\Zd + \text{counterterms}\right] .
\end{split}
\end{equation}
In terms of this data, the full grSK solution is given using the ingoing Green's function \eqref{eq:Ginexp} as\footnote{
		The average/difference basis is defined by a Keldysh rotation: 
		$\phi_a = \frac{1}{2}\left(\phi_\skR +\phi_\skL \right)$ and $\phi_d = \phi_\skR-\phi_\skL $.  }
\begin{equation}\label{eq:ZVDSK}
\begin{split}
\Vd^{\SKs}(\zeta,\omega,\bk) 
&= 
	\Gin{\Vd} \,  \PoV_a + \left[\left(\nB+\frac{1}{2}\right) \Gin{\Vd} \, - \nB \,e^{\beta \omega(1-\zeta)} \, \Grev{\Vd} \, \right]\PoV_{d}\,, \\
\Zd^{\SKs}(\zeta,\omega,\bk) 
&= 
	\Gin{\Zd} \, \PoZ_a + \left[\left(\nB+\frac{1}{2}\right) \Gin{\Zd}- \nB\,  e^{\beta \omega(1-\zeta)} \, \Grev{\Zd}\right]\PoZ_{d}\,,
\end{split}
\end{equation}
where $\Grev{\Vd,\Zd} (r,\omega,\bk) = \Gin{\Vd,\Zd}(r,-\omega,\bk)$, and $\nB$ is the Bose-Einstein distribution:
\begin{equation}\label{eq:nBdef}
\nB = \frac{1}{e^{\beta\omega}-1}\,.
\end{equation}

The boundary sources for the fields $\Vd$ and $\Zd$, which we denote as 
$\JoV$ and $\JoZ$, respectively, are in fact the asymptotic values of the conjugate momenta, which as discussed in \cref{sec:dynamics} have finite limits. We define
\begin{equation}\label{eq:sources}
\lim_{r\rightarrow \infty\pm i0} \PiV  \equiv - \JoV_{\skL/\skR}\,,
\qquad 
\lim_{r\rightarrow \infty\pm i0} \PiZ \equiv -k^2\, \JoZ_{\skL/\skR}\,.
\end{equation}
Using the grSK solution \eqref{eq:ZVDSK}, together with the asymptotic behaviour of the functions $\Mser{\Vd,\Zd}{m,2n}$ worked in \cref{sec:ZWasym}, we can determine the relation between the sources and the expectation values. We will give the expressions below once we write down the boundary influence functional.

\subsection{The Wilsonian influence functional of designer fields}
\label{sec:ZVWIF}

We now have all the data necessary to evaluate the Wilsonian influence functional for charge diffusion and sound propagation from the designer fields. Evaluating the on-shell action on the grSK solution, and implementing the Legendre transform to parameterize it in terms of the boundary operator expectation values, we find two contributions: a Schwinger-Keldysh factorized contact term  and an influence functional. To wit, 
\begin{equation}\label{eq:SVZall}
\frac{1}{c_\text{eff}} \, S\left[\PoV,\PoZ\right] 
	= S_{\text{contact}}\left[\PoV, \PoZ\right] + S_{\text{WIF}}\left[\PoV,\PoZ\right] .
\end{equation}

Let us first examine the influence functional contribution; it is given by
\begin{equation}\label{eq:WIF}
\begin{split}
S_{\text{WIF}}\left[\PoV, \PoZ\right] 
&= 
	-\int_k \,  k^2 \left[
	 \Np{\Vd}(\BQT)\, \PoV_d^\dagger\,  \Kin{\Vd} \left(\PoV_a + \left(\nB+\frac{1}{2}\right)\PoV_d   \right) + \text{cc} \right]\\
&\qquad
	-\int_k   \left[ \Np{\Zd}(\BQT)\, \PoZ_d^\dagger \, \Kin{\Zd} \left(\PoZ_a + \left(\nB+\frac{1}{2}\right)\PoZ_d\right) +\text{cc} \right] .
\end{split}
\end{equation}
We have introduced here the inverse  boundary Green's functions for the fields $\Kin{\Vd, \Zd}(\omega, \bk) $ and its time-reversed counterpart $\Krev{\Vd,\Zd}(\omega,\bk) = \Kin{\Vd, \Zd}(-\omega, \bk) $. These are themselves given by the dispersion functions, up to some dimension and charge dependent factors as
\begin{equation}\label{eq:DispKer}
\begin{split}
\Kin{\Vd}(\omega,\bk)
&= 
	\frac{\Kc(\omega,\bk)}{(1-\sdc)^2\,r_+^{d-4}} 
		= \frac{r_+}{\sigma_\text{dc}}\,  \Kc(\omega,\bk)\\
\Kin{\Zd}(\omega,\bk)
&=
	 \frac{\KS(\omega,\bk)}{2\,d\,(d-1)^2\,r_+^{d-2}\, (1+Q^2)}
	= \frac{r_+^2}{2\,d\,(d-1)^2\, \Pbg}\, \KS(\omega,\bk)  \,, 
\end{split}
\end{equation}
where we have employed the background data \eqref{eq:bgPQs}.

The function $\Kc(\omega,\bk)$ characterizes the dispersion locus for charge diffusion. It is given in terms of the horizon values of the gradient expansion functions characterizing the solution\footnote{
	 Functions with subscript $\MY$ and $d-1$ correspond to the basis used to solve for the Markovian transverse vector photon, and transverse tensor graviton polarizations, respectively, and were introduced in \cite{He:2021jna}. For completeness, we give integral expressions and asymptotic expansions of these functions in \cref{sec:gradexpfns}.} 
for the field $\Vd$
\begin{equation}\label{eq:Kc3}
\begin{split}
\Kc(\bwt,\bqt) 
&= 	
		-i  \bwt +\left(1-\frac{\sdc}{d-1}\right)\frac{\bqt^2}{d-2} - \Dfn{\MYd}{2,0}(r_+)\,   \bwt^2
		-2i\, \Mser{\Vd}{0,2}(r_+)\, \bwt\, \bqt^2 \\
&\qquad \qquad 
		- i\left(2\,\Mser{\MYd}{2,0}(r_+)-\Dfn{\MYd}{2,0}(r_+)^2\right) \bwt^3 	 + \cdots\,.
\end{split}
\end{equation}	
The reader can find the expression accurate to quartic order in \eqref{eq:Kc4}. 

Similarly, the function $\KS(\omega,\bk)$ characterizes the dispersion locus for sound propagation with attenuation. Once again it can be expressed in the boundary gradient expansion, and takes the form:
\begin{equation}\label{eq:KS2}
\begin{split}
\KS(\bwt,\bqt) 
&= 
	-\bwt^2 + \frac{\bqt^2}{d-1} +  \frac{\nu_s}{1+Q^2} \,\bqt^2  \, \Gatt(\bwt,\bqt) \,, \\
\Gatt(\bwt,\bqt) 
&=  
	-i \bwt + \Dfn{d-1}{2,0}(r_+) \,\bwt^2 + \frac{d-3}{(d-1)(d-2)}\bqt^2 + \frac{\sdc}{d-2}\left(\bwt^2-\frac{\bqt^2}{d-1}\right) + \cdots \,.
\end{split}
\end{equation}
We have quoted here the sound attenuation function $\Gatt$ accurate to quadratic order (and thus the dispersion locus to quartic order). It is possible to extract a prediction for this function to quartic order in gradients (even though we are only solving the equations accurately to that order) as we explain in \cref{sec:gradexpfns}. We give our prediction for this function in \cref{sec:ZWasym}; see \eqref{eq:OstDef} and the discussion surrounding it.

From $S_{\text{WIF}}\left[\PoV, \PoZ\right]$ given in \eqref{eq:WIF}, we can determine the boundary Schwinger-Keldysh sources corresponding to the designer fields to be
\begin{equation}\label{eq:SVeVRel}
\begin{split}
\JoV_a 
&= 
	\Kin{\Vd} \, \PoV_a + \left(\nB+\frac{1}{2}\right) \left(\Kin{\Vd} - \Krev{\Vd}\right)\PoV_d\,, 
	\qquad \;
	\JoV_d = \Krev{\Vd} \, \PoV_d\,, \\
\JoZ_a 
&= 
	\Kin{\Zd} \, \PoZ_a + \left(\nB+\frac{1}{2}\right) \left(\Kin{\Zd}-\Krev{\Zd}\right)\PoZ_d\,,
	\qquad 
	\JoZ_d =\Krev{\Zd}\,\PoZ_d\,.
\end{split}
\end{equation}
One can check that this relation is consistent with our identification \eqref{eq:sources}, which employed the standard asymptotic fall-off conditions to extract the sources.

Before turning to the contact term contribution, it is worth noting that we recover pre-existing results when we switch off the charge $Q\to 0$. The charge diffusion function $\Vd$ simplifies considerably, and reduces to a non-Markovian probe Maxwell field of index $\ann =3-d$ analyzed in \cite{Ghosh:2020lel} (see their Section 8; the field $\Vd$ reduces to $\sen{3-d}$ studied there). Likewise,  the designer phonon field $\Zd$ limits to the sound field $\MZ$ analyzed in \cite{He:2022jnc} (compare with their Eq.~(5.2)). The normalization for $\Np{\Zd}(\BQT)$ specified below \eqref{eq:NCdef} ensures that this works out naturally, modulo the fact that the prefactor is given here in terms of the dimensionless momentum $\bqt^2$.  Moreover, in the charged plasma, the normalization factors $\Np{\Vd}(\BQT)$ and $\Np{\Zd}(\BQT)$ are closely related to those obtained in \cite{He:2021jna} while decoupling the vector perturbations. This is not altogether surprising given the unified origin of the perturbations from a fluid/gravity perspective, but serves as simple sanity check of our results.

To write the contact term contribution it will be necessary to record the boundary metric and gauge field source, parameterized by the sources for the fields $\Vd$ and $\Zd$. This information can be extracted using the asymptotic behaviour of the designer fields, as we explain in  \cref{sec:ZWasym}. We introduce the induced boundary chemical potential
\begin{equation}\label{eq:CbdyLR}
\begin{split}
\mathbf{C}_{\skL/\skR}
&=
	 \lim_{r\to \infty \pm i0} \, \AGR_v \\
&
=	-  r_+ \,  \frac{2+\BQT^2}{4}\, \JoV_{\skL/\skR} + \frac{d\,(d-1)}{2}\, \frac{\Pbg}{\Qbg} \,  \BQT^2  \, \JoZ_{\skL/\skR} \,,
\end{split}
\end{equation}	
and the boundary metric  induced on the two boundaries of the grSK geometry 
\begin{equation}\label{eq:gammaLR}
\begin{split}
(\gamma_{\skL/\skR})_{\mu\nu}
&=
	 \lim_{r\to\infty \pm i0} \left[ \frac{\Dz_+\MW}{r^{d-1} }\, dv^2   +\left(1+ \frac{\PHW}{r^{d-2}}\right) \, \eta_{\mu\nu}\, dx^\mu\, dx^\nu \right]  \\
&=
	-\left(1- (d-3) \, \mathbf{W}_{\skL/\skR} \right) dv^2 + \left(1+\mathbf{W}_{\skL/\skR}\right) d\vb{x}^2\,.
\end{split}
\end{equation}	
We have parameterized the above in terms of the boundary value of the metric fields $\PHW$ and $\Dz_+ \MW$  using the asymptotic relation $\PHE = \Dz_+ \MW + \cdots$, and 
\begin{equation}\label{eq:WbdyLR}
\begin{split}
\mathbf{W}_{\skL/\skR}
&=  
	\frac{1}{d-2}\, \lim_{r\to \infty \pm i0} \, \frac{\Dz_+ \MW}{r^{d-1}}
	=  \lim_{r\to \infty \pm i0} \, \frac{\PHW}{r^{d-2}} \\\
& =
	 \frac{\JoV_{\skL/\skR}  }{\bRQ} 
           + \, \frac{8}{d\,\nu_s}\, \JoZ_{\skL/\skR} \,.
\end{split}
\end{equation}

With the boundary sources identified, we can write down the contact term 
 \begin{equation}\label{eq:Scontact}
\begin{split}
\frac{1}{c_\text{eff}} \, S_{\text{contact}}\left[\PoV, \PoZ\right] 
&= 
	 \Pbg  + \Pbg \, \int_k \frac{(d-1)\,(d-2)}{2} \left(\mathbf{W}_\skR- \mathbf{W}_\skL\right) 
	 + 2\,\Qbg\, \int_k \left(\mathbf{C}_\skR - \mathbf{C}_\skL\right) \\
&\qquad
	+ \frac{(d-1)\, (d-2) \,(d-6)}{8} \,\Pbg\,  \int_k
	 \left(\mathbf{W}_\skR^\dag\, \mathbf{W}_\skR
	 	- \mathbf{W}_\skL^\dag\, \mathbf{W}_\skL 
	 	\right) .
\end{split}
 \end{equation}
This contact term includes the background free energy density, which is proportional to the pressure $P$ introduced in \eqref{eq:bgP}. This entire contact term originates from the  on-shell ideal fluid action evaluated on the background \eqref{eq:gammaLR}. The ideal fluid additionally contributes to part of the Wilsonian influence phase.  The details of this calculation are similar to the analogous analysis in \cite{He:2022jnc}. We will add some further commentary in \cref{sec:hydrocurrents}, but encourage the reader to consult the aforementioned reference for a more complete discussion. The main change is that we have mixing between the phonon mode and the charge diffusion mode, but once we have diagonalized the modes, at Gaussian order, we have decoupled effective dynamics, which allows us to infer the result directly from the boundary metric deformation.

\subsection{Conserved currents}
\label{sec:currents}

The conserved charge current and energy-momentum tensor  can be extracted from our solution using standard techniques. We sketch the basic derivation of these results in \cref{sec:JBYT}. There are two contributions that are relevant: the background \RNAdS{d+1} black hole has a non-vanishing free energy and charge current, which is corrected at linear order in amplitudes by contributions which can be expressed in terms of the field operators $\PoV$ and $\PoZ$. 

\paragraph{Charge current:} To express the  $U(1)$ conserved current on the boundary, let us first introduce the boundary charge diffusion operator
\begin{equation}\label{eq:ChOp}
\QOp_{\skL/\skR}
= \left[\frac{r_+}{4}\, (2+\BQT^2)\, \OpV - \frac{2\,\Qbg}{d\,(d-1)\,\Pbg}\, \frac{1}{2+\BQT^2}\, \OpZ\right]_{\skL/\skR} \,.
\end{equation}	
This is a particular linear combination of the operators $\OpV$ and $\OpZ$, which are the boundary duals of the bulk fields $\Vd$ and $\Zd$, respectively. It can be identified as  
the renormalized boundary operator dual to the field $\MV$ which parameterizes the bulk  gauge field fluctuations. The factors of $\BQT$ in its definition originate from the diagonalization of the bulk equation, and can be traced back to \eqref{eq:MZVdiagonal}. 

In terms of this operator we have a simple expression for the current, 
\begin{equation}\label{eq:JLRcurrent}
\begin{split}
\frac{1}{2\,c_\text{eff}} \; (\Jcft_\mu )_{\skL/\skR} \,dx^\mu
&= 
	(J_\mu^\text{contact})_{\skL/\skR} \, dx^\mu
	+  \int_k\left( k^2 \, \ScS \, dv + i\omega\, k\,\ScS_i\, dx^i\right) \; \QOp_{\skL/\skR} \,,\\
(J_\mu^\text{contact})_{\skL/\skR}  \, dx^\mu
&= 	
	\Qbg \left[ -1+ (d-2)\, \int_k\, \ScS\, \mathbf{W} \right]_{\skL/\skR} dv\,.
\end{split}
\end{equation}	
We have written the current in terms of the background charge density \eqref{eq:rho0def}.

The operator content is captured by the specific combination of the bulk charge diffusion and sound modes introduced above.   Once we identify the energy operator, we will be able to re-express the operators dual to the decoupled gravitational designer fields in terms of the physical boundary operators.  The second line includes the background charge density, and an additional polarization contribution arising from the change of the boundary conformal frame (the $\mathbf{W}$ piece). 

Current conservation demands that $\tensor[^\gamma]{\nabla}{_\mu} J^\mu =0 $ on both the left and right boundaries of the grSK geometry. If we examine the operator part of this statement, viz., the contribution from $\QOp$, we find that the  Ward identity is trivially satisfied, since $J^v \propto \, -k^2 \, \ScS$ and $J^i \propto  - \omega\, k_i\, \ScS$.  It is straightforward to check that the source term proportional to $\mathbf{W}$ is also covariantly conserved (it cancels against the connection term for the background \eqref{eq:gammaLR}). As a result we should view the current as a sum of a polarization  term on the deformed boundary  geometry, which is the second line of \eqref{eq:JLRcurrent}, and a physical operator part, proportional to $\QOp$, as we have indicated.

\paragraph{Energy-momentum tensor:} The boundary stress tensor is computed by the usual Brown-York result supplemented with suitable counterterms. We give the details of how to write this in terms of fields we are using to parameterize the metric in \cref{sec:JBYT}. The result after some simplifications can be expressed in terms of an energy operator:
\begin{equation}\label{eq:EnOp}
\EOp_{\skL/\skR} 
\equiv 
	\left[ \OpZ + k^2  \,\frac{\OpV }{\bRQ} \right]_{\skL/\skR}  \,.
\end{equation}	
Once again, this operator is the boundary value of the renormalized field $\MZ$, which is the distillation of gravitational perturbations. One infers the relation above directly from the definition \eqref{eq:MZVdiagonal}, with the factor of $k^2$ originating from the modulation function $\Lk$. 

In terms of this field, we can give a simple expression for the energy-momentum tensor 
density.\footnote{
	Working with the densities leads to simpler expressions and is also more natural from the computation using the renormalized Brown-York tensor. Note that, $\sqrt{-\gamma} = 1 + \int_k\, \mathbf{W}\, \ScS$.
} We decompose it again into separate contact and operator contributions:
\begin{equation}\label{eq:Tmnfull}
\left(\TcftU^{\mu\nu}\right)_{\skL/\skR}
=  
	2\,  \left[\fdv{S}{\gamma_{\mu\nu}}\right]_{\skL/\skR}
=
	c_\text{eff}\,  \left(T_\text{contact}^{\mu\nu}\right)_{\skL/\skR} +
	c_\text{eff}\, \widehat{T}^{\mu\nu}_{\skL/\skR} \,.
\end{equation}	
The physical operator part of the stress tensor is quite simple and can be shown to be
\begin{equation}\label{eq:TmnLR}
\begin{split}
\left(\widehat{T}\indices{_v^v}\right)_{\skL/\skR}
&=
	 -   \int_k\, \frac{k^2}{d-1}  \, \ScS \; \EOp_{\skL/\skR}   \,, \\
\left(\widehat{T}\indices{_v^i}\right)_{\skL/\skR}
&= 
	i\, \int_k\, \frac{\omega\, k}{d-1}\,\ScS^i  \; 
		\EOp_{\skL/\skR}  \,,\\
\left(\widehat{T}\indices{_i^j}\right)_{\skL/\skR}
&=
	\frac{1}{d-1} \int_k\, \left[ \frac{k^2}{d-1}\, \ScS\, 	 \delta\indices{_i^j}
	+\frac{d-1}{d-2}\, \left( \frac{k^2}{d-1} - \omega^2 \right) (\ScST)\indices{_i^j}  \right] \EOp_{\skL/\skR} \,.
\end{split}
\end{equation}
We have decomposed the spatial part of the stress tensor into a diagonal `pressure' and a traceless `shear-stress' part, utilizing the invariant Kronecker delta and the traceless derived scalar harmonic.  The contact term is obtained as
\begin{equation}\label{eq:Tcontact}
\begin{split}
\left(T^\text{contact}_{\skL/\skR}\right)\indices{_v^v}
&= 
	  (d-1)\, \Pbg \left[-1 +  \frac{d-2}{2}\int_k\, \ScS\, \mathbf{W}\right]_{\skL/\skR} \,, 
	 \\
\left(T^\text{contact}_{\skL/\skR}\right)\indices{_i^j}
&=
		- \frac{1}{d-1} \, \left(T^\text{contact}_{\skL/\skR}\right)\indices{_v^v}
	-  \frac{d\, (d-1)}{2}\, \Pbg\, \int_k\, (\ScST)\indices{_i^j}  
		\left[\mathbf{W}  + \frac{4}{d\,(d-2)} \, \frac{\Qbg}{\Pbg} \, \mathbf{C} \right]_{\skL/\skR} \,.	  
\end{split}
\end{equation}
We notice that the stress tensor is manifestly traceless (note that $(\ScST)_i^i =0$). Furthermore, the operator part and the contact terms are separately  conserved.  Note that since we have a background charge density, the conservation law includes a Joule heating term\footnote{
	Note that we have a bulk Maxwell term with unconventional normalization. The boundary charge current defined in \eqref{eq:JLRcurrent} consequently has an extra factor of $2$, as indicated in \cref{fn:J2ceff}. This ensures that the Joule heating term has the conventional normalization, consistent with what one would derive from a diffeomorphism Ward identity. \label{fn:Jouleheat}
}
\begin{equation}\label{eq:TJcons}
\tensor[^\gamma]{\nabla}{_\mu} \TcftU^{\mu\nu} =   \tensor[^\gamma]{F}{^{\nu\rho}} \, \Jcft_\rho\,.
\end{equation}	
Since we have a non-trivial boundary chemical potential, $\tensor[^\gamma]{A}{_\mu} \neq 0 $,  from \eqref{eq:CbdyLR}, it follows that there is a non-trivial boundary field strength $\tensor[^\gamma]{F}{_{\mu\nu}}$ which contributes to r.h.s.\ of the conservation equation 
\eqref{eq:TJcons}.  Nevertheless, it turns out that the contact terms in the currents conspire to cancel each other out, effectively implying that the $\ScST_{ij}$ component of the contact term is the gravitational response to the background charge density. We will return to analyzing the currents  from a hydrodynamic viewpoint in \cref{sec:hydrocurrents}.
 
\subsection{Current correlators}
\label{sec:corrTJ}

For the purposes of computing physical correlation functions, it will be useful to record the Green's functions for the boundary charge diffusion and phonon operators, $\OpV$ and $\OpZ$, respectively. From the influence functional \eqref{eq:WIF} we infer that 
\begin{equation}\label{eq:VZretBare}
\begin{split}
\expval{\OpV(-\omega,-\bk) \, \OpV(\omega,\bk)}^\text{Ret} 
&=  
  - i 
  \,  \frac{1}{c_\text{eff}\, k^2\, \Np{\Vd}(\BQT)\, \Kin{\Vd}(\omega,\bk)}  \,, \\
\expval{\OpZ(-\omega,-\bk) \, \OpZ(\omega,\bk)}^\text{Ret} 
&= 
  - i 
  \,  \frac{1}{c_\text{eff}\, \Np{\Zd}(\BQT)\, \Kin{\Zd}(\omega,\bk)}  \,, \\
\end{split}
\end{equation}
The Keldysh Green's function  can be determined by the fluctuation-dissipation relation 
\begin{equation}\label{eq:VZkelBare}
\begin{split}
\expval{\OpV(-\omega,-\bk) \, \OpV(\omega,\bk)}^\text{Kel} 
= 
 -  \frac{1}{2\,c_\text{eff}}\,\coth\left(\frac{\beta\omega}{2}\right) \frac{\Im\left[\Kin{\Vd}(\omega,\bk)\right]}{\abs{\Kin{\Vd}(\omega,\bk)}^2} \,,\\
 \expval{\OpZ(-\omega,-\bk) \, \OpZ(\omega,\bk)}^\text{Kel} 
= 
 -  \frac{1}{2\,c_\text{eff}}\,\coth\left(\frac{\beta\omega}{2}\right) \frac{\Im\left[\Kin{\Zd}(\omega,\bk)\right]}{\abs{\Kin{\Zd}(\omega,\bk)}^2} \,.\\
 \end{split}
\end{equation}	
We will therefore only quote the retarded Green's functions below. 

To write the current correlators we  first pick a spatial direction for the propagating modes, setting $\bk = k\, \hat{x}$, decomposing $\mathbb{R}^{d-1}$ coordinates into $\{x,x_s\}$ with $s=2,\cdots,d-1$. One can check that with this choice $\ScST_{ij} = \frac{1}{d-1}\,\text{diag}\{-(d-2),1,\cdots, 1\}$.  We will write the correlation functions of the currents in terms of the corresponding answers for the 
operators $\QOp$ and $\EOp$, so let us therefore first record the retarded Green's functions for these charge diffusion and energy operators. From \eqref{eq:ChOp} and \eqref{eq:EnOp} we find
\begingroup
\allowdisplaybreaks
\begin{subequations}
\begin{align}
\hspace{-2cm}
i\, c_\text{eff}\, \expval{\QOp(-\omega,-\bk) \, \QOp(\omega,\bk)}^\text{Ret} 
&=
	\frac{\sigma_\text{dc}}{2\,r_+}\,
		\frac{2+\BQT^2}{1+\BQT^2} \, \frac{1}{k^2\, \Kc(\bwt,\bqt)} 
		\cr
&\qquad	
	+  (d-2)\,\frac{ r_+^{d-2}\, \sdc}{4\,\Np{\Vd}(\BQT)}\, \frac{1}{k^2\, \KS(\bwt,\bqt)}  \,,
\end{align}
\begin{align}\label{eq:EEc}	
i\,c_\text{eff}\, \expval{\EOp(-\omega,-\bk) \, \EOp(\omega,\bk)}^\text{Ret} 
&=
	\sigma_\text{dc} \, r_+\, \left(\frac{2\Qbg}{d(d-2)\, \Pbg}\right)^2
		\frac{1}{\Np{\Vd}(\BQT)} \, \frac{k^2}{\Kc(\bwt,\bqt)} \cr
&\qquad		 
	+ d\,(d-1)^2\,\frac{\Pbg}{r_+^2} \, \frac{2+\BQT^2}{1+\BQT^2}\, \frac{1}{k^2\, \KS(\bwt,\bqt)} \,,
\end{align}
\begin{align}\label{eq:QEc}	
i\,c_\text{eff}\, \expval{\QOp(-\omega,-\bk) \, \EOp(\omega,\bk)}^\text{Ret} 
&=
	\frac{2\,\sigma_\text{dc}}{r_+} \, \frac{2\Qbg}{d(d-2)\, \Pbg}\, 
		\frac{1}{(1+\BQT^2)} \, \frac{1}{\Kc(\bwt,\bqt)} \cr
& \qquad 		
	-\frac{2\,(d-1)\, \Qbg}{r_+^2\, (1+\BQT^2)}\,  \frac{1}{k^2\, \KS(\bwt,\bqt)} 
	\,. 	 
\end{align}
\label{eq:QEcorr}
\end{subequations}
\endgroup
In these equations  we have used \eqref{eq:DispKer} to convert $\Kin{\Vd}$ and $\Kin{\Zd}$ to the dispersion functions $\Kc$ and $\KS$, respectively.

The correlation functions of the $U(1)$ current and stress tensor are then compactly presented in terms of these correlators as 
\begin{equation}\label{eq:TJcorr}
\begin{split}
\expval{ J_{_\text{CFT}}^\mu (-\omega,-\bk) \, J_{_\text{CFT}}^\nu (\omega,\bk)}^\text{Ret}_{_\text{non-ideal}}
&= 
		4\,c_\text{eff}^2\, \mathfrak{J}^{\mu\nu}(\omega,k)\, \expval{\QOp(-\omega,-\bk) \, \QOp(\omega,\bk)}^\text{Ret}  \,, \\
\expval{ T_{_\text{CFT}}^{\mu\nu} (-\omega,-\bk) \, T_{_\text{CFT}}^{\rho\sigma}\nu (\omega,\bk)}^\text{Ret}_{_\text{non-ideal}}
&= 
	c_\text{eff}^2\, \frac{\mathfrak{G}^{\mu\nu,\rho\sigma}(\omega,k)}{(d-1)^2}\, \expval{\EOp(-\omega,-\bk) \, \EOp(\omega,\bk)}^\text{Ret}  \,, \\
\expval{ J_{_\text{CFT}}^{\mu} (-\omega,-\bk) \, T_{_\text{CFT}}^{\rho\sigma}\nu (\omega,\bk)}^\text{Ret}_{_\text{non-ideal}}
&= 
	2\,c_\text{eff}^2\, \frac{\mathfrak{K}^{\mu,\rho\sigma}(\omega,k)}{d-1}\, \expval{\QOp(-\omega,-\bk) \, \EOp(\omega,\bk)}^\text{Ret}	\,,
\end{split}	
\end{equation}	
where we have parameterized the tensor structure data into certain momentum and frequency dressings.

For the current-current correlators these are 
\begin{equation}\label{eq:JJtensor}
\mathfrak{J}^{vv}(\omega,k) = k^4\,, \qquad 
\mathfrak{J}^{vx}(\omega,k) = \omega\, k^3\,\,,  \qquad 
\mathfrak{J}^{xx}(\omega,k) = \omega^2\, k^2\,\,, \qquad 
\mathfrak{J}^{s\mu}(\omega,k) = 0\,.
\end{equation}	
The naive pole at $k=0$ from \eqref{eq:QEcorr} is cancelled by the factor of $k^2$ in this dressing tensor. 
Likewise, we parameterize the energy-momentum tensor  correlators using
\begin{equation}\label{eq:GGtensor}
\begin{aligned}
\mathfrak{G}^{vv,vv} &=	k^4 \,,  \qquad  \quad\;  
\mathfrak{G}^{vv,vx} =	\omega\, k^3 \,, \qquad\;
\mathfrak{G}^{vv,ss} =  \frac{k^2}{d-2}  \left(k^2-\omega^2 \right), \\ 
\mathfrak{G}^{vv,xx} &= \omega^2\,k^2 \,, \qquad
\mathfrak{G}^{vx,vx} =	\omega^2\, k^2 , \qquad
\mathfrak{G}^{vx,ss} =	 \frac{\omega\,k}{d-2} \left(k^2-\omega^2 \right), \\
\mathfrak{G}^{vx,xx} &=	\omega^3\,k \,, \qquad\;\;
\mathfrak{G}^{xx,xx} =    \omega^4 \,, \qquad\;\;\;  
\mathfrak{G}^{xx,ss} =  \frac{\omega^2}{d-2}  \left(k^2-\omega^2 \right), \\
& \qquad \qquad \qquad 
	\mathfrak{G}^{ss,ss} =    \frac{1}{(d-2)^2} \left(k^2-\omega^2 \right)^2.   
\end{aligned}
\end{equation}	
Finally,  the mixed current-stress tensor correlation functions depend on 
\begin{equation}\label{eq:KKtensor}
\begin{split}
\mathfrak{K}^{v,vv} 
&= 
	- k^4 \,, 
\qquad
	\mathfrak{K}^{v,vx} 
= 
	-\omega\, k^3\,,
\qquad
\mathfrak{K}^{v,xx} 
=
	-\omega^2\, k^2 \,,
\qquad
\mathfrak{K}^{v,ss} 
=
	-\frac{k^2}{d-2}\left(k^2-\omega^2\right)\,,	 \\ 
\mathfrak{K}^{x,vv} 
&= 
	-\omega\,k^3 \,, 
\quad\;
	\mathfrak{K}^{x,vx} 
= 
	-\omega^2\, k^2\,,
\quad\,\
\mathfrak{K}^{x,xx} 
=
	-\omega^3\, k\,,
\qquad\;
\mathfrak{K}^{x,ss} 
=
	-\frac{\omega\,k}{d-2}\left(k^2-\omega^2\right)\,,	 \\ 	
\end{split}
\end{equation}
with $\mathfrak{K}^{s,\mu\nu} =0$.

In writing these expressions, we have made a particular choice of the contact terms.  While this writing appears to suggest the presence of a double pole at $k=0$, stemming from the $k^2\, \KS$ factors in \eqref{eq:QEcorr}, these are spurious. They can be removed by adding suitable contact terms. The only physical poles are at the vanishing loci of $\KS$ and $\Kc$, which correspond to the dispersion relations for sound and charge diffusion. By adding suitable contact terms we can replace of $\omega^2 \to k^2\left(\frac{1}{d-1} + \frac{\nu_s}{1+Q^2}\, \Gatt\right)$  in $\mathfrak{G}^{\mu\nu,\rho\sigma}$ and $\mathfrak{K}^{\mu, \rho\sigma}$.\footnote{
	In \cite{Policastro:2002tn,He:2022jnc}  such a choice was made to simplify the stress tensor correlators of a neutral fluid.}  
This manipulation does not affect the residue at the sound pole, which is the physical data, one can  extract from these results.  We have refrained from implementing such contact terms  explicitly to retain the simplicity of the expressions above. 

\subsection{A fluid dynamical perspective}
\label{sec:hydrocurrents}

We have motivated our analysis by arguing that the low-lying modes in scalar sector of the charged plasma are the  phonon and a charge diffusion mode.  With the correlation functions at hand, we can see that they have poles at the dispersion loci characterizing these modes. We can obtain the explicit dispersion relations and confirm that they can be understood directly in terms of the transport data obtained in the earlier literature. 

First, solving $\Kc(\bwt,\bqt) =0$ we obtain the result for the dispersion function governing charge diffusion to be  
\begin{equation}\label{eq:CDiff}
\begin{split}
\bwt(\bqt) 
&= 
	-i\frac{d-1-\sdc}{(d-2)\, (d-1)} 
	\, \bqt^2 \left(1 + \mathfrak{c}_4\, \bqt^2 +\cdots \right)\,,\\
\mathfrak{c}_4 
&= 
	\frac{d-1-\sdc}{(d-2)\,(d-1)} \, \Dfn{\MY}{2,0}(r_+)  -2\, \Mser{\Vd}{0,2}(r_+)  \\
&\qquad 
	+ \frac{d-2}{d-1-\sdc}\, \left[
	 (d-1)\, (1-\sdc)^2\, \Dfn{\Vd}{0,4}(r_+) +\frac{2\,\nu_s\, \sdc^2}{1+Q^2}\, \Xi_{_\Vd}(r_+) \right] 
	 \,.
\end{split}
\end{equation}
The constants appearing here are horizon values of functions defined in \cref{sec:gradexpfns}.   
In obtaining this result we have used the expression for $\Kc$ accurate to quartic order in gradients \eqref{eq:Kc4}.  From the quadratic part of the dispersion we can obtain a prediction for the charge diffusion constant 
\begin{equation}\label{eq:Dcrate}
D_c = \frac{1}{(d-2)\,r_+}  \left(1-\frac{\sdc}{d-1}\right) \,.
\end{equation}	
One can check that this agrees with the result obtained in $d=4$ in \cite{Banerjee:2008th}.\footnote{
	Much of the literature computing charge diffusion rate focuses on the diffusion of a current in a neutral background. The general result for the charge diffusion constant in finite density medium  \eqref{eq:Dcrate}, insofar as we have been able to discern, is therefore new. 
} 

The charge diffusion constant  $D_c$ and  the dc conductivity \eqref{eq:sigmadc}  are proportional to each other:
\begin{equation}\label{eq:sigmaDcrel}
\sigma_\text{dc} = 
	 \frac{\delta \rho}{T\, \delta \left(\frac{\mu}{T}\right)}\bigg|_{P\, \text{fixed}}  \, D_c\,.
\end{equation}	
To see this, recall that the leading dissipative contribution to the hydrodynamic charge current can either be expressed as the gradient of the charge density or as the gradient of chemical potential (normalized by temperature), viz.,\footnote{
	For conformal fluids, we can express the results in a Weyl covariant manner using the Weyl covariant derivative $\mathcal{D}^{\scriptscriptstyle{\mathcal{W}}}$, which is defined using the Weyl connection for the fluid.  The charge density has Weyl weight $d-1$, implying that the  contribution to current takes the form $\mathcal{D}^{\scriptscriptstyle{\mathcal{W}}}_\nu \rho =  \left(\nabla_{\nu}+ (d-1)\, u^\alpha\, \nabla_\alpha u_\nu\right)\rho $.
} 
\begin{equation}
\frac{1}{2\,c_\text{eff}} \,J^\mu  = \rho\, u^\mu - D_c\, P^{\mu\nu}\,\mathcal{D}^{\scriptscriptstyle{\mathcal{W}}}_\nu \rho + \cdots 
\,, \qquad 
\frac{1}{2\,c_\text{eff}} \,J^\mu  = \rho\, u^\mu - \sigma_\text{dc}\, T\, P^{\mu\nu}\, \nabla_{\nu} \left(\frac{\mu}{T}\right) + \cdots  \,,
\end{equation}	
with $P^{\mu\nu}$ being is the spatial projector orthogonal to the velocity.  The two expressions are the canonical and the grand canonical expressions for the charge current, respectively, with the translation being made with fixed energy density or pressure (to ensure that we are not turning on the phonon).  Using the background charge density \eqref{eq:rho0def} and the thermodynamic data \eqref{eq:TRN}, we find 
\begin{equation}
 \frac{\delta \rho}{T\,\delta \left(\frac{\mu}{T}\right)} \bigg|_{P\, \text{fixed}}
	 = (d-2)\, r_+^{d-2}\, \frac{(1-\sdc)^2}{1-\frac{\sdc}{d-1}} \,.
\end{equation}	
This shows that  the results \eqref{eq:sigmadc} and \eqref{eq:Dcrate} are indeed related as claimed in 
 \eqref{eq:sigmaDcrel}.

For the sound dispersion locus, we solve $\KS(\bwt,\bqt) =0$ with $\KS$ given in \eqref{eq:KS2}  and obtain
\begin{equation}\label{eq:SProp}
\begin{split}
\bwt(\bqt) 
&= 
	\frac{\bqt}{\sqrt{d-1}} -i\,\frac{\nu_s}{2\,(1+Q^2)}\,\bqt^2 
	+ \frac{\nu_s\,  \mathfrak{h}_3}{2\, \sqrt{d-1}\, (1+Q^2)^2} \, \bqt^3 + \cdots \,, \\ 
\mathfrak{h}_3 
&=
	\frac{d\,(d-2)-4}{2\,d\,(d-2)}+\frac{d-3}{d-2}\,Q^2 + (1+Q^2)\, \Dfn{d-1}{2,0}(r_+)\,.
\end{split}
\end{equation}
For simplicity, we have only quoted the positive branch of solution. Since sound has a propagating mode there are two solutions which start off as $\omega = \pm \frac{k}{\sqrt{d-1}} + \cdots$. In \cref{sec:ZWasym} we conjecture the cubic and quartic corrections to the sound attenuation function and use it to derive the dispersion relation to quintic order, see \cref{eq:SPropA}.  This exploits the asymptotic expansion of the metric functions and how the stress tensor is assembled from the solution.

The quadratic term in \eqref{eq:SProp} captures the leading sound attenuation which originates from shear viscosity. The coefficient  of the $\bqt^2$ term, denoted $D_\eta$, arises from viscous damping of the sound mode. It is related to the shear viscosity of the conformal charged plasma via,
\begin{equation}\label{eq:Detarate}
D_\eta = \frac{d-2}{d-1}\, \frac{\eta}{\epsilon_0 + \Pbg} \;\; \Longrightarrow \;\;
\eta = c_\text{eff}\, r_+^{d-1} = \frac{s}{4\pi} \,,
\end{equation}	
reproducing the well-known result. The background energy density $\epsilon_0$ is given by the conformal equation of state $\epsilon_0 =(d-1)\, \Pbg$. 

The cubic term in the dispersion \eqref{eq:SProp}, which gives the correction to sound attenuation, arises from a second order contribution in the  constitutive relation for the currents. We can use the results of \cite{Banerjee:2008th,Erdmenger:2008rm} and the classification of second order transport in \cite{Haehl:2015pja} to argue that only a subset of second order transport data contributes to the linear dispersion relations. Specifically, we can only have contribution from two terms in the stress tensor, 
$T_{\mu\nu} \supset \tau_\pi\, u^\alpha \mathcal{D}^{\scriptscriptstyle{\mathcal{W}}}\alpha\sigma_{\mu\nu} + 
\tau_{_Q} \, \frac{1}{\rho}\, \mathcal{D}^{\scriptscriptstyle{\mathcal{W}}}_\mu \, \mathcal{D}^{\scriptscriptstyle{\mathcal{W}}}_\nu \rho $, and two terms in the charge current  
$J^\mu \supset C_\sigma\, P^{\mu\nu}\sigma_\nu  + C_\omega\, P^{\mu\nu}\omega_\nu $, as these are the only terms which are non-vanishing at linear order in the amplitude expansion. However, upon computing the equations of motion, one finds that only $\tau_\pi$ contributes to the dispersion relations at third order. The other three contribute at fourth order, but their contribution mixes with third order constitutive relations. In any event, solving the conservation equations, one finds:
\begin{equation}\label{eq:dispsoundgen}
\bwt = \frac{\bqt}{\sqrt{d-1}} - \frac{d-2}{d}\, \frac{\eta}{P}\, \frac{\bqt^2}{d-1} + \left(\frac{d-2}{2\,d}\, \frac{\tau_\pi}{P} - \frac{(d-2)^2}{2\,d^2}\, \frac{\eta^2}{P^2} \right)  \frac{\bqt^3}{(d-1)^\frac{3}{2}}+ \cdots \,.
\end{equation}	

Comparing the prediction from hydrodynamics with the gravity analysis, we deduce that $\tau_\pi$ is given in terms of the horizon value of the function   $\Dfn{d-1}{2,0}$. This coefficient also appears in the shear dispersion but only at fourth order \cite{He:2021jna}. Second order transport coefficients for a charged fluid have only been evaluated hitherto in the literature for $d=4$ in \cite{Banerjee:2008th,Erdmenger:2008rm}.  We have been able to evaluate $\Dfn{3}{2,0}(r_+)$ in closed form in this case, and find

\begin{equation}\label{eq:D204d}
\Dfn{3}{2,0}(r_+) = \frac{1}{2} \left(1+ \frac{1+Q^2}{\sqrt{1+4\,Q^2}} \, \log\left(\frac{3-\sqrt{1+4\,Q^2}}{3+\sqrt{1+4\,Q^2}}\right) \right).
\end{equation}	
We have confirmed that this result agrees with the values for the transport coefficients obtained earlier. The comparison with the result of \cite{Banerjee:2008th} is straightforward (they refer to the coefficient $\tau_\pi$ as $\mathcal{N}_1$, while that to \cite{Erdmenger:2008rm} (who instead use $\eta\tau_\pi$ for their coefficient) requires small algebraic manipulations to rewrite the temperature and chemical potential in terms of the black hole parameters. The numerical result for $\Dfn{d-1}{2,0}$ in other dimensions $3\leq d \leq 6$ can be found in Figure 4 of \cite{He:2021jna}.

Let us finally turn to understanding the boundary currents from a fluid dynamical perspective. We will examine the thermal one-point function of the stress tensor, which we express using \eqref{eq:ZVvevs}.  As explained in \cite{He:2022jnc}, there is an interesting complication with energy transport. The sound mode has a propagating degree of freedom, one that is already visible at the level of an ideal fluid. Therefore, any decomposition in fluid dynamical terms must treat the contributions from $\Zd$ and $\Vd$ asymmetrically. In particular, we will have to break up the simpler packaging in terms of the charge diffusion and energy operators $\QOp$ and $\EOp$, and treat the contributions from $\PoZ$ and $\PoV$ asymmetrically. 

We claim that the currents can be decomposed  into an ideal, a non-ideal,\footnote{
	We refrain from characterizing the non-ideal part as dissipative, since this part includes non-dissipative contributions as well starting at second order in gradients.
}  and a polarization contribution, respectively, in the following manner: 
\begin{equation}\label{eq:TJiddynpol}
\begin{split}
\frac{1}{2\,c_\text{eff}}\, \expval{\Jcft_\mu}
&= 
	J^\mu_\text{ideal}[\mathbf{W},\PoZ]  + J^\mu_\text{non-ideal}[\PoV, \PoZ] \,, \\
\frac{1}{c_\text{eff}}\, \expval{\TcftU^{\mu\nu}}  
&= 
	T^{\mu\nu}_\text{ideal}[\mathbf{W},\PoZ]  + T^{\mu\nu}_\text{non-ideal}[\PoV, \PoZ] + 
	T^{\mu\nu}_\text{pol}[\JoV,\JoZ] \,.
\end{split}
\end{equation}	
This separation works independently on both the R and L boundaries, so we will drop the corresponding subscripts in the expressions henceforth. The ideal fluid currents are of the standard form 
\begin{equation}\label{eq:Curideal}
T^{\mu\nu}_\text{ideal}  = 
	\sqrt{-\gamma}\, P \left(d\, u^\mu\, u^\nu  + \gamma^{\mu\nu}\right) \,, 
\qquad
J^\mu_\text{ideal}  =  \rho \, u^\mu\,.
\end{equation}	
We need to ascertain if we can isolate the deformed expressions for pressure $P$, charge density $\rho$, and velocity $u^\mu$, noting the background values $\Pbg$ and $\Qbg$ given in \eqref{eq:bgP} and \eqref{eq:rho0def}. 

To motivate this split, we first rewrite the stress tensor density as follows:
\begin{equation}\label{eq:Tmnalt}
\begin{split}
\expval{(\TcftU)\indices{_v^v}}_{\skL/\skR}
&=
	  (d-1)\, \Pbg \left[-1 +  \frac{d-2}{2}\int_k\, \ScS\, \mathbf{W}_{\skL/\skR}\right]
	  -   \int_k\, \frac{k^2}{d-1}  \, \ScS \left(\PoZ_{\skL/\skR} + k^2\, \frac{\PoV_{\skL/\skR}}{\bRQ}\right)  \,, \\
\expval{(\TcftU)\indices{_v^i} }_{\skL/\skR}
&= 
	i\, \int_k\, \frac{\omega\, k}{d-1}\,\ScS^i  \left(\PoZ_{\skL/\skR} + k^2\, \frac{\PoV_{\skL/\skR}}{\bRQ}\right)  \,,\\
\expval{(\TcftU)\indices{_i^j} }_{\skL/\skR}
&=
	- \frac{1}{d-1} \, \expval{(\TcftU)\indices{_v^v}}_{\skL/\skR} \, \delta\indices{_i^j}
 	- \int_k \frac{k^2}{d-2} \frac{\nu_s\, \Gatt(\bwt,\bqt)}{1+Q^2}\, (\ScST)\indices{_i^j} \, \PoZ_{\skL/\skR}\, \\
 &
  + \int_k \,  (\ScST)\indices{_i^j}  
  	\left[\frac{k^2}{d-2}\left(\frac{k^2}{d-1} -\omega^2\right)  \frac{\PoV_{\skL/\skR}}{\bRQ} \right.\\
&\left.\qquad \qquad \qquad 
 	 + \frac{d-1}{2\,(d-2)} \,  \BQT^2 	\,
		\left(\Qbg\, r_+\, \JoV_{\skL/\skR} - 2\,d\,(d-1)\, \Pbg\, \JoZ_{\skL/\skR} \right)
 	  \right] \,.	  
\end{split}
\end{equation}
In obtaining this expression we have separated the contributions from $\PoV$ and $\PoZ$ (and their sources). Furthermore, in the spatial part of the stress tensor we have replaced $\JoZ $ in terms of $\PoZ$ using \eqref{eq:SVeVRel}, and used the explicit form of $\KS$ \eqref{eq:KS2}. The upshot is that the contribution of $\PoZ$ to the stress tensor is now considerably simpler with the spatial shear-stress part being proportional to the sound attenuation function. 

In the limit where we switch off sound attenuation $\Gatt\to 0$, we see that the entire contribution of $\PoZ$ to the stress tensor density can be understood as that of an ideal fluid with 
\begin{equation}\label{eq:idealPu}
\begin{split}
P_{\skL/\skR}
&= 
	\Pbg\, \left(1-\frac{d}{2}\, \int_k\, \ScS\, \mathbf{W}_{\skL/\skR} \right) + \int_k\, \frac{k^2}{(d-1)^2}\, \ScS\, \PoZ_{\skL/\skR} \,,\\
u_\mu\, dx^\mu
&=
	-\left[1-  \frac{d-3}{2}\, \int_k\, \ScS\, \mathbf{W}_{\skL/\skR}\right]  dv - i\,\int_k\, \frac{\omega\,k}{d\,(d-1)\,\Pbg} \, \ScS_i \,  \PoZ_{\skL/\skR}\, dx^i \,.
\end{split}
\end{equation}	
In the absence of charges, this result coincides with that obtained for neutral fluid in \cite{He:2022jnc}. Note that the temporal component of the velocity is arising from the background red-shift $u_v = - \sqrt{-\gamma_{vv}}$. 

In the present case we have additional contributions from the charge diffusion mode. These are captured in the dynamical and polarization terms. To wit,
\begin{equation}\label{eq:Tdyndef}
\begin{split}
\expval{(T_\text{non-ideal})\indices{_v^v}}_{\skL/\skR}
&= 
	-\int _k\, \frac{k^4}{d-1}\, \ScS \, \frac{\PoV_{\skL/\skR}}{\bRQ} \,, \\
\expval{(T_\text{non-ideal})\indices{_v^i}}_{\skL/\skR}
&= 
	i\, \int _k\, \frac{\omega\, k^3}{d-1}\, \ScS_i \, \frac{\PoV_{\skL/\skR}}{\bRQ} \,, \\
\expval{(T_\text{non-ideal})\indices{_i^j}}_{\skL/\skR}
&= 		
	\int _k\, \frac{k^2}{(d-1)^2} \left[k^2\, \delta\indices{_i^j} \, \ScS+ \frac{(d-1)^2}{d-2} \left(\frac{k^2}{d-1} -\omega^2\right) (\ScST)\indices{_i^j} \right]\, \frac{\PoV_{\skL/\skR}}{\bRQ}   \\	
&\qquad 
	- \int_k \frac{k^2}{d-2} \frac{\nu_s\, \Gatt(\bwt,\bqt)}{1+Q^2}\, (\ScST)\indices{_i^j} \, \PoZ_{\skL/\skR}\,,		
\end{split}
\end{equation}
and 
\begin{equation}\label{eq:Tpoldef}
\expval{(T_\text{pol})\indices{_i^j}}_{\skL/\skR}
= 
	\frac{d-1}{2\,(d-2)} \, \int_k \, \BQT^2 	\,  (\ScST)\indices{_i^j}\,  
		\left[\Qbg\, r_+\, \JoV_{\skL/\skR} -2\, d\,(d-1)\, \Pbg\, \JoZ_{\skL/\skR} \right] .
\end{equation}	
Since the charge diffusion is dissipative, it follows that the contribution from $\PoV$ cannot be included in the ideal fluid part. For similar reasons the $\Gatt, \PoZ$ contribution is also non-ideal. The novel feature is the additional polarization in the spatial part of the stress tensor proportional to the sources. This term is present to account for the Joule heating effect, but we do not have a heuristic explanation for its presence.

Given the decomposition of the stress tensor, we can examine what it implies for the charge current. We have the modified charge density 
\begin{equation}\label{eq:cdenideal}
\rho_{\skL/\skR} = \Qbg \left[1-(d-1)\, \int_k\, \ScS \left(\frac{1}{2}\, \mathbf{W}_{\skL/\skR} - \frac{k^2}{d\,(d-1)^2\, \Pbg}\, \PoZ_{\skL/\skR} \right)  \right] ,
\end{equation}	
with  the  background density \eqref{eq:rho0def}. Along with the velocity field determined from the stress tensor in \eqref{eq:idealPu} these charge densities reproduce the contact term of the current as an ideal fluid contribution. The additional piece is purely dynamical and  can be shown to be:
\begin{equation}\label{eq:Jnonideal}
\begin{split}
\expval{J^\text{non-ideal}_v}_{\skL/\skR}
&=
	\int_k k^2\, \ScS \,  \expval{\QOp^\text{non-ideal}_{\skL/\skR} } \,,\\ 
\expval{J^\text{non-ideal}_i}_{\skL/\skR}
&=
	i\, \int_k \omega\, k\, \ScS_i \, \expval{\QOp^\text{non-ideal}_{\skL/\skR} } \,,
\end{split}
\end{equation}
where (using the identity \eqref{eq:bRQPQreln})
\begin{equation}\label{eq:ChopDyn}
\begin{split}
\QOp^\text{non-ideal}_{\skL/\skR} =
	 \left[\frac{r_+}{4}\, (2+\BQT^2)\, \OpV + \frac{\Qbg}{d\,(d-1)\, \Pbg}\, \frac{\BQT^2}{2+\BQT^2}\, \OpZ\right]_{\skL/\skR} \,.
\end{split}
\end{equation}
 Owing to the fact that the ideal fluid already supports a propagating sound mode, we see that the non-ideal part of the charge diffusion operator is not quite \eqref{eq:ChOp}, but a modification thereof  to $\QOp^\text{non-ideal}_{\skL/\skR}$.

Not only can one give a simple interpretation for the conserved currents, but it also turns out to be possible to give a similar decomposition of the on-shell action. First of all one can check that
\begin{equation}\label{eq:Scwdecompose}
 	\int_k \left[\frac{1}{2} \, \TcftU^{\mu\nu}\, \delta \gamma_{\mu\nu} +  \sqrt{-\gamma}\, J_{_\text{CFT}}^\mu\, \delta\, \tensor[^\gamma]{\!A}{_\mu}\right]
	\;\; 
	\;\; \overset{\text{\tiny{Legendre}}}{\underset{\text{\tiny{transform}}}{\longrightarrow}} \;\; 
\frac{1}{c_\text{eff}} \,S[\PoV,\PoZ] \,.
\end{equation}	
In writing this expression we are assuming that the currents are correctly normalized as indicated in \cref{fn:Jouleheat}. The ideal fluid contribution is isolated by setting $\Gatt \to 0$ and dropping the contribution from $\PoV$. 
The contributions from $\PoV$ are associated with charge diffusion, which is non-ideal. On the other hand,  the phonon, owing to its propagating degree of freedom, contributes non-trivially to the ideal fluid part. Switching off $\Gatt$ ensures that we ignore the non-ideal sound attenuation piece.  Specifically, turning off all the sources of dissipation, we anticipate  that the surviving part of the WIF can be obtained by a Legendre transformation of the ideal fluid action (the integrated pressure), viz.,  
\begin{equation}\label{eq:SconWif}
S_\text{ideal,LT} = S\left[\PoV=0,\PoZ\right]  \bigg|_{\Gatt \to \,0} \,.
\end{equation}	

The ideal fluid action for a charged plasma is given by the zeroth order Class L action of \cite{Haehl:2015pja}. We can view this as a functional of the  background metric $\gamma_{\mu\nu}$, the vector $\vb{b}^\mu$, and scalar $\bm{\Lambda}_{\vb{b}} \equiv \mu/r_+$, which are the related to the thermal vector and thermal twist introduced in the aforementioned reference. The variation of this ideal fluid action has pieces that lead to the stress tensor and the current, as in the r.h.s.\ of \eqref{eq:Scwdecompose}, but additionally it also has contributions proportional to 
$\delta \vb{b}^\mu$ and $\delta \bm{\Lambda}_{\vb{b}} $, which are the change of these variables from their background value. Once we account for this effect, the Legendre transform of the ideal fluid action should reproduce the contact term and the propagating part of the sound mode.  We have not checked this explicitly, but the fact that \eqref{eq:Scwdecompose} holds provides strong evidence for our statement.

In recent years, various groups have constructed effective actions for non-linear, dissipative hydrodynamics using symmetry arguments. These actions are argued to be universal, governing the low-energy behaviour, with the dynamical equations of motion being the energy-momentum and charge conservation. The dynamical fields in this approach are target maps from an auxiliary space, the worldvolume, onto the physical fluid (cf., \cite{Haehl:2018lcu}). The physical temperature and chemical potential are pushforwards of reference data (fixed thermal vector and twist) from the worldvolume into the fluid target under these maps. These non-linear analyses, while interesting in their own right, are per se not optimal for understanding loop effects in the effective field theory. For these it suffices to have an action which isolates the physical modes. For a charged plasma, these are the shear and charge diffusion waves, and the sound mode. An effective action parameterized by the associated operators suffices to understand the dynamics in an amplitude expansion. 

In \cite{He:2021jna} we have obtained the Gaussian effective action for the shear modes, while the results \eqref{eq:WIF} herein give the corresponding results for the charge diffusion and sound modes. Using the bulk graviton vertices one can work out non-Gaussian corrections systematically. This data is useful since thus far there does not exist in the literature a precise proposal for a charged plasma. Even the adiabatic Class L action for such systems has not been constructed, though  \cite{Haehl:2015pja} classified the various tensor structures under the eightfold classification of transport up to second order in gradients. Moreover, the non-linear fluid/gravity dual of a charged plasma has also not been constructed in general dimensions. The original works \cite{Banerjee:2008th,Erdmenger:2008rm} focused on $d=4$ with no background sources, while \cite{Plewa:2012vt} turned on general boundary metric sources (also in $d=4$) but not background gauge field sources. The combined results of the current work and \cite{He:2021jna} are the first steps towards a better understanding of charged fluids from holography.

\section{Discussion}
\label{sec:discuss}

We have used holography to understand the dynamics of charge diffusion and sound propagation in a charged plasma. Both of these physical effects arise from the longitudinal scalar modes of the energy-momentum tensor  and charge current, but owing to the presence of background charge density they mix non-trivially. One advantage of the holographic modeling is that it provides a clear mechanism to decouple the modes and indicates which combination of the currents leads to the diffusive dynamics of charges, and which leads to the phonon mode. The fact that one can decouple the modes at the Gaussian order is not a surprise per se, since one only needs to find a suitable diagonal basis of the physical modes. The gravitational modeling simplifies this considerably, since one is essentially diagonalizing a pair of coupled differential equations. While this result dates back to the analysis of \cite{Kodama:2003kk}, it has thus far not been employed to extract general lessons for a charged plasma.

During our analysis, we have simplified the derivation of the linearized Einstein-Maxwell equations. In addition, we have undertaken a careful analysis of the variational problem for the decoupled modes, an exercise that had not been carried out earlier. In particular, we note that the decoupled fields parameterizing the perturbations of the \RNAdS{} black hole are to be quantized with Neumann boundary conditions for purposes of computing the generating function of boundary current correlators. Furthermore, we have solved these equations in a gradient expansion to quartic order. Note that the non-linear fluid/gravity analysis for charged \AdS{} black holes has only been analyzed to quadratic order in gradients, and that too only in $d=4$. Thus, our results for the most part are largely new, a surprising fact considering the extensive investigation into hydrodynamics using holography. 

For the boundary theory we are not interested in computing the generating function of current correlators as a functional of the sources as one is wont to do conventionally in the AdS/CFT context. The conserved current correlators have hydrodynamic poles, and the resulting generating function does not admit a well-behaved low-energy expansion.  Rather, we are interested in developing a systematic low energy effective field theory parameterized by the physical low-lying degrees of freedom: the phonon and the charge diffusion mode. This is achieved by a suitable Legendre transform of the generating function, but operationally, as argued in \cite{Ghosh:2020lel}, it amounts to quantizing the dual bulk fields with Dirichlet boundary conditions. We have obtained the Gaussian part of the real-time effective action, and the Wilsonian influence phase for the charge diffusion and sound modes. This can then be used  to recover (and extend) the Schwinger-Keldysh two-point functions of the conserved currents. 

The physical results of the boundary currents and the Wilsonian influence phase can be nicely understood in hydrodynamic language. We have demonstrated that the currents can be separated into an ideal part, a non-ideal part (which includes both dissipative and non-dissipative contributions), and background polarization terms. A similar statement holds for the effective action.  The non-ideal part includes  non-dissipative contributions in the functions $\Kc$ and $\Gatt$, which can be extracted. These should be matched against the higher order gradient terms in the Class L charged fluid action. The latter is easy to construct following the analysis of \cite{Haehl:2015pja}. However, it has not yet been analyzed explicitly in the literature; 
we therefore leave a detailed hydrodynamic analysis to the future.

Together with our earlier analysis of momentum diffusion and charge propagation modes in \cite{He:2021jna}, the analysis of this paper completes the derivation of the effective action for a charged plasma. The rapidly decaying Markovian modes are the  $\frac{d\,(d-3)}{2}$  modes of the tensor sector  and the $(d-2)$ modes that correspond to transverse photons. The hydrodynamic modes are the $(d-2)$ momentum diffusion modes, and the two modes analyzed herein. Altogether these make up the set of a stress tensor and charge current, modulo  tracelessness, and the $d+1$ conservation equations. 

An interesting physical output of our analysis is the identification of the combination of currents which correspond to the physical charge diffusion mode and the sound mode. The charge current is parameterized by an operator $\QOp$ and the stress tensor by $\EOp$, cf., \eqref{eq:ChOp} and \eqref{eq:EnOp}, respectively. Inverting these, we find 
\begin{equation}\label{eq:VZmaps}
\begin{split}
\OpV 
&= 
	 \frac{2}{(1+\BQT^2)\, r_+} \left[\QOp + \frac{2}{d\,(d-1)}\, \frac{\Qbg}{\Pbg} \, \frac{\EOp}{2+\BQT^2} \right] , \\
\OpZ  
&=
	\frac{2+\BQT^2}{2\,(1+\BQT^2)}\, \EOp  - \frac{d\,(d-1)}{4}\, \frac{\Pbg}{\Qbg} \, \frac{\BQT^2\, (2+\BQT^2)}{1+\BQT^2}\, \QOp \,.
\end{split}
\end{equation}
We employed the identity noted in \cref{fn:BQTprops} to simplify the above. What is curious in these expressions is the presence of the deformed momentum factor $\BQT$ defined in \eqref{eq:diagDefs}. These factors, we emphasize, are exact, arising as they do from the decoupling of the gravitational equations of motion. This quantity $\BQT$ is an interesting prediction of gravity, and its origins deserve to be better understood.	The occurrence of such momentum factors is, in fact, quite ubiquitous in the analysis of probe fields in a black hole background. They have been encountered in various explorations in the AdS/CMT literature. While it is not therefore a surprise to see them in the context of linearized photon and graviton fluctuations, one ought to understand their presence more clearly. One could speculate, without any evidence, that such factors appear even in weakly coupled charged plasmas, a possibility that should be investigated further.

While we have focused herein on obtaining the Gaussian  part of the real-time effective action, the grSK geometry, as explained in \cite{Jana:2020vyx}, is sufficient to allow one to obtain the non-Gaussian terms in the influence functional, which can be computed by standard Witten diagram techniques suitably adapted to this two-sheeted geometry. The only ingredient one needs are the bulk vertices for the various fields. These can be obtained from the Einstein-Maxwell action. Note however that there is mixing between the scalar, vector and tensor modes, from the cubic order onward. One can however combine the parameterization of the scalar modes in this paper, together with that adapted for the tensor and vector modes in \cite{He:2021jna} to compute the vertices. This computation is straightforward, albeit technically challenging, and has the potential to shed further light on the nature of hydrodynamic effective field theories.

\section*{Acknowledgements}
It is a pleasure to thank Veronika Hubeny  and  Akhil Sivakumar for helpful discussions. 

TH was supported by  U.S. Department of Energy grant DE-SC0020360 under the HEP-QIS QuantISED program. RL acknowledges support of the Department of Atomic Energy, Government of India, under project no. RTI4001, and would also like to acknowledge his debt to the people of India for their steady and generous support to research in the basic sciences. MR and JV were supported  by U.S.\ Department of Energy grant DE-SC0009999. TH, MR, and JV also acknowledge support from the University of California. 

\appendix

\section{Dynamics}
\label{sec:dynamicsderive}

The dynamics we study is governed by the Einstein-Maxwell action with its Gibbons-Hawking variational boundary term and appropriate counterterms. As we work  to quartic order in boundary gradients, we expect that we would need counterterms accurate to that order. While it will transpire that we can make do with the quadratic gravitational counterterms, we nevertheless include below the full set for completeness. 

The gravitational dynamics  we consider is\footnote{
	In the appendices we will drop the normalization factor $16\pi G_N$ to keep the expressions simple. We also work in units where $\lads =1$. Boundary quantities can be easily obtained by scaling up the results by a factor of  $c_\text{eff} = \frac{\lads^{d-1}}{16\pi G_N}	$ at the end.} 
\begin{equation}\label{eq:SEMaxA}
\begin{split}
S_\text{grav} 
&=  \int d^{d+1} x\, \sqrt{-g} \left(R + d(d-1) - \frac{1}{2} \,F_{AB}\, F^{AB}\right) + 
	 \int d^d\,x \sqrt{-\gamma} \, \left(2 K +L_\text{EH,ct}  + L_\text{Max,ct} \right) , \\
L_\text{EH,ct} 
&= 	-2 (d-1) - \frac{1}{d-2} \, \tensor[^\gamma]{R}{} 
	- \frac{1}{(d-4)\,(d-2)^2} \left(
		\tensor[^\gamma]{R}{_{\mu\nu}} \, \tensor[^\gamma]{R}{^{\mu\nu}}  - \frac{d}{4(d-1)}\, \tensor[^\gamma]{R}{}^2
	\right) , \\
L_\text{Max,ct}
&=
	\frac{1}{4(d-4)} \left(\tensor[^\gamma]{F}{_{\mu\nu}} \, \tensor[^\gamma]{F}{^{\mu\nu}} 
	+ \frac{1}{(d-4)(d-6)}\, \tensor[^\gamma]{F}{_{\mu\nu}} \Box_\gamma\, \tensor[^\gamma]{F}{^{\mu\nu}}\right) 	.
\end{split}
\end{equation}
The counterterm action accurate to quartic order was originally obtained in \cite{Emparan:1999pm,deHaro:2000vlm}. The equations of motion have been recorded in \eqref{eq:EMeqns}. We will first analyze them, before turning to understanding the variational principle.

The perturbations of the background are parameterized in terms of the metric functions and gauge potentials recorded in  \eqref{eq:hscalarpert}. As explained in \cite{Kodama:2003kk} and in our earlier works \cite{He:2021jna,He:2022jnc}, we can work with suitable gauge invariant combinations made out of curvatures. We work in the two-dimensional $\{v,r\}$ spacetime,  \emph{the orbit space}, which has a connection\footnote{
	In this appendix, we use lowercase early alphabet Latin characters to indicate orbit space tensors in addition to the conventions specified in \cref{fn:conventions}.}
\begin{equation}\label{eq:OrbitC}
\begin{split}
\gOrb &\equiv \dv{r}(r^2 f) \,.
\end{split}
\end{equation}

The discussion of the metric perturbations parameterized in terms of $\HH_{AB}$, as presented in \eqref{eq:hscalarpert}, is unchanged from \cite{He:2022jnc}, since the organization into $SO(d-2)$ harmonics and orbit space transformations are unaffected. The main change is the addition of data from the Maxwell potential, but this has also been previously explored in \cite{Ghosh:2020lel}, who analyzed the dynamics of probe Maxwell fields in the \SAdS{d+1} background. Much of that can again be imported with some slight modifications. Indeed, we shall see that the dynamical equations can be suitably understood by combining the analysis of the aforementioned two papers.

\subsection{Dynamics in the Debye gauge}
\label{sec:Debye}

A generic  perturbation of the background is captured by ten functions, encoded in $\HH_{AB}$ and $\AGR_A$, for we can parameterize them in terms of $SO(d-2)$ harmonics as
\begin{equation}\label{eq:hscalarpert}
\begin{split}
ds_{(1)}^2 
&= 
        \int_k\,  \Bigg\{\left(2\,\HS  \, ds_{(0)}^2 	+ 
        \HH_{vv}\, dv^2+2\,\HH_{vr} \, dv dr+\HH_{rr} \,dr^2\right)\ScS \\
& \qquad \qquad 
    -\left[ 2\,r\, (\HH_{vx} \,dv+\HH_{rx} \,dr)\, \ScS_i\, dx^i- 
 2\,r^2 \,\HHT\, \ScST_{ij}\ dx^i dx^j \right] \Bigg\} \,, \\
\AGR_A\, dx^A
&=
	\int_k\bigg\{ 
		\left(\HA_v \, dv + \HA_r\, dr \right)\, \ScS + \HA_x\, \ScS_i\, dx^i
		\bigg\} \,.
\end{split}
\end{equation}  
Here $\ScS=e^{i\bk\cdot\bx-i\omega v}$ is the scalar plane wave on $\mathbb{R}^{d-1,1}$ and $\ScS_i$ and $\ScST_{ij}$ are derived harmonics, defined as 
\begin{equation}\label{eq:SiSij}
\ScS_i = \frac{1}{k}\, \partial_i\, \ScS \,, \qquad \ScST_{ij} = \frac{1}{k^2}\left(\partial_i\, \partial_j - \frac{\delta_{ij}}{d-1}\, \partial^2 \right)\, \ScS\,.
\end{equation}	
$\ScST_{ij}$ is traceless but not transverse, $\partial_i\ScST_{ij} =\frac{d-2}{d-1}\, \partial^2\,\partial_j$. It is to be distinguished from the derived harmonic $\ScS_{ij}$ defined in \cite{Ghosh:2020lel}, which is neither transverse nor traceless.

 We can use the gauge invariant combinations to help choose a specific gauge and simplify the linearized equations. The natural choice is to use an analog of the \emph{Debye gauge} employed in \cite{He:2022jnc}. The idea is that we set to zero, a priori, the perturbations that depend on the derived harmonics $\ScS_i$ and $\ScST_{ij}$. To wit, we have the gauge conditions:
\begin{equation}\label{eq:Debye}
\text{Debye Gauge :}\quad \HH_{vx}=\HH_{rx}=\HHT=0\,, \qquad \HA_x = 0\,.
\end{equation}
 This immediately reduces us to six functions  $\{\HS,\HH_{vv}, \HH_{vr},  \HH_{rr}, \AGR_v,\AGR_r\}$.\footnote{
 	As described in \cite{He:2022jnc} there is a subtlety with the spatially homogeneous modes, but we will not analyze them in any detail, focusing instead only on the spatially inhomogeneous modes.  One issue is that \eqref{eq:eeqT} no longer holds at $k=0$, so one has to treat $\HH_{rr}$ or $\PHB$ as an independent degree of freedom.
 } 

 To proceed further, we introduced a certain parameterization of the metric perturbations inspired by the discussion of  sound modes in neutral fluids \cite{He:2022jnc}. Let us introduce rescaled variables $\PHE, \PHO,\PHW$:
\begin{equation}\label{eq:rEOSdef}
\begin{split}
\PHE = r^{d-3}\, \HH_{vv}\,, \quad \PHO = r^{d-3} \left(\HH_{vv} + r^2f\, \HH_{vr}\right) \,,
\quad 
\PHW = 2\, r^{d-2}\, \HS \,.
\end{split}
\end{equation}  
Furthermore, we express $\HH_{rr}$ in terms of these functions and a new function $\PHB$ as
\begin{equation}\label{eq:rBTdef}
\HH_{rr}	= - \frac{1}{r^{d+1} f^2} \left[2(\PHO-\PHE) + r f \,(d-1)\, \PHW + \PHB\right] .
\end{equation}	
All told the perturbative corrections now take  the following form:
\begin{equation}\label{eq:EOWDebA}
\begin{split}
 ds_{(1)}^2
 &=  
 	\frac{\PHE-r f\,\PHW}{r^{d-3}}\, dv^2 
	 +\frac{2}{r^{d-1}f} \left(\PHO-\PHE +rf\,\PHW\right) dv\,dr
	 + r^2\, \frac{\PHW}{r^{d-2}}\, d\vb{x}^2 \\
& \qquad \qquad 
	-\frac{1}{r^{d+1}f^2}\left[2(\PHO-\PHE) + r f \,(d-1)\, \PHW + \PHB\right] dr^2  \,,\\
\AGR_A\, dx^A 
&=
	  \HA_v\, dv + \HA_r\, dr	\,.
\end{split}
\end{equation}	
For simplicity this has been written directly in position space. Importantly, the metric perturbation is unchanged from the analysis of \cite{He:2022jnc}. We will now present the linearized Einstein-Maxwell equations in terms of these fields by decomposing \eqref{eq:EMeqns} into plane waves. 

\paragraph{Maxwell equations:} The analysis of Maxwell's equations parallels the discussion in \cite[Section 8]{Ghosh:2020lel}. We find a constraint equation
\begin{equation}
r^2f\, \dv{r}(r^{d-3}\,\HA_v) + \Dz_+ \left(r^{d-1}\,  f\, \HA_r\right) =0 \,.
\end{equation}	
This can be trivialized by introducing a designer scalar $\MV$, in terms of which
\begin{equation}
\HA_v =  \frac{1}{r^{d-3}}\Dz_+\MV\,, 
\qquad 
\HA_r = -\frac{1}{r^{d-3}} \dv{r}\MV \,.
\end{equation}	
The remaining equation is the radial component of the Maxwell equation, which can then be shown to be equivalent to 
\begin{equation}\label{eq:E0}
\begin{split}
\EMax_r 
&= 
	-\frac{i\omega}{r^{d-1}\,f} \, \EEq_0 \,, \\
\EEq_0
&=
	r^{d-3}\, \Dz_+ \left(\frac{1}{r^{d-3}}\, \Dz_+ \MV\right) - (k^2\,f-\omega^2)\, \MV  + (d-2)^2\, a\, \left[ f\, \PHW + \frac{1}{2\,(d-2)\, r} \PHB\right] .
\end{split}
\end{equation}	

\paragraph{Einstein's equations:}  It will be helpful to assemble the equations of motion \eqref{eq:EMeqns} into time-reversal invariant orbit space tensor combinations. The scalar equation  involves only $\PHB$ and takes a simple form:
\begin{equation}\label{eq:eeqT}
\EEqT
 =- \frac{k^2}{2\,r^{d-1}}  \PHB  \,.
\end{equation}	
This allows us to eliminate another function algebraically.

The orbit space tensor equations assembled again into time-reversal invariant combinations, which when judiciously combined with the Maxwell equation $\EEq_0$ result in
\begin{equation}\label{eq:eeq12B}
\begin{split}
\EEq_1 
&\equiv 
	\frac{2}{d-1}\, \left(r^{d-1}\, \EEq_{vv} - a'\, r^2\, \EEq_0 \right)\\ 
&=
	\Dz_+ \left(\Tone - r\, \PHB\right) + \frac{k^2}{d-1}\, \left(\PHE - \PHB  \right)  + \frac{2\,a'}{d-1}\, r^2\, k^2\, f\, \MV \,, \\ 
\EEq_2
&\equiv
	\frac{2\,r^{d-1}}{d-1}\,  (\EEq_{vv} + r^2f\, \EEq_{vr}) \\
&=
	-i\omega 	\left(\Tone - r\, \PHB\right) + \frac{k^2}{d-1}\PHO\,, \\  
\EEqB
&\equiv
	\frac{2\,r^{d+1}f}{d-1}\, \left(\EEq_{vr} + \frac{1}{2}\, r^2f\, \EEq_{rr} \right)  + \frac{2\,a'}{d-1}\,r^2\, \EEq_0\,
	\\
&= 
	-\Dz_+ \left(\Tone - \frac{r}{2}\, \PHB \right)
	 - i\omega\, r\, \PHO 
	 - \frac{r}{2} \left(\Dz_+ - \gOrb + r f \right) \left[\Dz_+ \PHW - (d-2) \, rf\, \PHW \right] \\ 
&\qquad \quad
	+ \frac{r}{2}\,(\omega^2-k^2f)\, \PHW + \frac{k^2+ d(d-1)\, r^2- (d-2)^2\, a^2}{2\,(d-1)}  \, \PHB  - \frac{2\,a'}{d-1}\, k^2\,r^2\,\MV  \,.
\end{split}
\end{equation}	
The function $\Tone$ introduced above is defined as 
\begin{equation}\label{eq:T1def}
\Tone
\equiv 
	r \, \PHE - \left(\Dz_+ - \frac{\gOrb}{2}\right) (r\,\PHW) 
=	
	r\left[\PHE - \Dz_+ \PHW + \frac{r^2f'}{2}\PHW\right] .
\end{equation}	
This leaves us with the vector equations, which are the coefficients of $\ScS_i$ and have an explicit momentum factor. We find
\begin{equation}\label{eq:eeq45}
\begin{split}
\EEq_4\
&\equiv
	2\,r^{d-1} f\, \EEq_{vi}  = i k_i\, \widetilde{\EEq}_4 \\ 
&=
	ik_i\left[\Dz_+ \PHO + i\omega \left(\PHE - \PHB \right) +2\,i\omega\, r^2\ f\, a'\, \MV \right] , \\
\EEq_5
&\equiv
	2\,r^{d-1} f\, \left(\EEq_{vi}  + r^2f\, \EEq_{ri} \right) =i k_i\, \widetilde{\EEq}_5\\ 
&=
	ik_i\left[\Dz_+ \PHE+ i\omega\, \PHO - (d-1) \left(\Dz_+ - \frac{1}{2}\, \gOrb\right)(rf\PHW) 
		-\frac{r}{2}\left((d-1) f + \frac{rf'}{2}\right) \PHB \right. \\
&\qquad \qquad \left.	
	 - 2 \, a'\,r^2f\,\Dz_+\MV\right] .
\end{split}
\end{equation}	
The tilded equations strip out the momentum factor. The remaining equations, which are orbit space scalars picking out the trace and the $\ScS_{ij}$ part of the spatial harmonics, can be naturally expressed in terms of them as 
\begin{equation}\label{eq:eeq67}
\begin{split}
\EEq_6
&\equiv 
	- \frac{2\,r^{d-1} f}{d-1}\, \sum_{i=1}^{d-1}\, \EEq_{ii} \\	
&=
	\Dz_+ \left(\frac{\widetilde{\EEq}_5}{f}\right) 	 + i\omega \, \frac{\widetilde{\EEq}_4}{f} +2 \,\frac{d-2}{d-1} \, r^{d-1}\, \EEqT \,, \\
\EEq_7
&\equiv 
	f\,\EEq_{ij} = \frac{k_i\,k_j}{k^2}\, \EEqT \,.
\end{split}
\end{equation}
Finally, a natural way to combine the equations  involves taking a particular combination of $\EEqB$ and $\EEq_5$:  
\begin{equation}\label{eq:eeq3}
\begin{split}
\EEq_3 
&=
	 \frac{2}{r} \,\EEqB+ \widetilde{\EEq}_5  \\
&=
	(\Dz_+ + 2 rf )\left[\Dz_+ \PHW - \PHE +\PHB\right] - i\omega\, \PHO 	+\frac{\Lk}{(d-1)\,  r} \PHB \\
& \qquad 
	+ \left(\omega^2 - k^2 f -\frac{1}{2}\,r^3\,f \left[(d+5)\,f' + 2r\, f''\right] \right) \PHW  -2\,a'\,rf\, \left( r \Dz_+ + \frac{2k^2}{d-1}\right) \MV\,.
\end{split}
\end{equation}	

Apart from the contribution from the Maxwell potential, the equations for the most part are identical to the ones obtained for the scalar polarizations of the gravitons presented in \cite{He:2022jnc}.  As noted earlier, the expressions in terms of the function $f$ and its derivatives are in fact unchanged.\footnote{
	The main difference is that \cite{He:2022jnc} used an identity specific to the \SAdS{d+1} background to simplify expressions. \label{fn:fidentity}}

\subsection{Parameterizing the solution space: \texorpdfstring{$k\neq 0$}{inhomogeneous}}
\label{sec:psolspace}

There are five physical functions $\{\PHE,\PHO,\PHW, \PHB, \MV\}$, which implies that the equations of motion are not all independent.  Some of the relations are manifest (e.g., $\EEq_6$ and $\EEq_7$), but we can check that there are only four independent Einstein's equations.  An efficient choice turns out to be the set $\{\EEq_0,\EEqT, \EEq_1, \EEq_2, \EEq_3\}$; satisfying them will ensure that the reminder are also upheld (for $k\neq0$). We will now analyze the equations introducing $\MW$ and $\MZ$ to simplify the dynamics in the process. 

We first use  $\EEqT =0$ to set  $\PHB(r,\omega,k) = 0$ for non-zero $k$ and simplify the Einstein's equations to the set 
\begin{equation}\label{eq:ssindL}
\begin{split}
\mathbb{E}_1 
&= 
	\Dz_+ \Tone+ \frac{k^2}{d-1}\, \PHE +\frac{2\,a'}{d-1}\, k^2\, r^2f\, \MV \,, \\
\mathbb{E}_2
&=
    -i \omega\, \Tone+ \frac{k^2}{d-1}\, \PHO \,,\\
\mathbb{E}_3
&= 
       \left( \Dz_+  + 2\,rf \right)\left[\Dz_+ \PHW-\PHE \right] - i\omega\,\PHO   + \left(\omega^2-k^2f -\frac{1}{2}\,r^3\,f \left[(d+5)\,f' + 2r\, f''\right] \right) \PHW\\
&\qquad 
	-2\,a'\,rf\, \left( r \Dz_+ + \frac{2k^2}{d-1}\right) \MV\,.
\end{split}
\end{equation}
These three together with $\EEq_0$ (with now the $\PHB$ term set to zero) comprise the full dynamical content of the action \eqref{eq:SEMaxA}.

\paragraph{The Weyl factor and momentum flux fields:} To solve these equations we proceed to analyze the constraint equations, trivializing them by a suitable field redefinition. Let
\begin{equation}\label{eq:T123Xpar}
\Tone= -\frac{k^2}{d-1}\, \MW \,, \qquad \PHO= -i\omega\, \MW\,, \qquad 
\PHE =   \Dz_+ \MW -2\, a'\,r^2f\, \MV\,.
\end{equation}  
This ensures that the first two equations in \eqref{eq:ssindL} are satisfied. The third equation $\mathbb{E}_3$, which is a relation between $\PHW$ and $\MW$, can be converted into a dynamical equation. We implement one more redefinition motivated by observing  that $\PHW$ and $\MW$ are not independent but related to each other through the relation  (notice now a shift compared to \cite{He:2022jnc} due to the function determining the Maxwell potential $\MV$)
\begin{equation}\label{eq:PHEXpar}
\PHE- \frac{1}{r} \, \Tone= \left(\Dz_ + - \frac{1}{2}\, r^2\, f'\right) \PHW = \left(\Dz_+ +  \frac{k^2}{d-1}\, \frac{1}{r}\right)\, \MW -2\, a' \, r^2f\, \MV\,.
\end{equation}  
The first equality follows from \eqref{eq:T1def} and the second from \eqref{eq:T123Xpar}. We solve this system by introducing a field $\MZ$ as in \cite{He:2022jnc}: 
\begin{equation}\label{eq:MWXpar}
\begin{split}
\MW 
&= 
	\frac{r}{\Lk}\left[\Dz_+  -\frac{r^2\, f'}{2}\right]   \MZ + \frac{2\,(d-1)\, r^3f\, a'}{\Lk}\, \MV \,, \\
\PHW 
&= 
	   \frac{1}{ \Lk}\left[r\,\Dz_+ + \frac{k^2}{d-1} \right]\, \MZ+ \frac{2\,(d-1)\, r^3f\, a'}{\Lk}\, 
	   \MV\,.
\end{split}
\end{equation}
Note that these two fields satisfy a linear relation 
\begin{equation}\label{eq:linearXSZ}
\MW = \PHW- \frac{1}{(d-1)} \, \MZ \,.
\end{equation}  
Armed with this reparameterization we can further simplify the system. Using the relation $\Dz_+ \PHW-\PHE =\Dz_+\PHW - \Dz_+\MW +2\, a'\,r^2f\,\MV = \frac{1}{d-1}\,\Dz_+\MZ +2\,a'\,r^2f\, \MV$, we find that the remaining Einstein's equation, $\EEq_3$, reduces to a second order ODE for $\MZ$.

\paragraph{The coupled Einstein-Maxwell dynamics:}
After the dust settles, our redefinitions have resulted in reducing the dynamical content to two  fields: $\MZ$ and $\MV$. The dynamics of these fields is described by a system of two coupled equations, which after some simplification read 
\begin{equation}\label{eq:ZVOff}
\begin{split}
&
	r^{d-3}\Dz_+\left(r^{3-d}\Dz_+\MV\right) + \left(\omega^2-k^2 f -\frac{2(d-3)^3(d-1)}{\Lk}r^2 a^2 f^2\right)\MV \\
&\qquad\quad 
	+ \frac{(d-2)^2a f}{\Lk}\left(r \Dz_+ \MZ+\frac{k^2}{d-1}\MZ\right)=0\, ,\\
&r^{d-3}\Lk^2 \Dz_+\left(\frac{1}{r^{d-3}\Lk^2}\Dz_+\MZ\right) + \left(w^2-k^2f\left(1-\frac{2(d-2)^2}{\Lk}a^2\left(\frac{2h-1}{1-h}\right)\right)\right)\MZ\\
&+4(d-2)a f  \left[\Lk +(d-1)r^2\left((d-3)f-r f'\right) - (d-1)^2(d-2)^2 \frac{r^2 a^2 f}{\Lk}\left(\frac{2h-1}{1-h}\right)\right]\MV = 0\,.
\end{split}
\end{equation}
For purposes of computing real-time correlators on the grSK geometry, it is important that the dynamics is governed by time-reversal invariant equations. As explained in \cite{Jana:2020vyx}, it is this fact that allows one to construct smooth solutions on the grSK geometry. 
 
\subsection{Decoupling the scalar modes}
\label{sec:ZVdiagonaleqs}

The equations \eqref{eq:ZVOff} can be decoupled by the following field redefinitions:  
\begin{equation}\label{eq:diagRed}
\MZ = 
	\frac{\Lk}{h} \, \Vd + h \, \Zd \,, 
	\qquad
\MV = \frac{d-2}{2}\frac{a}{\Lk}\MZ+ \frac{r_+}{4\,\Lk}
		\left(\bRQ\,\BQT^2\, h\, \Zd 
		- (2+\BQT^2)\frac{\Lk}{h} \, \Vd\right) \,,
\end{equation}
where $\BQT^2$ was defined in \eqref{eq:diagDefs}. Using \eqref{eq:diagRed} in \eqref{eq:ZVOff} we obtain the equations \eqref{eq:ZVDiagEom}. While the kinetic terms work out straightforwardly, we note that the equations have potential functions which are quite involved. We have simplified them to the extent possible, grouping together terms by powers of $\Lk$.  

For the field $\Vd$ we find the following expression for the potential $\VVd$:
\begin{equation}\label{eq:VdPot}
\begin{split}
\VVd 
&= 
	(d-2) \, \frac{k^2\, f}{(1+\BQT^2)\,\Lk^2}\,(1-h)\; \VVd^{(1)} 
	+(d-2)\, \frac{r^3 f' f}{4\,(1+\BQT^2)\,h^2\,\Lk^2} \, \BQT^2\; \VVd^{(2)}\,,\\
\VVd^{(1)} 
&=
	-\frac{4}{d-1}\, \Lk^2+(d-2)\,(d-1)\,r^5 f' f \left(\frac{1-2\,h}{h}\right)^2 - \BQT^2 \left(2f(h-(d-2))+r f'\,h\right)\frac{r^2\,\Lk}{h^2}\\
&\quad
	+2\left[\left(d-2-\left(1+2(d-3)\,h\right)h\right)f+(2h-1)\,r f'\, h\right]\frac{r^2\,\Lk}{h^2}\, ,\\
\VVd^{(2)} 
&= 
	2\left(2h-1\right)h \,\Lk^2  + 2(d-1)\left[2\,(d-2)\,f-(5d-9)\,f h+\left(4\,(d-2)\,f-r f'\right)h^2\right]
	r^2 \, \Lk\\
&\quad
-(d-2)\,(d-1)^2\,(2h-1)^2\, r^5 f' f\,.
\end{split}
\end{equation}
On the other hand the potential for $\Zd$ takes the form
\begin{equation}\label{eq:ZdPot}
\begin{split}
\VZd 
&= 
	\frac{d-2}{(1+\BQT^2)\,\Lk^2}\, k^2\, (1-h)f \;  \VZd^{(1)} - \frac{(d-1)\, (d-2)\, r^5 f'}{2\, (1+\BQT^2)\, \Lk^2\, h^2}\, \BQT^2f \; \VZd^{(2)}  \,,\\
\VZd^{(1)} 
&= 
	\frac{4}{d-1}\, \Lk^2 + 2\left[2\left((d-3)\, f-r f'\right)h-(d-3)\,(1+\BQT^2) f\right]
	\frac{r^2\,\Lk}{h} \\
&\quad
	+ \frac{(d-1)\,(d-2)\,r^5 f f'}{h^2}\left(1+\BQT^2-2\,h\right)\left(2\,h-1\right) \,,\\
\VZd^{(2)} 
&= 
	\left((d-3)\, f-\frac{1}{2}r f'\right)h\,\Lk - \frac{1}{2}\, (d-1)\,(d-2)\,r^3 f f'\,(2h-1)\, ,
\end{split}
\end{equation}
As noted earlier, these expressions were derived in \cite{Kodama:2003kk}, but our presentation and rewriting of them should make some of the structure more transparent. It is important that the modulation function $\Lk$ only appears in the potential terms in the dynamics of  $\Vd$.

\section{Variational principle in the scalar sector}
\label{sec:actionderive}
 
The dynamics of the gravitational system is governed by the Einstein-Maxwell system with the appropriate Gibbons-Hawking boundary term and counterterms, as indicated in \eqref{eq:SEMaxA}. We now describe how the effective action for the decoupled designer fields, and the variational principle that leads to their equations of motion \eqref{eq:ZVDiagEom}, can be derived. As in the neutral case discussed in \cite{He:2022jnc}, we will find it helpful to proceed in stages: first we work out the action for the fields parameterizing the perturbations in the Debye gauge, then pass to the $\MV$ and $\MZ$ variables (where we can already decipher the variational principle), before finally explaining the desired result in terms of the decoupled designer field $\Vd$ and $\Zd$.

\subsection{Action in the Debye gauge}
\label{sec:SDebye}
 
We can write the action first in terms of  \eqref{eq:EOWDeb}. The bulk piece of the action can be decomposed as
\begin{equation}\label{eq:EHMAct}
\sqrt{-g}\left(R + d(d-1)- \frac{1}{2}F_{AB}F^{AB}\right) 
= 	
	 L_{_\text{EOW}}^\text{\tiny{bulk}} + \pdv{r}L_{_\text{EOW}}^\text{\tiny{bdy}}  
	  + L^\text{\tiny{bulk}}_{_\MV}+ \partial_r L^{\text{\tiny{bdy}}}_{_\MV}  
	   +\frac{1}{r^{d-1}\,f^2}\,\EEq_0^2\,.
\end{equation}
The final term on the decomposition  has been written in terms of the simplified Maxwell equation \eqref{eq:E0}. Since it does not contribute to the variational principle or the on-shell action,  we shall ignore it in all future analysis. The bulk contributions are
\begin{equation}\label{eq:EOWVGrav}
\begin{split}
L_{_\text{EOW}}^\text{\tiny{bulk}}
&=
	\frac{d-1}{4\, f \,r^{d}} 
	\Bigg\{
	 d\, r \left(\Dz_+\PHW \right)^2
		- \frac{2}{f}\, \Dz_+\PHW\Dz_+\PHE 
		+2\, (d-3)\, r \, \PHE \Dz_+\PHW  \\
&\quad
	+2\,(d-2)\, r\, \PHW \, \Dz_+\PHE + 2\, r^2\left(d-(d-2)\, (d+1)\, f\right)\PHW\, \Dz_+\PHW  \\
&\quad
	+\frac{4}{(d-1)\,f}\left[(d-2)\,\partial_v \PHW\,\Dz_+\PHO
		+ \partial_v \PHO\, \Dz_+\PHW 
	+ \frac{1}{r \,f}\left(\partial_v\PHO\, \Dz_+\PHE
		-\partial_v\PHE\, \Dz_+\PHO\right)\right]  \\
&\quad
	 +\frac{2\,r}{(d-1)\,f}\left(d\,(d-3)+(d^2-5d+8)\,f\right)\PHW \,\partial_v\PHO 
	 	- \frac{8\left(d-f\right)}{(d-1)\, f^2}\, \PHE\, \partial_v\PHO  \\
&\quad
	+\frac{2\,k^2}{(d-1)\,r \,f}\left(\PHO^2-\PHE^2\right)  
	- \frac{2}{f}\left[\omega^2 - k^2 \,f +(d-3)\,(d-2)\,r^2\, 
		f^2\right]\PHE\, \PHW \\
&\quad
	+\;r \left[(d-2)\, (\omega^2-k^2 f)
		+r^2 f \left(-2\,d\,(2\,d-5)+(d+2)\,(d-2)^2\,f \right)\right]\PHW^2\\
&\quad
	-\left(\frac{(d-2)}{d-1}\, \frac{a}{r f}\right)^2
		\bigg[(d-2)\,(d-1)(4\,d-7)\, r^3 f^3\, \PHW^2 \\	
& \hspace{3.5cm}
		 + 2\, (d-3)\, r f\,  \PHW\, \partial_v \PHO- 8\, \PHE\,\partial_v\PHO\bigg]
		 \Bigg\}\, ,\\
L^\text{\tiny{bulk}}_{_\MV} 
&=
  - \frac{k^2}{r^{d-1}f}\left(\Dz_+\MV\, \Dz_+\MV - (\omega^2-k^2 f)\, \MV^2 
  	- 2\,(d-2)^2\, a \,f\, \MV\,\PHW\right)\, .
\end{split}
\end{equation}

Similarly, we can expand the Gibbons-Hawking boundary term as
\begin{equation}
2\sqrt{-\gamma}\,K = L_{_\text{EOW}}^{\text{\tiny{GH}}} -  L_{_\text{EOW}}^{\text{\tiny{bdy}}} \,.
\end{equation}
The second term $  L_{_\text{EOW}}^{\text{\tiny{bdy}}}$ precisely cancels with the boundary term arising from the Einstein-Hilbert part of the bulk action in \eqref{eq:EHMAct}. We will therefore not indicate this term explicitly, as it plays no role in our analysis, apart from providing a consistency check of our evaluation.  The rest of the contribution evaluates to:
\begin{equation}\label{eq:EOWGH}
\begin{split}
 L_{_\text{EOW}}^{\text{\tiny{GH}}}
 &=
	 \frac{1}{r^{d-1}\,f^2}\left(
	 	(d-2)\,r\, f\, \PHW\, \partial_v\PHO - \PHE \,\partial_v \PHO\right) 
	 	- \frac{(d-1)\,(d-2)}{r^{d-3}}\, \PHW\, \PHE\\
 &\qquad
	 +\frac{d-1}{4\,r^{d-4}}\left((d-2)\,r \,f'+(2\,d+1)(d-2)\,f -d\right)\PHW^2\, .
\end{split}
\end{equation}
The remaining boundary term in \eqref{eq:EHMAct} comes from integration by parts of the Maxwell term and reads
\begin{equation}\label{eq:MaxBdy}
\begin{split}
L^{\text{\tiny{bdy}}}_{_\MV}  
&=
  \frac{2\, k^2}{r^{d-3}} \, \MV \, \Dz_+ \MV\, .
\end{split}
\end{equation}

We note that the structure of this action can be pretty much guessed by judiciously combining the corresponding analysis in the \SAdS{d+1} case. In particular, the Einstein-Hilbert and Gibbons-Hawking terms work very similarly to the analysis described in Appendix B.1 of \cite{He:2022jnc}, while the Maxwell part can be inferred from the probe designer gauge field analysis reported in  Section 8.3 of \cite{Ghosh:2020lel}. The main changes are a contribution from the background Maxwell field strength to the gravitational action and the difference in the background function $f(r)$ (see also \cref{fn:fidentity}). 

Finally, we have the counterterms  for the Einstein-Hilbert and Maxwell actions:
\begin{equation}
L^\text{ct} = 
	L_{_\text{EOW}}^\text{ct}  + L_{_\text{Max}}^\text{ct}\, .
\end{equation}
We consider the counterterms to leading and next-to-leading order in the asymptotic expansion, i.e., to quartic order in boundary gradients. The gravitational contributions are 
\begin{equation}\label{eq:GravCT}
\begin{split}
L_{_\text{EOW}}^\text{ct}
&= 
	-\sqrt{-\gamma}\left[
		2\, (d-1) + \frac{1}{d-2}{}^{\gamma}R
	+ \frac{1}{(d-4)\,(d-2)^2} \left(
		\tensor[^\gamma]{R}{_{\mu\nu}} \, \tensor[^\gamma]{R}{^{\mu\nu}}  - \frac{d}{4(d-1)}\, \tensor[^\gamma]{R}{}^2
	\right) \right] \\
&=
	 \frac{d-1}{4\,r^{d-2}\, f^{3/2}}
	 	\left[1 - \frac{1}{(d-1)^2\, (d-2)\, (d-4)}\, \frac{k^4}{r^4}\right]	 \PHE^2 \\
&\qquad 
	+ \frac{d-1}{4\,r^{d-2}\sqrt{f}}\left(\omega^2-k^2 \,f - d\,(d-2)\, r^2 f\right)\PHW^2\\ 
&\qquad 
	+ \frac{1}{2\,r^{d-1}\, \sqrt{f}}\left(k^2 + (d-1)\, (d-2)\, r^2\right)\PHE \,\PHW\,,
 \end{split}
\end{equation}
while the Maxwell contributions are 
\begin{equation}\label{eq:MaxCT}
\begin{split}
 L_{_\text{Max}}^\text{ct} 
 &=
 	\frac{1}{4\,(d-4)}\sqrt{-\gamma}\left({}^{\gamma}F_{\mu\nu} {}^{\gamma}F^{\mu\nu}
	+\frac{1}{(d-4)\,(d-6)} {}^{\gamma}F_{\mu\nu} \Box_{\gamma} {}^{\gamma}F^{\mu\nu}\right)\\
&= 
	-\frac{k^2}{2\,(d-4)\,r^{d-2}\,\sqrt{f}}\left(1+ \frac{\omega^2-k^2 \,f}{(d-4)\,(d-6)\,r^2 f}\right)(\Dz_+\MV)^2\, .
\end{split}
\end{equation}

This completes the basic analysis for the action. From the nature of the boundary terms, it is clear that $\PHE, \PHO, \PHW$ have Dirichlet boundary conditions, since \eqref{eq:EOWGH} is a functional of these fields alone.  On the other hand, $\MV$ obeys Neumann boundary conditions, owing to the boundary term \eqref{eq:MaxBdy}.

\subsection{Action for designer scalars}
\label{sec:SVsZs}

Now we would like to write the action in terms of the designer field $\MZ$ eliminating the gravitational fields $\PHE, \PHO, \PHW$. To this end we simply plug in \eqref{eq:T123Xpar}, \eqref{eq:PHEXpar}, and \eqref{eq:MWXpar} into the action obtained hitherto. For reasons explained in \cite{He:2022jnc}, despite the fact that the redefinitions above involve $\MZ$, $\Dz_+\MZ$, and even $\Dz_+^2\MZ$, we obtain a simple two-derivative action in the bulk. After a series of simplifications, we find that the bulk action takes the form 
\small
\begin{equation}\label{eq:DesBulk}
\begin{split}
L_{_{\MZ, \MV}}^{\text{\tiny{bulk}}} 
&=
	-A_1\left\{
		\left(\Dz_+\MZ\right)^2 
		- \left[\omega^2- k^2\,f \left(1-\frac{2\,(d-2)^2}{\Lk}\frac{2h-1}{1-h}\,a^2\right) \right]\MZ^2\right\}
	-\frac{(d-2)^3a^2k^4}{2(d-1)r^{d-1}\Lk^2}\MZ^2
	\\
&\quad
	-\frac{k^2}{r^{d-1}f}\left\{\left(\Dz_+\MV\right)^2 - \left(\omega^2-k^2\,f - \frac{2\,(d-2)^3\,(d-1)}{\Lk}r^2\, a^2\, f^2\right)\MV^2\right\}
	\\
&\quad
	- \frac{2\,(d-2)^2\,a^2\,k^2}{r^{d-1}\,\Lk}\left[\Lk + (d-1)\,r^2\left((d-3)\,f-r\, f'\right) 
	- (d-2)^2\,(d-1)^2\left(\frac{2h-1}{1-h}\right)\frac{a^2\, r^2\, f}{\Lk}\right]\MV^2\\
&\quad
	+\frac{(d-2)^2 \,a \,k^2}{r^{d-1}\,\Lk^2}\left(k^2 + (d-2)\,(d-1)\,a^2\right)
		\left(r\, \MV \,\Dz_+\MZ+ \frac{2\,k^2}{d-1}\, \MV \, \MZ\right)\\
&\quad
	+\frac{(d-2)\, a \,k^2}{r^{d-1}\, \Lk\, f}\left(\Dz_+\MV\, \Dz_+\MZ 
	- (\omega^2 -k^2 f)\,\MV\,\MZ\right) ,
\end{split}
\end{equation}
\normalsize
where 
\begin{equation}\label{eq:A1def}
A_1  = \frac{(d-2)\left(k^2+(d-2)\,(d-1)\,a^2\right)k^2}{4\,(d-1)\,r^{d-1}f\, \Lk^2} \,.
\end{equation}	
One can check that the  variation of this action reproduces the equations of motion  \eqref{eq:ZVOff}.

All the higher derivative terms arising from the field redefinitions can be  packaged into  boundary terms. This is now complicated when written in terms of $\MV$ and $\MZ$. We find it efficacious to write the boundary action in terms of fields $\MW$ and $\PHW$. The final result can be written in the form
\begin{equation}\label{eq:DesBdy}
\begin{split}
L_{_{\MZ, \MV}}^{\text{\tiny{bdy}}}
&=
	-\frac{d-1}{4\,r^{d-2}\,f}\left[(\Dz_+\MW)^2 + c_1 \, \PHW \, \Dz_+\MW 
	+ c_2\,  \MZ\, \Dz_+\MW + c_3 \,\PHW^2 + c_4\, \MZ \,\MW + c_5\, \MZ^2 \right]\\
&\quad
	+  \frac{2\,k^2}{r^{d-3}}\, \MV\, \Dz_+\MV - \frac{1}{2}\, (d-2)^2\, (d-1)^2\, 
	\frac{a^2\, f}{r^{d-4}\, \Lk} \left(\frac{4\,k^2}{d-1} + r^3 f'\right)\MV^2\\
&\quad
	-\frac{(d-2)\,(d-1)\,a}{2\, r^{d-2}}\left[2\,r \MV\,\Dz_+\MW + \left(\frac{4\,k^2}{d-1}
		+ r^2\left(2\,(d-2)\, f+r \,f'\right)\right)\MW \,\MV \right.\\
&\quad
\left. 
	+\frac{r^2}{\Lk}\left(
		\frac{2\left((d-2)\,f+r\, f'\right)\Lk}{d-1}-\frac{1}{2}\,r^4 f'^2\right)\MV\, \MZ\right]\, ,
\end{split}
\end{equation}
with
\begin{equation}\label{eq:DesBdy2}
\begin{split}
c_1 
&= 
	\frac{2\,k^2}{(d-1)\,r} +r  \left(2\,(d-2)\, f+r f'\right)  , \\
c_2 
&=
	-\frac{2\,k^2}{(d-1)^2\, r} \,,\\
c_3 
&=
	 \frac{r f'}{2\,(d-1)}\, \Lk + \frac{d}{2} \, r^2 f\left(2\, f+ r f'-2\, (d-1)\right) 
	 + (\omega^2 - k^2 \,f)\,,\\
c_4 
&= 
	-\frac{k^2}{(d-1)^2}\, r f'\,,\\
c_5 
&=
	 \frac{(d-2)\, k^4\left(2\,(d-2)\, a^2-r^3\, f'\right) f}{2\,(d-1)^2\, \Lk^2} - \frac{k^2\left(k^2+(d-1)\,r^3f'\right)r f'}{2\,(d-1)^3 \,\Lk}-\frac{k^2}{(d-1)^2\,\Lk}(\omega^2-k^2 \,f)\,.
\end{split}
\end{equation}

Finally, the counterterms \eqref{eq:GravCT}, \eqref{eq:MaxCT} can themselves be combined and written as (retaining only quadratic counterterms) 
\small
\begin{equation}\label{eq:MasterCT}
\begin{split}
 L_{_{\MZ, \MV}}^{\text{ct}} 
&= 
	\frac{d-1}{4\,r^{d-2}f^{3/2}} \left(\left(\Dz_+\MW\right)^2 
	+ b_1 \, \PHW \Dz_+\MW + b_2\, \PHW^2\right)
	+\frac{(d-1)\,(d-2)\,a}{2\,r^{d-3}\sqrt{f}}\left(2\,\MV \,\Dz_+\MW +b_1\, \PHW\,  \MV\right)\\
&\quad
 	-\frac{k^2}{2\, (d-4)\, r^{d-2}\sqrt{f}}\left(1+ \frac{\omega^2-k^2\, f}{(d-4)\,(d-6)\,r^2 f}\right)(\Dz_+\MV)^2 + \frac{(d-1)\,(d-2)^2\,a^2 \sqrt{f}}{r^{d-4}} \, \MV^2\,,
\end{split}
\end{equation}
\normalsize
where 
\begin{equation}\label{eq:MasterCT2}
\begin{split}
b_1 &= \frac{2\,k^2 \,f}{(d-1)\,r} + 2\,(d-2)\,r f\,,\\
b_2 &= (\omega^2- k^2 \,f)f - d\,(d-2)\,r^2\, f^2\, .
\end{split}
\end{equation}

While the boundary terms and the counterterms are complicated, we can achieve a great deal of simplification by combining them. This leads to no effect on the variational principle, but has the advantage of making clear what are the terms that dominate in the asymptotic region, which then inform us what the nature of the boundary condition is. In any event, putting the two terms in \eqref{eq:DesBdy} and \eqref{eq:MasterCT},  we find the combination simplifies to
\begin{equation}
\begin{split}
L_{_{\MZ, \MV}}^{\text{\tiny{bdy}}}+ L_{_{\MZ, \MV}}^{\text{ct}} 
&=
-	\frac{d-1}{4\,r^{d-2}f}\left[
		\left(1-\frac{1}{\sqrt{f}}\right)(\Dz_+\MW)^2 
		+ \left(c_1 -\frac{b_1}{\sqrt{f}}\right) \PHW \, \Dz_+\MW 
		+ c_2\, \MZ\, \Dz_+\MW\right. \\
&\quad
\left. 
	+\left(
		 c_3 - \frac{b_2}{\sqrt{f}}\right) \PHW^2 
		 + c_4 \,\MZ \, \MW + c_5\, \MZ^2 \right] 
		 	+ 2\,k^2\,  \frac{1}{r^{d-3}} \, \MV\, \Dz_+\MV\\
&\quad
 	-(d-1)^2\, (d-2)^2\, \frac{a^2 f}{r^{d-4}\Lk}\left(
 		\frac{2\,k^2}{d-1} + \frac{1}{2}\, r^3 f'-\frac{\Lk}{(d-1)\,\sqrt{f}}\right)\MV^2 \\
&\quad
	-2\, \frac{(d-1)\,(d-2)\,a}{2\, r^{d-3}}\left(1-\frac{1}{\sqrt{f}}\right) \MV\, \Dz_+\MW \\
&\quad
-	\frac{(d-1)\,(d-2)\,a}{2\, r^{d-3}}\left[ \left(
	\frac{4\,k^2}{(d-1)\,r}+ r\left(2\,(d-2)\, f+r f'\right) - \frac{b_1}{\sqrt{f}}\right) \PHW\,  \MV 
	\right.\\
&\quad
\left. 
	-\frac{k^2}{(d-1)\,r\Lk} \left(\frac{4\, k^2}{d-1}+ r^3\, f' \right) \MZ\,\MV\right] \\
&\quad
 	-\frac{k^2}{2\, (d-4)\, r^{d-2}\sqrt{f}}\left(1+ \frac{\omega^2-k^2\, f}{(d-4)\,(d-6)\, r^2 f}\right)
 	(\Dz_+\MV)^2 \, .
\end{split}
\end{equation}

Written in this form, we can evaluate the coefficients in front of various terms at large $r$. Additionally, we can also use an asymptotic relation obtained from our solutions in \cref{sec:ZWasym}: 
\begin{equation}\label{eq:ThetaAsymp}
\frac{1}{r^{d-1}}\Dz_+\MW = (d-2) \frac{\PHW}{r^{d-2}}\,.
\end{equation}
With these data, we can then show that the total boundary action can be whittled down to
\begin{equation}\label{eq:BdyTotal}
\begin{split}
L_{_{\MZ, \MV}}^{\text{\tiny{bdy}}}+L_{_{\MZ, \MV}}^{\text{ct}}  
&=
	\frac{(d-6)(d-1)}{8(d-2)} \, \Pbg\, \left(\frac{\Dz_+\MW}{r^{d-1}}\right)^2
	+\frac{k^2}{2\,(d-1)} \,\MZ \,\frac{\Dz_+\MW}{r^{d-1}}
	+ \frac{2\,k^2 }{r^{d-3}}\, \MV\,\Dz_+\MV \\
&\qquad
  + \text{subleading}\,.
\end{split}
\end{equation}
The essential part of the action is pretty simple in this form, and the subleading pieces capture all the terms which have subleading divergences, starting at order $r^{d-4}$. It is not hard to check from this action that the fields $\MV$ and $\MZ$ obey Neumann boundary conditions.

\subsection{Variational principle  for diagonal fields}
\label{sec:varZdVd}

We conclude our analysis of the action by finally obtaining the effective action in terms of the diagonal fields $\Vd$ and $\Zd$. Our starting point is the bulk action \eqref{eq:DesBulk}, into which we plug in the field redefinitions  \eqref{eq:diagRed} and then simplify to obtain
\begin{equation}\label{eq:ZVDBulk}
\begin{split}
L_{_{\Zd, \Vd}}^{\text{\tiny{bulk}}} 
&=
	-\frac{d \,\nu_s}{8 \, r^{d-1}}\, 
		\frac{h^2\, k^2}{f \,\Lk^2} \Np{\Zd} 
		\left[\left(\Dz_+\Zd\right)^2 -\left(\omega^2 - \left(1-\frac{(d-2)}{2}\frac{2+\BQT^2}{1+\BQT^2}\frac{r^3 f'}{h\Lk}\right)k^2f+ \VZd\right)\Zd^2 \right]\\
&\quad
-	\frac{k^2\, \Np{\Vd}}{r^{d-1}\, f h^2}\left[\left(\Dz_+\Vd\right)^2-\left(\omega^2 -  k^2 f +\VVd\right)\Vd^2\right] .
\end{split}
\end{equation}
As expected this action is diagonal, and its variation reproduces the equations \eqref{eq:ZVDiagEom}. In deriving this part we have carried out a set of integration by parts to remove some total derivative terms (terms of the form $\Vd \, \Dz_+ \Vd$ can be rewritten this way).
The coefficients $\Np{\Zd}$ and $\Np{\Vd}$ were introduced in \eqref{eq:NCdef} and originate from this final simplification. 

Next, we simplify the boundary term \eqref{eq:BdyTotal}, including the  terms that arise from integrating by parts to bring bulk action into the form given in 
\eqref{eq:ZVDBulk}. We find 
\begin{equation}\label{eq:BdyTotalD}
\begin{split}
L_{_{\Zd, \Vd}}^{\text{\tiny{bdy}}}
&=
	\frac{(d-1)\,(d-6)\,\Pbg}{8\,(d-2)}
		\left(\frac{\Dz_+\MW}{r^{d-1}}\right)^2 
	+ \frac{h\, k^2}{2\, (d-1)}\, \Zd\,  \frac{\Dz_+\MW}{r^{d-1}}
	+ \frac{k^2\,\Lk}{2\,(d-1)\,h\,\bRQ} \,\Vd \, \frac{\Dz_+\MW}{r^{d-1}} \\
&\quad
	+\frac{d_1\, k^2}{d-2} \, \Zd \, \frac{\Dz_+\MW}{r^{d-1}} 
	+ \frac{d_2\, k ^2}{d-2} \,\Vd \,  \frac{\Dz_+\MW}{r^{d-1}} 
	+ d_3\, k^2 \, \Vd \,\frac{\Dz_+\Vd}{r^{d-3}} + d_4\, k^2\,\Zd \, \frac{\Dz_+\Vd}{r^{d-3}}
  + \text{subleading}\, ,
\end{split}
\end{equation}
where
\begin{equation}
\begin{split}
d_1 
&= 
	\frac{ h\, \bRQ^2}{8\,\Lk}\left(\BQT^2+2\,(1-h)\right)^2\,,\\
d_2 
&= 
	- \frac{\bRQ}{8\, h}\left((2+\BQT^2)\BQT^2+4\, (1-h)\, h\right)\, ,\\
d_3 
&= 
	\frac{(1+\BQT^2)}{4\,h^2}\left(\BQT^2+2 \,h\right)\, ,\\
d_4 
&= 
	-\frac{(1+\BQT^2)\, \bRQ}{4\,\Lk}\left(\BQT^2+2\, (1-h)\right)\, .
\end{split}
\end{equation}

At this point, we have gathered all the pieces necessary to begin deciding what boundary conditions the designer fields $\Zd$ and $\Vd$ satisfy. Recall that  their conjugate momenta were  defined in \eqref{eq:PiD}. What is of import is that they satisfy a simple asymptotic relation:
\begin{equation}\label{eq:PiWRel}
\begin{split}
\frac{k^2}{r^{d-1}}\, \Dz_+\MW 
&=
	 (d-2)\, k^2\, \frac{\PHW}{r^{d-2}}  \\ 
&=
	-4(d-1)\, \PiZ -(d-2)\,\frac{k^2}{\bRQ} \, \PiV\,.
\end{split}
\end{equation}
Using this relation, we can further distill the boundary action into the form
\begin{equation}\label{eq:BdyDFinal}
\begin{split}
L_{_{\Zd, \Vd}}^{\text{\tiny{bdy}}}
&=
	\frac{(d-1)\,(d-1)\,\Pbg}{8\, (d-2)}\left(
		\frac{\Dz_+\MW}{r^{d-1}}\right)^2
 		- \frac{2\, \Np{\Zd}}{\Lk} \, \Zd\,  \PiZ
	 -2 \,k^2\, \Np{\Vd}\, \Vd\, \PiV 
  + \text{subleading}\,.
\end{split}
\end{equation}

Putting all the pieces together we can write the full action for the designer fields as
\begin{equation}\label{eq:ZVactionfinal}
S[\Vd,\Zd] = \int dr\, \int_k\,  L_{_{\Zd, \Vd}}^{\text{\tiny{bulk}}} + \int_k L_{_{\Zd, \Vd}}^{\text{\tiny{bdy}}} 
\end{equation}	
Now we have all the necessary ingredients to obtain the boundary conditions for the diagonal master fields. Making use of the definitions \eqref{eq:PiD}, the total variation of the action, up to equations of motion, is
\begin{equation}\label{eq:deltaSZV}
\begin{split}
\delta S\left[\Zd, \Vd\right] 
	&=\int_k\left[  
		 -\frac{2\,\Np{\Zd}}{\Lk}\,\Zd\,\delta\PiZ  
		 -2\, k^2\, \Np{\Vd} \, \Vd \,  \delta \PiV + \frac{(d-1)\, (d-6)\,\Pbg}{4\,(d-2)}
		 	\left( \frac{\Dz_+\MW}{r^{d-1}}\right) \, \delta\left(\frac{\Dz_+\MW}{r^{d-1}}\right) \right] ,
\end{split}
\end{equation}
which together with \eqref{eq:PiWRel} implies that, when computing the gravitational on-shell action, the master fields must obey Neumann boundary conditions:
$\delta\PiV = \delta\PiZ = 0$.

\section{Solutions in the boundary gradient expansion}
\label{sec:gradexpfns}

We describe the general solution in a boundary gradient expansion (in powers of $\bwt $ and $\bqt$, working in the limit $\bwt, \bqt\ll 1$) for the designer fields $\Vd$ and $\Zd$.  We find it convenient to use the basis of functions that appears in the $SO(d-2)$ tensor and vector perturbations of the \RNAdS{d+1} black hole. This has the advantage of not only utilizing functions that have been understood in \cite{He:2021jna}, but it also helps demonstrate the unified origin of these results from a fluid/gravity perspective. We try to be self-contained below, but the reader might find it useful to consult Appendices B--D of \cite{He:2021jna} in parallel. 

It will be convenient to introduce a dimensionless inverse radius coordinate  
\begin{equation}\label{eq:ridef}
\ri= \frac{r_{+}}{r}\,,
\end{equation}	
and to parameterize the  solution in terms of boundary to bulk Green's functions for the fields. We will solve for the ingoing Green's function, parameterizing it by unit boundary expectation values of the corresponding field and demanding regularity at the future horizon. We therefore adopt an ansatz:
\begin{equation}
\Gin{\Zd,\Vd}(\ri,\bwt, \bqt)  
= 
	\exp\left(\sum_{n,m=1}^\infty \,  (-i)^m \,\bwt^m\, \bqt^{2n}\, 
	\Mser{\Zd,\Vd}{m,2n}(\ri) \right) .
\end{equation}

We utilize the following two integral transforms to give compact expressions. The first of these depends only on the metric function $f$: 
\begin{equation}\label{eq:MItransform}
\mathfrak{T}\big[\mathfrak{g}\big](\ri) \equiv \int_0^\ri\, \frac{d\rib}{f(\rib)}\, \mathfrak{g}(\rib) , \ \hspace{0.5cm} 
\hat{\mathfrak{T}}\big[\mathfrak{g}\big](\ri) \equiv \int_1^\ri\, \frac{d\rib}{f(\rib)}\, \mathfrak{g}(\rib) \,.
\end{equation}	
On the other hand, the second transform also depends on the ohmic function $h$ and is defined to be 
\begin{equation}\label{eq:MItransformH}
\mathfrak{H}\big[\mathfrak{g}\big](\ri) \equiv \int_0^\ri\, \frac{d\rib}{f(\rib)\, h(\rib)^2}\, \mathfrak{g}(\rib) , \ \hspace{0.5cm} 
\hat{\mathfrak{H}}\big[\mathfrak{g}\big](\ri) \equiv \int_1^\ri\, \frac{d\rib}{f(\rib) \, h(\rib)^2}\, \mathfrak{g}(\rib) \,.
\end{equation}	

\renewcommand{\arraystretch}{1.2}
\begin{table}[th!]
\centering
\begin{tabular}{|c|c|c|}
\hline
\shadeR{$\mathfrak{H}\big[\mathfrak{g}\big]$}   &   \shadeB{$\mathfrak{g}$ }   &		\shadeB{Asymptotics}\\     \hline
$ \Mser{\MYd}{1,0}$ 	&	 $h(\ri)^2 -h(1)^2\ri^{d-3}$ 	& 		$\ri - \frac{h(1)^2}{d-2}\ri^{d-2}$ \\ \hline
$\Mser{\MYd}{2,0} $ 	&	 $-h(1)^2\ri^{d-3}\Dfnh{\MYd}{2,0}(\ri)$		& 	$ -\frac{\ri^2}{2(d-4)} +\frac{h(1)^2 \Dfn{\MY}{2,0}(1)}{d-2}\ri^{d-2} - \frac{(d-2)\sdc}{d(d-4)}\ri^{d}$\\	\hline
$\Mser{\MYd}{3,0}$ 	&	 $2h(1)^2\ri^{d-3}\Mserh{\MYd}{2,0}(\ri) $ 	& 	$ -\frac{2h(1)^2\Mser{\MYd}{2,0}(1)}{d-2}\ri^{d-2} - \frac{h(1)^2}{d(d-4)}\ri^d$ \\ \hline
$\Mser{\MYd}{4,0}$ 	&	 $2h(1)^2 \ri^{d-3}\left(\Mserh{\MYd}{3,0}(\ri)+ \frac{1}{2}\Dfnh{\MYd}{3,0}(\ri)\right)$ 	& 	$\frac{\ri^{4}}{2(d-6)(d-4)^2} + \frac{\Dfn{\MY}{0,4}(1)}{d-2}\ri^{d-2}$\\ 
		& 			& 		$+\left(\frac{1}{d(d-4)}\left(\frac{1}{d-4}+\frac{\sdc}{2}+\frac{\sdc^2}{d}-\frac{\sdc^3}{2(d-1)}\right) - \frac{\sdc}{4d\bRQ^2}\right)\ri^{d}$ \\ \hline
$\Dfn{\MYd}{2,0}$ 	&	 $h(1)^2\ri^{d-3} - \frac{h(\ri)^4}{h(1)^2}\ri^{3-d}$ &  $\frac{1}{(d-4)h(1)^2}\ri^{4-d} + \frac{h(1)^2}{d-2}\ri^{d-2}$ \\ \hline
$\Dfn{\MYd}{3,0}$ 	&	 $-h(1)^2 \ri^{d-3}\Dfnh{\MYd}{2,0}(\ri)^2$ & 	$\frac{1}{(d-6)(d-4)^2 h(1)^2}\ri^{6-d} - \frac{h(1)^2\Dfn{\MYd}{2,0}(1)}{d-2}\ri^{d-2}$\\	
		&		& $-\frac{2 h(1)^2}{d(d-4)(d-2)}\ri^d + \frac{\Dfn{\MYd}{2,0}(1)}{d-4}\ri^2$ \\ \hline
\end{tabular}
\caption{Functions used to parameterize the gradient expansion of the Markovian transverse photon  modes in \cite{He:2021jna}. We have collected the essential functions defined using the integral transform (\ref{eq:MItransform}). We also indicate the leading asymptotic behavior; note that only the last two functions diverge as $\ri\rightarrow 0$.}
\label{tab:Ygradsol}
\end{table}

\begin{table}[th!]
\centering
\begin{tabular}{|c|c|c|}
\hline
\shadeR{$\mathfrak{T}\big[\mathfrak{g}\big]$}	&	\shadeB{$\mathfrak{g}$ }		&		\shadeB{Asymptotics} \\		\hline

$\Mser{\ann}{1,0}$ 	& 
	$1 - \ri^{\ann}$ 	&  
		$\ri - \frac{\ri^{1+\ann}}{1+\ann} $ \\ \hline

$\Mser{\ann}{0,2}$ 	& 
	$\frac{\ri}{1-\ann}\left(1-\ri^{\ann-1}\right)$  	&  
			$\frac{1}{1-\ann}\left(\frac{\ri^2}{2} -\frac{\ri^{1+\ann}}{1+																																	\ann}\right)$ \\ \hline

$\Mser{\ann}{2,0}$ 	& $-\ri^{\ann}\Dfnh{\ann}{2,0}(\ri)$ 						  	& $\frac{1}{2(1-\ann)}\ri^{2} + \frac{\Dfn{\ann}{2,0}(1)}{1+\ann}																																	\ri^{1+\ann}$ \\ \hline

$\Mser{\ann}{3,0}$	& $2\ri^{\ann}\Mserh{\ann}{2,0}(\ri)$							& $-\frac{2\Mser{\ann}{2,0}(1)}{1+\ann}\ri^{1+\ann}+																																					\frac{\ri^{3+\ann}}{(3+\ann)(1-\ann)}$ \\ \hline

$\Mser{\ann}{1,2}$ 	& $2\ri^{\ann}\Mserh{\ann}{0,2}(\ri)$ 						&  $-\frac{2\Mser{\ann}{0,2}(1)}{1+\ann}\ri^{1+\ann}+																																					\frac{\ri^{3+\ann}}{(3+\ann)(1-\ann)}$ \\ \hline

$\Mser{\ann}{4,0}$	& 	$2\ri^{\ann}\left(\Mserh{\ann}{3,0}(\ri)+\frac{1}{2}\Dfnh{\ann}{3,0}(\ri)\right)$ 
	& 
		\scriptsize $-\frac{\ri^4}{4(3-\ann)(1-\ann)^2}-\frac{\Dfn{\ann}{3,0}(1)+
		2\Mser{\ann}{3,0}(1)}{1+\ann}\ri^{1+\ann}$\normalsize 
											\scriptsize $-\frac{\Dfn{\ann}{2,0}(1)}{(3+\ann)(1-\ann)}																																			\ri^{3+\ann}$\normalsize	\\ \hline

$\Mser{\ann}{2,2}$	& \scriptsize$2\ri^{\ann}\left(\Mserh{\ann}{1,2}(\ri)-\frac{1}{1-\ann}\left(\Dfnh{\ann}{1,2}(\ri)-\Mserh{\ann}{2,0}(\ri)\right)\right)$\normalsize																				&	\scriptsize$\frac{1}{2(\ann-3)(1-\ann)^2}\ri^{4}+\frac{1-(1-																																		\ann)\Dfn{\ann}{2,0}(1)}{(3+\ann)(1-\ann)^2}\ri^{3+\ann}$																																			\normalsize \\ 
										&																					& \scriptsize $-\frac{2\left(\Mser{\ann}{2,0}(1)-\Dfn{\ann}{1,2}																																		(1)+(1-\ann)\Mser{\ann}{1,2}(1)\right)}{(1+\ann)(1-\ann)}																																			\ri^{1+\ann}$ \normalsize\\ \hline

$\Mser{\ann}{0,4}$	& $\frac{1}{1-\ann}\left(\Mserh{\ann}{0,2}(\ri)-\Mserh{2-\ann}{0,2}(\ri)\right)$ 
	& 
		\scriptsize $-\frac{\ri^4}{4(3-\ann)(1-\ann)^2}-\frac{\Mser{\ann}{0,2}(1)
			-\Mser{2-\ann}{0,2}(1)}{1-\ann^2}			\ri^{1+\ann}+\frac{\ri^{3+\ann}}{(3+\ann)(1-\ann)^2}$																																			\normalsize \\ \hline
$\Dfn{\ann}{2,0}$	& $\ri^{\ann}-\ri^{-\ann}$ &  \scriptsize $-\frac{1}{1-\ann}\ri^{1-\ann} + \frac{1}{1+\ann}\ri^{1+\ann}$\normalsize \\ \hline
$\Dfn{\ann}{3,0}$	& $-\ri^{\ann}\Dfnh{\ann}{2,0}(\ri)^2$ 
&  \scriptsize $-\frac{1}{(3-\ann)(1-\ann)^2\ri^{3-\ann}} + \frac{2}{(3+\ann)(1+\ann)}\ri^{3+\ann}$\normalsize \\
	& 	& \scriptsize $- \frac{\Dfn{\ann}{2,0}(1)}{1-\ann}\ri^2 - \frac{\Dfn{\ann}{2,0}(1)^2}{1+\ann}\ri^{1+\ann}$  \normalsize \\ \hline
$\Dfn{\ann}{1,2}$ 		& 		$-\ri \Dfnh{\ann}{2,0}(\ri)$  & 	\scriptsize $\frac{1}{(3-\ann)(1-\ann)}\ri^{3-\ann} - \frac{1}{(3+\ann)(1+\ann)}\ri^{3+\ann} + \frac{\Dfn{\ann}{2,0}(1)}{2}\ri^2$ \normalsize \\ \hline
\end{tabular}
\caption{Functions appearing in the gradient expansion of probe Markovian fields of index $\ann$ in the \RNAdS{d+1} background from \cite{He:2021jna}. We have expressed the solutions using the integral transform (\ref{eq:MItransform}).  The two cases of interest are $\ann = d-1$ which corresponds to minimally coupled scalars (the transverse traceless tensor gravitons) and $\ann= d-3$ which analytically continues to the non-Markovian solutions of index $\ann =3-d$ (with appropriately chosen constants to ensure the quoted asymptotic fall-offs). }
\label{tab:Margradsol}
\end{table}

\subsection{Gradient expansion of the decoupled fields}
\label{sec:cdsound}

Modulo introducing a few new functions, we will for the most part  employ a function basis comprising of solutions to the Markovian minimally coupled scalar field (index $\ann =d-1$). Additionally, we will also use the functions appearing in the solution for the physical photons in the \RNAdS{d+1} background. These were captured by a designer scalar $\MY$ in \cite{He:2021jna}. We tabulate the function basis obtained in the aforementioned reference in \cref{tab:Ygradsol,tab:Margradsol}.   

In writing the asymptotic expansions for the auxiliary functions appearing in the gradient expansion, we have set $f(u)=h(u)=1$ in the integrals, thus only indicating the leading contribution from each piece of the integrand. In general, there are subleading pieces, since $f(u) = 1 +\order{u^d}$ and $h(u) = 1+ \order{u^{d-2}}$, which lie betwixt the terms we have quoted. We have refrained from indicated these to keep the expressions simple; these terms have been carefully accounted for in the computation of the physical data. Later when we give the expressions for the physical fields, we will indicate the full asymptotic expansion.

\paragraph{The charge diffusion field $\Vd$:}  The solutions appearing in the gradient expansion at zero spatial momentum are given by the corresponding solutions to the field $\MY$:
\begin{equation}\label{eq:VDsol1}
\begin{split}
\Mser{\Vd}{1,0}(\ri) 
&=  \Mser{\MY}{1,0}(\ri)	 + \Dfn{\MY}{2,0}(\ri) \, ,\\
\Mser{\Vd}{2,0}(\ri) 
&=
	 - \Mser{\MY}{2,0}(\ri) +  \Dfn{\MY}{2,0}(1) \Dfn{\MY}{2,0}(\ri)-\frac{1}{2} \Dfn{\MY}{2,0}(\ri)^2 \, ,\\
\Mser{\Vd}{3,0}(\ri) 
&=
	  - \Mser{\MY}{3,0}(\ri)  - \Dfn{\MY}{3,0}(\ri) + 2\Dfn{\MY}{2,0}(\ri)\Mserh{\MY}{2,0}(\ri)  \\
		&\quad
		+ \Dfn{\MY}{2,0}(1)\left( \Dfn{\MY}{2,0}(1) \Dfn{\MY}{2,0}(\ri)- \Dfn{\MY}{2,0}(\ri)^2-2\Mser{\MY}{2,0}(\ri)\right) + \frac{1}{3} \Dfn{\MY}{2,0}(\ri)^3 \, ,\\
\Mser{\Vd}{4,0}(\ri) 
&=
	 -\Mser{\MY}{4,0}(\ri)  +2 \Dfn{\MY}{2,0}(\ri)\Mserh{\MY}{3,0}(\ri) -2 \Dfn{\MY}{2,0}(1)\Mser{\MY}{3,0}(\ri) 
		+ \Dfn{\MY}{2,0}(\ri)\Dfnh{\MY}{3,0}(\ri) \\
		&\quad
		+2\Dfn{\MY}{2,0}(\ri)\Mserh{\MY}{2,0}(\ri)\left(\Dfn{\MY}{2,0}(1)-\Dfnh{\MY}{2,0}(\ri)\right) 
		- \Dfn{\MY}{2,0}(1)\, \Dfn{\MY}{3,0}(\ri) + \Dfn{\MY}{2,0}(1)^3\, \Dfn{\MY}{2,0}(\ri) \\
		&\quad
		+\Dfn{\MY}{2,0}(1)^2\left(\Dfn{\MY}{2,0}(1)\Dfn{\MY}{2,0}(\ri)-\frac{3}{2}\Dfn{\MY}{2,0}(\ri)^2-2\Mser{\MY}{2,0}(\ri)\right) - \frac{1}{4}\Dfn{\MY}{2,0}(\ri)^4\,.
\end{split}
\end{equation}

Once we consider solutions with non-zero spatial momentum, the potentials will modify the solution drastically.  Remarkably, at leading order we can still write the solution in terms of previously defined functions as (nb: $h(1) = 1-\sdc$)
\begin{equation}\label{eq:VDsol2}
\begin{split}
h(1)\,\Mser{\Vd}{0,2}(\ri) 
&=
	-\frac{d-4}{d-2}\Mser{d-3}{0,2}(\ri) + \frac{1}{d-2}\Dfn{d-3}{2,0}(\ri)+\sdc \frac{h(1)}{d-1} \Dfn{\MYd}{2,0}(\ri)\\
&\quad
	+\sdc\left[-\Mser{d-1}{0,2}(\ri) + \frac{h(1)}{d-1}\left( \Mser{\MYd}{1,0}(\ri) -\Mser{d-1}{1,0}(\ri)\right)\right]-\frac{\nu_s}{d-2}\, \frac{\sdc^2}{1+Q^2}\, \frac{h(1)}{h(\ri)}\, \ri^{d-2}\,.
\end{split}
\end{equation}

Higher order solutions can be found recursively as 
\begin{equation}\label{eq:VDsol3}
\begin{split}
\Mser{\Vd}{1,2}(\ri) 
&= 
	\frac{2}{h(1)^2}\mathfrak{T}\big[h(\ri)^2\ri^{3-d}\Mserh{\Vd}{0,2}(\ri)\big]\,,\\
\Mser{\Vd}{2,2}(\ri) 
&=
	2\, \mathfrak{T}\left[h(\ri)^2\ri^{3-d}\left(\frac{1}{h(1)^2}\Mserh{\Vd}{1,2}(\ri)-\Dfnh{\Vd}{2,2}(\ri)\right)\right] , \\
\Mser{\Vd}{0,4}(\ri) 
&=  
	\hat{\mathfrak{T}}\left[h(\ri)^2\ri^{3-d}\left(\Dfnh{\Vd}{0,4}(\ri)  + \frac{4(d-2)}{d(d-1)^2}\frac{\sdc^2}{1+Q^2}\hat{\Xi_{_\Vd}}(\ri) \right)\right], \\
\end{split}
\end{equation}
where we define some additional auxiliary functions:
\begin{equation}\label{eq:VDs3Ax}
\begin{split}
\Dfnh{\Vd}{2,2}(\ri) 
&= 
	\hat{\mathfrak{T}}\left[\frac{f(\ri)^2}{h(\ri)^2}\ri^{d-3}\frac{d}{d\ri}\Mser{\Vd}{0,2}(\ri)\frac{d}{d\ri}\Mser{\Vd}{2,0}(\ri)\right]\,, \\
\Dfnh{\Vd}{0,4}(\ri) 
&= 
	-\int_{1}^{\ri} d\rib\frac{f(\rib)\rib^{d-3}}{h(\rib)^2}\left(\frac{d}{d\rib}\Mser{\Vd}{0,2}(\rib)\right)^2\,, \\
 \hat{\Xi}_{_\Vd}(\ri) 
 &= 
 	\int_{1}^{\ri} \frac{d\rib}{h(\rib)^4}\rib^{2d-7}\left(\rib^2\left(h(\rib)^2-2\right)-2\sdc f(\rib)\frac{\left((d-1)h(\rib)^2+h(\rib)-3(d-2)\right)}{d(1+Q^2)h(\rib)^2}\right) .
\end{split}
\end{equation}

\paragraph{The sound field $\Zd$:} We begin with the solutions of $\Zd$, which, as noted above, satisfies a second order equation that changes character from being Markovian of index $\ann = d-1$ to a non-Markovian field of index $\ann = 3-d$. This change of character owes to the function $\Lk$, which being only dependent on the spatial momentum indicates that  the frequency dependent parts are captured by previously encountered solutions. In fact, the situation is even better, since the solutions for non-vanishing spatial momentum can themselves be given in terms of solutions of the minimally coupled scalar solutions.

First, note that the functions appearing at $\order{\bqt^0}$, i.e., those that determine the frequency dependent pieces, are in fact  very simple. They are the Markovian minimally coupled scalar solutions 
\begin{equation}\label{eq:ZDsol1}
\Mser{\Zd}{m,0}(\ri) = \Mser{d-1}{m,0}(\ri) \,.
\end{equation} 
The momentum dependent pieces start to involve additional complications owing to the nature of the potential in the wave equation for $\Zd$, but at low orders one can significantly simplify the analysis. We will write the answers in a manner suggested by the analysis of the neutral plasma. Let us introduce a class of functions $\Mser{\Zd}{n,m}(\ri)$, which are charged analogs of the functions seen in the gradient expansion of the scalar graviton mode in \cite{He:2022jnc}. These functions solve ODEs with specified sources -- the only distinction from \cite{He:2022jnc} is that the function $f(\ri)$ appearing in the equations is the one appropriate for the \RNAdS{d+1} geometry.

At second order in momenta, we find for instance 
\begin{equation}\label{eq:ZDsol2}
\Mser{\Zd}{0,2}(\ri) =\Mser{\MZ}{0,2}(\ri) + \frac{2}{d(d-1)}\frac{\sdc^2}{1+Q^2}\frac{\ri^{d-2}}{h(\ri)}\, ,
\end{equation}
where the first term is simply the solution found in \cite{He:2022jnc} (with the charge dependent $f(\ri)$), while the second term accounts for the additional modification coming from the coupling to the charged background. At higher orders we have 
\begin{equation}\label{eq:ZDsol3}
\begin{split}
\Mser{\Zd}{1,2}(\ri) 
&=
	\Mser{\MZ}{1,2}(\ri) - \frac{2(d-2)}{(d-1)^2}\sdc \Mser{d-1}{0,2}(\ri)\, , \\
\Mser{\Zd}{2,2}(\ri) &= 
	\Mser{\MZ}{2,2}(\ri) +\frac{\sdc}{(d-1)^2}\left[\frac{\ri^{4-d}}{(d-2)(d-4)}-2\Dfn{d-1}{1,2}(\ri)-(d-2)(3+Q^2)\Mser{d-1}{1,2}(\ri)+\frac{\Dfn{\Zd}{2,2}(\ri)}{d-2}\right] , \\
\Dfn{\Zd}{2,2}(\ri) &=
	\mathfrak{T}\left[\ri^{d-1}(1-\ri^2)\right] ,\\
\Mser{\Zd}{0,4}(\ri) &= 
	\Mser{\MZ}{0,4}(\ri)+\sdc \frac{(d-4)\Mser{d-3}{0,2}(\ri)-\Dfn{d-3}{2,0}(\ri)}{(d-1)^3(d-2)} - \frac{2\,d-5}{(d-1)^3}\sdc \Mser{d-1}{0,2}(\ri)\\
&\qquad
	- \frac{2\sdc^2}{d^2(d-1)^2(1+Q^2)^2}\left(\frac{1+h(\ri)}{h(\ri)}\right)^2 +\frac{8\sdc^2}{d^2(d-1)^2(1+Q^2)^2} \,.
\end{split}
\end{equation}

The functions $\Mser{\Zd}{n,m}$ themselves can be expressed in terms of the minimally coupled scalar solutions and a few additional auxiliary functions. For completeness, we write them below; the interested reader can find further details in Appendix D of \cite{He:2022jnc}, though we again caution that the functions appearing herein require $f(\ri)$ to be the \RNAdS{d+1} function.   
\begin{equation}\label{eq:NSound}
\begin{split}
\Mser{\MZ}{0,2}(\ri) 
&=  
   	 - \frac{d-3}{d-1}\, \Mser{d-1}{0,2}(\ri) \,, \\
\Mser{\MZ}{1,2}(\ri) 
&=  
     - \frac{d-3}{d-1}\, \Mser{d-1}{1,2}(\ri) + \frac{4\, (d-2)}{d\,(d-1)} \, \Mser{d-1}{0,2}(\ri)\,, \\
\Mser{\MZ}{2,2}(\ri) 
&= 
       - \frac{d-3}{d-1}\,  \Mser{d-1}{2,2}(\ri) - \frac{4}{d\,(d-1)} \, \Mser{d-1}{2,0}(\ri)+ \frac{4}{d\,(d-1)} \, \Mser{d-1}{0,2}(\ri) \\
&
    + \frac{2}{d\, (d-1)} \,\Dfn{\MZ}{2,2}(\ri) + \frac{4}{d\,(d-1)}\,\Dfn{d-1}{1,2}(\ri)- \frac{2}{d\,(d-1)\,(d-2)\,(d-4)} \, \ri^{4-d} \,,\\
\Mser{\MZ}{0,4}(\ri) 
&=  
       \frac{(d-3)^2}{(d-1)^2}\, \Mser{d-1}{0,4}(\ri)  + \frac{4\,(d-3)}{d(d-1)^2} \, \Mser{d-1}{0,2}(\ri) + \frac{2}{d\,(d-1)^2(d-2)}\, \Dfn{\MZ}{0,4}(\ri)\,,\\
  \Dfn{\MZ}{0,4}(\ri) 
&= 
  - \mathfrak{T}\big[ \ri^{3-d} -\ri^{d-1} \big]  , \\
\Dfn{\MZ}{2,2}(\ri)
&=
    \mathfrak{T} \left[\ri^{d-1}\left(2\,\Dfnh{\MZ}{0,4}(\ri) -\frac{1}{d-2}\right) +\frac{1}{d-2}\, \ri^3\right] \,.
\end{split}
\end{equation}
%

\subsection{Asymptotic data of designer fields and physical functions}
\label{sec:ZWasym}
The asymptotics of  the charge diffusion mode $\Vd$ and the designer sound field $\Zd$ can be captured by giving the behaviour of the ingoing Green's function. Note that we are normalizing the solution so that the physical solution on the grSK contour is parameterized by the moduli $\PoV_{\skL,\skR}$ and $\PoZ_{\skL,\skR}$, which, in particular, demands that the constant part of $\Gin{\Vd,\Zd}$ is unity.

\paragraph{The charge diffusion mode $\Vd$:}  Let us begin with the charge diffusion field $\Vd$, which is explicitly non-Markovian of index $\ann =3-d$ at every order in the gradient expansion. It therefore diverges as $\ri^{4-d}$, with a coefficient that is related to the charge diffusion dispersion function $\Kc$. There are several subleading divergences, and additionally, some convergent terms. We collect here the terms that are necessary to compute the near-boundary expansion of the metric and gauge potential to the order where they contribute to the physical conserved currents.
\small
\begin{equation}\label{eq:Vasym}
\begin{split}
\Gin{\Vd}(\ri,\omega,\bk) 
&=
	\frac{\Kc }{(1-\sdc)^2}\, \frac{\ri^{4-d}}{d-4} \, \left(1-i\,\bwt\, \ri\right) \left[1-\frac{\bqt^2+ (d-7)\,\bwt^2}{2\,(d-6)}\,  \ri^2 \right]-\frac{\bwt^4\, \ri^{7-d}}{3\,(1-\sdc)^2\, (d-4)}\\
&\qquad
	+ \left(1-i\,\bwt \, \ri\right)\left[1+ \left(1-\frac{d-3}{(d-2)\,(d-4)}\, \frac{\nu_s\, \sdc}{1+Q^2}\,\bqt^2\right)
	\left(\frac{\sdc\,\Kc}{(1-\sdc)^2}  \right.\right. \\
&\left.\left.\qquad \qquad \qquad\qquad 
	+ \frac{d-1}{2\,(d-2)} \left(\frac{\bqt^2}{d-1} -\bwt^2\right)\right)\ri^2\right]
	- \frac{i\,\bwt^3}{3}\, \ri^3 + \order{\ri^4} \,.
\end{split}
\end{equation}
\normalsize
In writing this expression, we have assumed $d>7$  for simplicity, so that we can disambiguate the constant term from the divergent terms. For low dimensions we run into the issue that both the fast and slow fall-off modes can be normalizable with appropriate boundary conditions. This is clear from the fact that the leading $\ri^{4-d}$ divergence can be rendered normalizable with alternate boundary conditions in $d\leq 4$ \cite{Witten:2003ya,Marolf:2006nd}. The subleading divergences with $d-6$ factors signal the   higher order counterterms. We have also assembled the terms factoring out $1-i\,\bwt \, \ri$ factors, which indicates that the terms arise from  a $\Dz_+$ action and again makes it clear that $\Vd$ is non-Markovian (since the source term is the conjugate momentum given by $\Dz_+ \Vd$). 

The charge dispersion function $\Kc$ can be computed accurate to quartic order in gradients,  extending the result quoted in the main text in \eqref{eq:Kc3}. The resulting answer is
\begin{equation}\label{eq:Kc4}
\begin{split}
\Kc(\bwt,\bqt) 
&= 	
	-i  \bwt +\left(1-\frac{\sdc}{d-1}\right)\frac{\bqt^2}{d-2} - \Dfn{\MYd}{2,0}(1)\,   \bwt^2 
		- i\left(2\,\Mser{\MYd}{2,0}(1)-\Dfn{\MYd}{2,0}(1)^2\right) \bwt^3 \\
&\qquad  
	-2i\, \Mser{\Vd}{0,2}(1)\, \bwt\, \bqt^2
    -2 \left(\Mser{\Vd}{1,2}(1) - h(1)^2\, \Dfn{\Vd}{2,2}(1)\right)  \bwt^2\,\bqt^2 \\
&\qquad
    + \left[h(1)^2\, \Dfn{\Vd}{0,4}(1) + \frac{2\,\nu_s}{d-1}\, \frac{\sdc^2}{1+Q^2}\, \Xi_{_\Vd}(1) \right]\bqt^4 \\ 
&\qquad 
	- \left(2\, \Mser{\MYd}{3,0}(1) + 4\, \Mser{\MYd}{2,0}(1) \, \Dfn{\MYd}{2,0}(1) - \Dfn{\MYd}{2,0}(1)^3 + \Dfn{\MYd}{3,0}(1)\right)\, \bwt^4    \,.
\end{split}
\end{equation}	

\paragraph{The sound mode $\Zd$:} We can similarly work out the asymptotics of the sound mode $\Zd$. The main novelty here, as in the neutral case, is that the solution is Markovian at low orders in the gradient expansion, and picks up non-Markovian behaviour only at fourth order, in $\order{\bqt^4}$ and $\order{\bwt^2\,\bqt^2}$. Therefore, the low order solutions and all the frequency dependent pieces are simple. We also find a simpler structure of subleading divergences.  Retaining the pieces that enter into the corresponding asymptotics of the boundary metric and gauge potential, we find 
\begin{equation}\label{eq:Zasym}
\begin{split}
\Gin{\Zd}(\ri,\omega,\bk) 
&=
    \frac{\nu_s \,\bqt^2\, \KS}{(d-2)^2\, (1+Q^2)}  \, \frac{\ri^{4-d}}{d-4}
            + 1- i\bwt\, \ri+ \sum_{i=2}^4\, \mathfrak{a}_i \, \ri^i    
            + \mathfrak{a}_{d-2}\, \ri^{d-2} \\
&\qquad \qquad 
           + \mathfrak{a}_{d-1}\, \ri^{d-1}
           	- \frac{\mathfrak{a}_d}{d}\,  \ri^d  +\cdots 
           	+\mathfrak{a}_{2d-4}\, \ri^{2d-4} 
           	+\order{\ri^{2d-3}}\,,
\end{split}
\end{equation}
where  (we use $\frac{\sdc}{1+Q^2} = \frac{d(d-2)}{2\, \bRQ^2}$ to keep some terms simple)
\begin{equation}\label{eq:Zasymcfs}
\begin{split}
\mathfrak{a}_2
&=
    \frac{d-3}{2\,(d-2)}\, \KS -   \frac{\bqt^2}{d \, (1+Q^2)} \left(\Gatt + \frac{2\,\sdc}{(d-1)\,(d-2)} \left(\frac{\bqt^2}{d-1}-\bwt^2\right)\right) , \\ 
\mathfrak{a}_3
&=
	\frac{i\bwt}{6\,(d-2)}\left[(d-5)\, \bwt^2 - 3\frac{d-3}{d-1}\,\bqt^2 \right] 
		+ \frac{d\,\nu_s}{2\, \sdc\, \bRQ^2}\, \bwt^2\, \bqt^2\,,  \\
\mathfrak{a}_4
&= 
	\frac{d-1}{8\,(d-2)\,(d-4)} \left(\frac{16}{d\,(d-1)}+d-5\right)\, \KS^2 - \frac{d+3}{12\, d}\, \bwt^4\,,\\ 
\mathfrak{a}_{d-2}
&=
		\frac{d\,\nu_s}{2\,\bRQ^2}\,\sdc\, \bqt^2 \left(1-\frac{d\,\nu_s}{\bRQ^2}\, \bqt^2\right)\,, \\
\mathfrak{a}_{d-1}
&=
	- \frac{d\,\nu_s}{2\,\bRQ^2} \, \sdc\, (i\bwt  \bqt^2) \,, \\
\mathfrak{a}_d 
&= 
		\Gatt + \Omega_s - \frac{\sdc}{d-2}\left(\bwt^2-\frac{\bqt^2}{d-1}\right) - \frac{d\,(d-3)}{2\,(d-1)}\, \frac{\sdc}{\bRQ^2}\, \, \bqt^2\, \KS\,, \\
 \mathfrak{a}_{2d-4}
 &=
 	\frac{d\, \nu_s}{2\,\bRQ^2}\,\sdc^2  \left(1-\frac{5\,d\,\nu_s}{4\,\bRQ^2} \, \bqt^2\right) .
\end{split}
\end{equation}

The coefficient of $\ri^d$ has been singled out above. It  in fact agrees  with the sound attenuation function $\Gatt$ defined in \eqref{eq:KS2} up to quadratic order in gradients (since $\bwt = \pm\frac{\bqt}{\sqrt{d-1}}$ on the dispersion locus defined by $\KS =0$) . However, it has higher order corrections;  we have parameterized these by $\Omega_s$, which is
\begin{equation}\label{eq:OmegaS}
\begin{split}
\Omega_s
&=
    2 i \,   \Mser{d-1}{2,0}(1)\, \bwt^3     
   + 2i\, \left(\frac{\nu_s}{d-2} + \frac{d-3}{d-1}\,  \Mser{d-1}{0,2}(1) -\frac{\sdc}{(d-1)^2}\right) \bwt\, \bqt^2  \\
&\qquad \qquad 
   + \lambda_{\omega}\, \bwt^4  + \lambda_k\, \bqt^4 
  +   \lambda_{\omega k}\, \bwt^2\, \bqt^2\,.
 \end{split}
\end{equation}	
  We defined the coefficients $\lambda_\omega$, $\lambda_k$, and $\lambda_{\omega k} $ to simplify the expressions above. They are
\begin{equation}\label{eq:lamdefs}
\begin{split}
\lambda_\omega 
&=
	2\,\Mser{d-1}{3,0}(1) + \Dfn{d-1}{3,0}(1) \,,  \\
\lambda_k 
&=
	-\frac{2\,d-5}{\sdc} \, \frac{1}{(d-1)^2\,\bRQ^2} + \frac{(d-3)^2}{(d-1)^2\, (d-2)} \left[\Mser{d-1}{0,2}(1) - \Mser{3-d}{0,2}(1)\right]\,, \\
\lambda_{\omega k}
&= 
	\frac{\nu_s}{d-2 }\bigg[
			d\, (d-3)\, \left( 
				\frac{\Dfn{d-1}{1,2}(1) - \Mser{d-1}{2,0}(1) }{d-2}+ \Mser{d-1}{1,2}(1)  \right) 
				-2\, \left( \Dfn{d-1}{2,0}(1) +\Dfn{\MZ}{0,4}(1) \right) \\
&\qquad \qquad \quad 
				+ \frac{d\,(d-2)}{d-1}\, (3+Q^2)\, \sdc \, \Mser{d-1}{0,2}(1)
				+\frac{d \,(1+2\,Q^2)}{2\,\bRQ^2\, \sdc}
	 \bigg]\,.
\end{split}
\end{equation}
In  \cite{He:2022jnc}, it was argued that this function has the correct properties to define the sound attenuation to quartic  order; we will return to this point when we compute the asymptotics of $\PHW$, where we will be able to make a clear statement.

\paragraph{The gauge potential:} The first physical function we will examine is the gauge potential for the Maxwell field $\AGR$. Our convention is to parameterize the solution using the boundary sources for $\Vd$ and $\Zd$, which will allow us to read off the boundary gauge potential $A_\mu\, dx^\mu$ directly. For simplicity, we focus on the temporal component (which has a limiting boundary value), and refrain from writing similar expressions for the radial component. Given our parameterization of the gauge field in \eqref{eq:EMpert} and \eqref{eq:EOWDeb} implies that the boundary gauge potential is\footnote{
	The expressions below are valid both on the left and right boundary of the grSK geometry, but we do not indicate that explicitly in our notation to keep the expressions clean. } 
\begin{equation}\label{eq:Avbdy}
\lim_{\ri\to 0}\, \AGR_A\, dx^A = \ri^{d-3}\, \Dz_+ \MV\, dv \equiv A_v\, dv\,.
\end{equation}	

We obtain upon direct evaluation
\begin{equation}\label{eq:GinAv}
\begin{split}
\AGR_v 
&=
	\mathbf{C } \, (1-i\,\bwt\,\ri) + \order{\ri^2}
\end{split}
\end{equation}	
 with
\begin{equation}\label{eq:Csource}
\begin{split}
\mathbf{C } 
&=
	-\frac{2+\BQT^2}{4}\, \frac{r_+\, \Kc}{(1-\sdc)^2}\PoV  +  \frac{r_+\, \bRQ\, \sdc}{(1+Q^2)^2}\, \frac{\nu_s\, \bqt^2\, \KS}{d\,(d-1)\,(d-2)^2}  \, \PoZ \\
&=
	r_+\, \left[ -\frac{2+\BQT^2}{4}\,   \JoV  
	+  \frac{2}{d\,\nu_s}\, \bRQ\, \BQT^2\, \JoZ \right] 	.
\end{split}
\end{equation}	
We can view $\mathbf{C}$ as a chemical potential induced on the boundary from the solution when we are off the dispersion locus. The contribution from the charge diffusion, encoded in $\JoV$, is intuitive, but we also have a contribution proportional to $\JoZ$, which arises from the sound mode. The latter can be understood as being due to the coupling between the two modes. Indeed, when $Q =0$ we can see that this piece switches off, as it must for a probe charge field in a black hole background.

\paragraph{The metric functions:}
The rescaled metric functions $\{\PHE,\PHO, \PHW\}$ and the field $\MW$ can be similarly recovered from the solutions for $\Vd$ and $\Zd$.  We start with $\PHW$, focusing on terms that enter computation of physical quantities, eschewing subleading divergences that are canceled by counterterms.  It is helpful to first write an expression for $\PHW$ directly in terms of the fields $\Vd$ and $\Zd$. Using \eqref{eq:EOWMZ} and \eqref{eq:MZVdiagonal} one can check that
\begin{equation}\label{eq:PhiWDiag}
\begin{split}
\PHW 
&= 
	\frac{h(r)}{\Lk(r)}\left(
		r\,\Dz_+ + \frac{k^2}{d-1}\right)\Zd + \frac{1}{h(r)}\left(r\,\Dz_++\frac{k^2}{d-1}\right)\frac{\Vd}{\bRQ}\\
&\quad
 +(d-2)\left(
 	\bqt^2 \,\sdc - \frac{d\,(d-1)}{4}(1+Q^2)\,\BQT^2\, h(r)\right)\frac{r_+^d\, f(r)}{r^{d-4}\,\Lk(r)}\left(\frac{\Zd}{\Lk(r)} - \frac{\Vd}{\bRQ\, h(r)^2}\right) .
\end{split}
\end{equation}
This form is particularly helpful to ascertain the  asymptotics.  The modulation function $\Lk$ has terms divergent near the boundary when expanded in gradients, but the structure is such that these are cancelled between the $\Dz_+\Zd$ and $\frac{\Zd}{\Lk} $ terms (this is not an issue for $\Vd$). Carrying through the computation, we find at the end of the day  
\begin{equation}\label{eq:PHWasym}
\begin{split}
 \PHW 
&= 	  
	\mathbf{W}\, r^{d-2} + \sum_{j=3}^{5} \, W_j\, r^{d-j} + \mathbf{G} + \cdots \,,
\end{split}
\end{equation}
where the ellipses are the decaying terms in $\PHW$ (behaving as $\ri^k$ with $k>0$). We have checked that  the subleading divergences in $W_k$ are taken care of by counterterms and do not enter in any physical answer. The quantities of relevance are the non-normalizable mode $\mathbf{W}$ and the constant piece $\mathbf{G}$.

The coefficient $\mathbf{W}$ is related to the boundary metric source, and can be parameterized in terms of the sources for the designer fields $\Vd$ and $\Zd$ as
\begin{equation}\label{eq:Wsource}
\begin{split}
\mathbf{W} 
 &=
	\frac{\Kc}{(1-\sdc)^2\,r_+^{d-4}} \, \frac{\PoV } {\bRQ}
  	+ \frac{2\,\KS}{d\,(d-1)\,(d-2)\,r_+^{d-2}\, (1+Q^2)}\, \PoZ   \\ 
&=
		\frac{\JoV }{\bRQ} \, + \frac{8}{d\,\nu_s}\, \JoZ   \,.
\end{split}
\end{equation}
Once again, the source which parameterizes the  induced boundary metric is an admixture of the charge diffusion and sound mode pieces. In this case we see that the contribution from the charge part vanishes when $Q\to 0$, recovering the phonon source obtained in \cite{He:2022jnc}. 

On the other hand the constant piece in $\PHW$ is  
\begin{equation}\label{eq:WGzeroA}
\begin{split}
\mathbf{G}
&=
	\left[
			\frac{1}{d-2}\left(\bwt^2-\frac{\bqt^2}{d-1}\right) - \left(1- \frac{d\,\nu_s}{2\,\bRQ^2}\,\bqt^2 \right)\frac{\sdc \,\Kc}{(1-\sdc)^2}
		\right]  \frac{r_+^2}{\bRQ} \, \PoV   \\
&\qquad
	+ \frac{d\,\nu_s}{2\,\bRQ^2\, \sdc} \left(1-\frac{d\,\nu_s}{2\,\bRQ^2}\,\bqt^2\right)  \left(\Gatt+ \Omega_s - \frac{\sdc}{d-2}\left(\bwt^2 - \frac{\bqt^2}{d-1}\right) \right)
    \PoZ  \,.	
\end{split}
\end{equation}
This expression can be rewritten in terms of the boundary sources for the fields $\PoV$ and $\PoZ$ as\footnote{
	In writing this expression we have made an educated guess that 
	$1-\frac{d\,\nu_s}{2\,\bRQ^2}\,\bqt^2$ resums into  $\frac{2}{2+\BQT^2}$, see \cref{fn:BQTprops}. }  
\begin{equation}\label{eq:WGZero}
\begin{split}
\mathbf{G}
&=
	\frac{1}{d-2}\left[
		\left(\bwt^2-\frac{\bqt^2}{d-1}\right) \, r_+^2\, \frac{\PoV}{\bRQ} + 
		 \frac{\nu_s}{1+Q^2}  \Ost\, \PoZ 
		\right]\\
&\qquad
	-  \left(1-\frac{d\,\nu_s}{2\,\bRQ^2}\,\bqt^2  \right) r_+^{d-2} \,\sdc \, \frac{\JoV }{\bRQ}
	+ \frac{2\,d\,(d-1)\,\Pbg}{r_+^2\, \bRQ^2} \, \frac{2}{2+\BQT^2}\, \JoZ
	\,. 
\end{split}
\end{equation}
We have defined a new function 
\begin{equation}\label{eq:OstDef}
\Ost
= 
	\left(1-\frac{d\,\nu_s}{\bRQ^2}\, \bqt^2\right) \Gatt(\omega,\bk) + \Omega_s(\omega,\bk) \,,
\end{equation}	
which we argue below is the sound attenuation function accurate to quartic order.

This constant piece $\mathbf{G}$ is important for, as we shall see, it enters into the expression for the spatial part of the stress tensor (all other components either care about the source terms, or about the normalizable part $\MZ_\text{ren}$, which we discuss below). The contribution from $\Vd$ is straightforward, and includes both a source piece (proportional to $\JoV$) and the field operator $\PoV$. The latter furthermore involves the non-dissipative part of the phonon kinetic term. 

On the other hand the contribution from $\Zd$, while superficially similar, is a bit more involved. The operator contribution proportional to $\PoZ$ picks up the coefficient $\mathfrak{a}_d$ in \eqref{eq:Zasymcfs}. We will try to justify below that we should treat $\Ost$ defined above in \eqref{eq:OstDef} as the  sound attenuation function accurate to quartic order in gradients.  We have written it in this particular manner to make the similarity with the neutral fluid analysis of \cite{He:2022jnc} manifest. 

The leading divergence in $\MW$ is the same as for $\PHW$, as can be inferred from \eqref{eq:EOWMZ}, since $\MZ$ diverges as $\ri^{4-d}$. From here we can quickly check that the leading divergence in $\Dz_+\MW$ and thence in $\PHE$ is also simply related to that of $\PHW$. One can, for instance, check that the following asymptotic relations hold:
\begin{equation}\label{eq:PhiEAsymp}
\begin{split}
\frac{\PHE}{r^{d-1}} 
&
=
	 \frac{ \Dz_+\MW}{r^{d-1}} + 2\,(d-2)\, a f\, \frac{\MV}{r^{d-2}}
=  
	\frac{ \Dz_+\MW}{r^{d-1}}+ \order{r^{2-d}}\\
&=
	(d-2)\, \frac{\PHW}{r^{d-2}} + \cdots \,.
\end{split}
\end{equation}
This information is all we will need for the evaluation of the boundary observables: induced metric, on-shell action, and conserved currents, so we will refrain from giving expressions for $\MW$ and $\PHE$.

\paragraph{Higher order sound attenuation:}
In the non-normalizable mode of $\PHW$ and $\Dz_+\MW$, the contribution from the sound field $\Zd$ is given by $\KS$, which to quartic order in gradients is given in \eqref{eq:KS2}. We note  that both the $\mathfrak{a}_d$ in $\Zd$ and the constant term $\mathbf{G}$ of $\PHW$ are given by a combination of $\Gatt$ and  $\Omega_s$ defined as $\Ost$ in \eqref{eq:OstDef}. The function  $\Omega_s$ starts out at cubic order in gradients, so it could indeed be viewed as updating $\Gatt$. 

Much of our intuition for identification here comes from the manner in which $\mathbf{G}$ enters the stress tensor (see discussion below \eqref{eq:TYsimp}). That will make clear that we should actually treat $\Omega_s$ as the cubic and quartic corrections to sound attenuation. In other words, we  conjecture $\Ost$ to be the sound attenuation function.  In fact, we can further simplify this function. We note 
\begin{equation}\label{eq:OstSimplify}
\begin{split}
\Ost 
&=
	\Gatt + \Omega_s - \frac{\sdc}{d-2}\left(\bwt^2-\frac{\bqt^2}{d-1} \right)  - \frac{2\,\sdc}{d-2}\,\KS+ \cdots \\
&= 	
	-i \bwt + \Dfn{d-1}{2,0}(r_+) \,\bwt^2 + \frac{d-3}{(d-1)(d-2)}\bqt^2 +\Omega_s + \cdots \,.
\end{split}
\end{equation}
In the first line, we rearranged terms, recognizing that the factors can be reassembled into the sound dispersion function $\KS$ (which contains a factor of $\Gatt$). In the second line we used the explicit form of $\Gatt$ from \eqref{eq:KS2} and dropped the dispersion function 
contribution (its vanishing sets the on-shell condition, so this contribution is of higher order). 
The correction to $\Gatt$ comes from non $\KS$ terms in the first line, but these we note are at least of cubic order. $\Omega_s$ is manifestly so from \eqref{eq:OmegaS}, but so is $ \frac{\sdc}{d-2}\left(\bwt^2-\frac{\bqt^2}{d-1} \right)$, since on the sound locus $\omega = \pm\frac{k}{\sqrt{d-1}} + \order{k^2}$. 

With these identifications we can solve for the sound dispersion locus accurate to quintic order. The crucial point is that written this way, $\Ost$ and $\mathfrak{a}_d$ agree on the sound dispersion locus.  In particular, our conjecture implies that the sound dispersion locus accurate to quintic order in gradients takes the form
\begin{equation}\label{eq:SPropA}
\begin{split}
\bwt(\bqt) 
&= 
	\frac{\bqt}{\sqrt{d-1}} -i\,\frac{\nu_s}{2\,(1+Q^2)}\,\bqt^2 
	+ \frac{\nu_s\,  \mathfrak{h}_3}{2\, \sqrt{d-1}\, (1+Q^2)^2} \, \bqt^3 
	+ \frac{i \nu_s \mathfrak{h}_4}{d\,(d-1)\,(1+Q^2)^2} \,\bqt^4  \\
&\qquad	
	+ \frac{\nu_s\,  \mathfrak{h}_5}{(d-1)^{3/2}} \,\bqt^5 + \cdots\,,\\
\mathfrak{h}_3 
&=
	\frac{d\,(d-2)-4}{2\,d\,(d-2)}+\frac{d-3}{d-2}\,Q^2 + (1+Q^2)\, \Dfn{d-1}{2,0}(1)\,,\\
 \mathfrak{h}_4
 &=
 	d\left((d-3)\,\Mser{d-1}{0,2}(1)+\Mser{d-1}{2,0}(1)\right)\left(1+Q^2\right) 
 	+ 2-(d-2)\,\Dfn{d-1}{2,0}(1) \,,\\
\mathfrak{h}_5
&= 
	\frac{(d-2)\left(3\,\Mser{d-1}{2,0}(1)+(d-3)\,\Mser{d-1}{0,2}(1)\right)}{d\,(1+Q^2)} + \frac{d\,(7+5\,Q^2)-1}{d^3\,(d-2)\,(1+Q^2)^4}\\
 &\quad
 +\frac{\left(d\,(2\,(d-3)\,Q^2-(d-6))-12\right)\Dfn{d-1}{2,0}(1)}{4\,d^4\,(1+Q^2)^3}
  +\frac{3\,(d-2)\,\Dfn{d-1}{2,0}(1)^2}{4\,d\,(1+Q^2)^2}\\
&\quad 
	+\frac{4\,d(9+Q^2(13+6\,Q^2))-4\,(31+Q^2\, (34+9\, Q^2))}{16\,d\,(d-2)\,(1+Q^2)^4} \\
&\quad 
 - \frac{(d-1)^2}{2\,(1+Q^2)}\lambda_k + \frac{\lambda_{\omega}}{2\,(1+Q^2)}+\frac{d-1}{2\,(1+Q^2)}\, \lambda_{\omega k}\,,
\end{split}
\end{equation}
where the functions $\lambda_\omega,\lambda_k,\lambda_{\omega k}$ were previously defined in \eqref{eq:lamdefs}.  This upgrades the expression  \eqref{eq:SProp} given in the main text. 

A priori, it is surprising that we can obtain results for the sound pole accurate to sextic order; we expect to find $\KS$ only to quartic order. Moreover, if we identify $\KS$ as the leading divergence in $\Zd$, we can only learn its behaviour to quadratic order, cf., \cref{sec:ZWasym}. It is when we translate this information to the metric functions that we see that non-normalizable part of 
 $ \PHW$ picks up contributions that assemble into the quartic order expression for $\KS$, which predicts $\Gatt$ as indicated in \eqref{eq:KS2}. But we are able to push this further, by exploiting the nature of the stress tensor, and how the Ward identities work, to glean information beyond the remit of our solution. While we do not prove that our identification is correct, we offer it is as an interesting conjecture for the higher order transport data damping sound propagation in the charged plasma.

\section{Boundary observables}
\label{sec:bdyobs}
 
 With the solutions for the designer fields $\Vd$ and $\Zd$ we can proceed to analyze the physical data of the dual charged fluid on the boundary.  Below we collect the information necessary to parametrize the boundary sources and expectation values of the conserved currents. 

To aid with this analysis it will be useful to have some intermediate results.  While we will most often directly use the fields $\Vd$ and $\Zd$, it is helpful to use their asymptotics to obtain the asymptotic behaviour of our intermediate variables $\MV$ and $\MZ$. We only need the renormalized value of these fields. To obtain these we can, for instance,  use the fact that the $\Dz_+ \MV$ term is the leading counterterm that renormalizes $\MV$, viz., 
\begin{equation}\label{eq:Vsren}
\MV_{_\text{ren}} = \MV - \frac{1}{(d-4)\, r\,\sqrt{f}}\, \Dz_+ \MV + \cdots  \,,
\end{equation}	
with the ellipses denoting higher order counterterms. Likewise, one can check that 
\begin{equation}\label{eq:Zsren}
\begin{split}
\MZ_{_\text{ren}} = \MZ - \frac{k^2}{(d-2)\,(d-4)}\, \frac{\Dz_+ \MW}{r^{d-1}} \, r^{d-4} + \cdots  \,.
\end{split}
\end{equation}
Using the relation to the designer fields \eqref{eq:MZVdiagonal} we can then infer  
\begin{equation}\label{eq:VEVRel}
\begin{split}
\MZ_{_\text{ren}}
&= 
	\frac{k^2}{\bRQ } \,\PoV + \PoZ \,, \\
\MV_{_\text{ren}}
&=  
	 - \frac{r_+}{4}\, (2+\BQT^2) \,\PoV+ \frac{d\,\nu_s}{2\,r_+\, \bRQ}\, \frac{1}{2+\BQT^2}\, \PoZ\,,
\end{split}
\end{equation}
where we are once again leaving implicit the boundary (L and R) labels to avoid clutter. 
This data will be helpful when we evaluate the conserved current operators.

\subsection{The boundary sources and operators}
\label{sec:bsourcevev}

Let us begin with the boundary metric. From \eqref{eq:EOWDeb} one can deduce the induced metric on the boundary to be
\begin{equation}\label{eq:indbdymetA}
\gamma_{\mu\nu}\, dx^\mu\, dx^\nu 
=
	 \left(1+\frac{\PHW}{r^{d-2}}\right) \, \eta_{\mu\nu}\, dx^\mu\, dx^\nu  
	 + \frac{\Dz_+\MW}{r^{d-1}}\, dv^2 \,.
\end{equation}	
We can now use the asymptotic data derived in \cref{sec:ZWasym}, especially \eqref{eq:PhiEAsymp} and \eqref{eq:PHWasym}, to infer that
\begin{equation}\label{eq:JMWWbdy}
\begin{split}
\lim_{r\to\infty\pm i0}\, \frac{\Dz_+\MW}{r^{d-1}} 
&=  
	\lim_{r\to\infty\pm i0}\, (d-2) \, \frac{\PHW}{r^{d-2}} \\
&=
	(d-2)\, \mathbf{W}_{\skL/\skR}  \\
&=
	(d-2)\, \frac{\JoV_{\skL/\skR}}{\bRQ}  + \left(1-\frac{d\,\nu_s}{2\,\bRQ^2}  \, \bqt^2\right) \, 4\, (d-1)\, \JoZ_{\skL/\skR}\,.
\end{split}
\end{equation}
As noted earlier,  in the limit $Q\to 0$, the charge contributions drop out and the last expression above limits to $4(d-1)\, \JoZ_{\skL/\skR}$, which was the result obtained in \cite[Eq.~(C.3)]{He:2022jnc}. Likewise, the boundary chemical potential is given by the result in \eqref{eq:Csource}, which we record here as 
\begin{equation}\label{eq:AvC}
\lim_{r\to \infty \pm i0} \, \AGR_A\,dx^A = \mathbf{C}_{\skL/\skR}\, dv\,.
\end{equation}	
%

\subsection{The boundary currents}
\label{sec:JBYT}

We can now compute the conserved currents from the boundary action. We will give the general expression for these in terms of the fields used to parameterize the perturbations, and then show that the result can be written in terms of the renormalized fields, up to source terms. 

\paragraph{Charge current:}  Let us begin with the charge current. It is given in terms of asymptotic behaviour of the bulk Maxwell field, or equivalently by varying the boundary action with respect to the (boundary) gauge potential. One has 
\begin{equation}\label{eq:Jcftgen}
\begin{split}
\frac{1}{2}\, \Jcft_v 
&= 
	- \lim_{r\to \infty}\, r^{d-1}\, \left[F_{rv} -\frac{1}{d-4}\, \frac{1}{r^3\,\sqrt{f}}\, \partial_i\, F_{vi} + \cdots \right] , \\
\frac{1}{2}\, \Jcft_i
&=
	-\lim_{r\to\infty} \, r^{d-3} \left[r^2 f\, F_{ri} + F_{vi} - \frac{1}{d-4}\, \frac{1}{r\,\sqrt{f}} \, \left(\partial_v F_{vi} 
	-f\, \partial_j F_{ji}\right) + \cdots\right] .
\end{split}
\end{equation}	
We have indicated here the leading order counterterm contribution given in \eqref{eq:SEMaxA} which suffice for $d\leq 6$; there are higher order terms necessary beyond that. Evaluating this on our linear perturbations we find a background contribution and a term proportional 
to $\MV$. 
 
The background contribution to the charge current is simple: it is just given by the background charge density $\Qbg$, cf., \eqref{eq:rho0def}.
Taking this into account the contribution from the scalar component of Maxwell dynamics assembles into the following:
\begin{equation}\label{eq:JcftA}
\begin{split}
\frac{1}{2}\, \Jcft_v
&= 
	- \Qbg + 
	\left(-\frac{\EEq_0}f + k^2 \left[\frac{1}{(d-4)\, r\, \sqrt{f}}\, \Dz_+ -1\right] \MV  + (d-2)^2\, a\, \PHW\right) \ScS \\
&=
	-\Qbg+ \left(- k^2 \,\MV_{_\text{ren}}    + (d-2)\, \bRQ\, \sdc\, r_+^{d-1}\,  \frac{\PHW}{r^{d-2}}\right) \ScS\,, \\
\frac{1}{2}\, \Jcft_i
&=
	i\omega \left[\frac{1}{(d-4)\, r\, \sqrt{f}}\, \Dz_+ -1 \right] \MV \; \ScS_i \\
&= 
	-i\omega\,k\, \MV_{_\text{ren}} \, \ScS_i	\,.
\end{split}
\end{equation}
We have assembled this in a particular fashion, using the Maxwell equation \eqref{eq:E0} to simplify the presentation (which is why $\PHW$ enters in the above). The renormalized field $\MV_\text{ren}$ is an admixture of the charge diffusion and the phonon fields, and is given in \eqref{eq:VEVRel}.

\paragraph{Stress tensor:}
The boundary stress tensor density is given by varying the boundary Gibbons-Hawking term and the counterterms given in  \eqref{eq:SEMaxA}. This leads to the following expression accurate to quartic order in gradients:
\begin{equation}\label{eq:TBYcft}
\begin{split}
 \TcftD_{\mu\nu} 
& =
 \lim_{r\to\infty}
 	\frac{2\sqrt{-\gamma}}{r^2} \left[ 
 	 K\, \gamma_{\mu\nu}  -K_{\mu\nu} 
 	 - (d-1)\, \gamma_{\mu\nu} + \frac{1}{d-2}  \, \tensor[^\gamma]{G}{_{\mu\nu}}
 	\right. 	\\
&\left.
	+\frac{1}{(d-2)^2\,(d-4)} \Bigg(\tensor[^\gamma]{\nabla}{^2}\, \tensor[^\gamma]{R}{_{\mu\nu}}
	+ 2\, \tensor[^\gamma]{R}{_{\mu\rho\nu\sigma}}\, \tensor[^\gamma]{R}{^{\rho\sigma}} 
	\right.\\
& \left. \quad	
	+ \frac{1}{2\,(d-1)}\, \left[-
		(d-2)\,\tensor[^\gamma]{\nabla}{_\mu}\, \tensor[^\gamma]{\nabla}{_\nu} \tensor[^\gamma]{R}{}  - 
		 d\, \tensor[^\gamma]{R}{} \, \tensor[^\gamma]{R}{_{\mu\nu}}
		  \right]
	\right.  \\
& \left. \quad
	-\frac{1}{2} \,\gamma_{\mu\nu} \left(
		 \tensor[^\gamma]{R}{_{\rho\sigma}}\,  \tensor[^\gamma]{R}{^{\rho\sigma}} 
		  -\frac{d}{4(d-1)}\,  \tensor[^\gamma]{R}{^2} 
		  +\frac{1}{d-1}\, \tensor[^\gamma]{\nabla}{^2}\, \tensor[^\gamma]{R}{}
		\right) \Bigg) \right] .
\end{split}
\end{equation}	

Once again the stress tensor has a background contribution and a correction arising from the perturbations. We  present the results for $(\TcftD)_\mu^{\ \nu}$ which makes it easier to see the traceless condition by inspection.  The  background ideal fluid contribution is determined by the pressure $\Pbg$ given in \eqref{eq:bgP}: 
\begin{equation}\label{eq:Tmnbg}
(T^\text{bg})\indices{_\mu^\nu} 
=  
	 r_+^d\, (1+Q^2)\, \text{diag}\{-(d-1), 1, \ldots, 1\} 
=
	\Pbg\, \text{diag}\{-(d-1), 1, \ldots, 1\} 	 \,.
\end{equation}	
At the first order in amplitudes of the perturbations one evaluates the components of the stress tensor from the Brown-York analysis supplemented with counterterms \eqref{eq:TBYcft}. We  quote the results for the individual components in turn. 

Firstly, the spatio-temporal pieces are
\begin{equation}\label{eq:Tupdnvi}
\begin{split}
\left(\TcftD\right)\indices{_v^i} 
 &=
 	-f \left(\TcftD\right)\indices{_i^v} 
 		=  -\lim_{r\to\infty} \,  ik\,\ScS_i\, T_1 \,, \\
T_1 &= 
		  i\, \PHO - \omega \,\sqrt{f}\, \PHW
		-\frac{\omega\,k^2}{(d-1)\,(d-2)\, (d-4)} \frac{1}{r^3\, \sqrt{f}}\, \PHE   \\
&=
 	- \frac{\omega}{d-1}  \, \MZ_{_\text{ren}}  + \omega\, \left(1-\sqrt{f}\right) \PHW    \\ 
&=
		- \frac{\omega}{d-1}  \, \MZ_{_\text{ren}} \,.
\end{split}
\end{equation}	
We have used \eqref{eq:linearXSZ} to write the expression in terms of $\MZ$. Additionally, the $\PHE$ term has been used to renormalize this field, and thus the final answer involves the renormalized field $\MZ_{_\text{ren}} $, which was obtained in \eqref{eq:VEVRel}.

The temporal component is a bit more complicated,  but it can be evaluated straightforwardly to be
\begin{equation}\label{eq:Tupdnvv}
\begin{split}
\left(\TcftD\right)\indices{_v^v}
&=  
		-(d-1)\, \Pbg + \lim_{r\to\infty}\, \ScS \, T_2\,, \\
T_2 
&=
	(d-1)\, r \left[ 
		\Dz_+ \PHW - \left(2 -\frac{1}{\sqrt{f}}\right)\, \PHE 
		- d\,  (1-\sqrt{f})\, r \sqrt{f}\,\PHW + \PHB\right]  \\
&\qquad \quad 
  - k^2\, \sqrt{f}\, \PHW - \frac{k^4}{(d-1)\,(d-2)\,(d-4)}\, \frac{1}{r^3\,\sqrt{f}}\, \PHE   \\
&=
	(d-1)\, \frac{1+Q^2}{2}\, \frac{\PHE}{r^{d-1}} -\frac{k^2}{d-1}\, \MZ_{_\text{ren}} \\
&=
	\frac{(d-1)\,(d-2)}{2}\, \Pbg\, \mathbf{W}  -\frac{k^2}{d-1}\, \MZ_{_\text{ren}} \,.
\end{split}
\end{equation}
We have written the coefficient of the source $\mathbf{W}$ in terms of the background pressure 
\eqref{eq:bgP}. 

Finally, the spatial stress tensor has contributions from two tensor structures, $\delta_{ij}$ and $\ScST_{ij}$, which we can think of as the pressure and shear-stress contribution. They are 
\begin{equation}\label{eq:Tupdnij}
\begin{split}
\left(\TcftD\right)\indices{_i^ j}  
&= 
	 \Pbg \, \delta\indices{_i^j} + \lim_{r\to\infty\pm i0} \, \left[\frac{1}{f}\, T_P\, \delta\indices{_i^j} \,\ScS+ T_Y\, (\ScST)\indices{_i^j} \right]  .
\end{split}
\end{equation}
The shear-stress part is actually quite simple, and is given by
\begin{equation}\label{eq:TYsimp}
\begin{split}
T_Y 
&=
	- \frac{k^2}{r\sqrt{f}} \left[rf \,\PHW - \frac{\PHE}{d-2}\right]  \\
&= 
	-k^2\, \mathbf{G}\,.
\end{split}
\end{equation}	
On the other hand, the pressure term, which has more intricate structure, can be shown to be
\begin{equation}\label{eq:TPsimpA}
\begin{split}
T_P
&=
	 (d-1) r \, \sqrt{f} (1-\sqrt{f}) \, \PHE 	-i\omega\, \PHO + \omega^2\, \sqrt{f}\, \PHW  + \frac{k^2}{d-1}\, \frac{\sqrt{f}}{r} \left(\PHE  -  (d-2) \, r f  \, \PHW  \right)	   \\
&\qquad
		+ \frac{r^2}{2} \left[d\,(d-1)  \left(1-\sqrt{f}\right)^2   f  - rf' + d\, (1-f)\right] \PHW -2a'\,r^2f\, \Dz_+ \MV - \widetilde{\EEq}_5 \\
&=
	\frac{d-1}{2}\, (1+Q^2)\, \frac{\PHE}{r^{d-1}} + \frac{\omega^2}{d-1}\, \MZ_{_\text{ren}}  - \frac{d-2}{d-1}\, k^2\, \mathbf{G} +2\, (d-2)\, \mu\, r_+^{d-2}\, \frac{1}{r^{d-3}} \,\Dz_+\MV   \\ 
&=
 	(d-2)\, \Pbg\, \left[ \frac{d-1}{2}\, \mathbf{W} + d\, \frac{\mathbf{C}}{r_+\,\bRQ} \right] \
	+ \frac{\omega^2}{d-1}  \MZ_{_\text{ren}}  - \frac{d-2}{d-1}\, k^2\, \mathbf{G} \,,
\end{split}
\end{equation}
where we used $\frac{2\,\mu}{r_+ (1+Q^2)} = \frac{d}{\bRQ}$ in the last line.  

We  refrained from writing out the quartic order counterterm term, but have used its presence to renormalize the field $\MZ$.  Another simplification we exploited is the fact that part of $T_P$ can be expressed in terms of the equation of motion $\EEq_5$.
The contribution  from the constant part in the asymptotic expansion of $\PHW$, $\mathbf{G}$, assembles cleanly into the sound attenuation function. While we find the contribution to be 
$k^2\, \mathbf{G}$, which to the order we are working in, truncates the result in \eqref{eq:WGZero} to $k^2\,\Gatt$, one can show using the Ward identities that the shear-stress part has to be the complete sound attenuation function. It is this observation that motivates our identification \eqref{eq:OstDef}.

We can simplify the spatial part of the stress tensor. Let us start with the term $k^2\, \mathbf{G}$ that enters in $T_P$ and $T_Y$. We first notice that the $\Ost$ contribution to $\mathbf{G}$ in \eqref{eq:WGZero}, when multiplied by $k^2$, can be simplified considerably:
\begin{equation}\label{eq:k2Gamreplace}
\begin{split}
k^2 \frac{\nu_s}{1+Q^2}\, \Ost\, \PoZ 
&=
	  r_+^2\, \KS\, \PoZ - \left(-\omega^2 + \frac{k^2}{d-1}\right) \PoZ  \\
&=
	\left(\omega^2 -\frac{k^2}{d-1}\right) \PoZ 
	+2\, d\, (d-1)^2\, \Pbg\,  \JoZ \,.
\end{split}
\end{equation}	
In the second line  we replaced $\KS \, \PoZ$ by the source term $\JoZ$. The above manipulation is independent of  our conjectured form for the corrected dispersion function. In computing $k^2\Ost$, we only care about the $\Gatt$ contribution since all the other pieces are at least of quintic order in gradients, and we can use the definition of $\KS$ in \eqref{eq:KS2}. 

Therefore, we can rewrite $k^2\, \mathbf{G}$ using the original definition \eqref{eq:WGZero} as
\begin{equation}\label{eq:k2Gsimplify}
\begin{split}
k^2\, \mathbf{G}
&=
		\frac{1}{d-2}  \left(\omega^2-\frac{k^2}{d-1}\right)  \left(\PoZ 
		+ k^2\,  \frac{\PoV}{\bRQ}  \right)  + \mathbf{G}_\text{sources}   \\
&=
	\frac{1}{d-2}  \left(\omega^2-\frac{k^2}{d-1}\right)  \MZ_{_\text{ren}} + \mathbf{G}_\text{sources} \,.
\end{split}
\end{equation}
We have simplified the operator contribution above, and isolated all the source contributions into $\mathbf{G}_\text{sources} $ which in turn is 
\begin{equation}\label{eq:k2GsimplifySources}
\begin{split}
 \mathbf{G}_\text{sources}
&=
		 \frac{4\,(d-1)}{\nu_s}\, \Pbg\, \JoZ 
	-  r_+^d\, \sdc\,  \bqt^2 
		 \left(1-\frac{d\,\nu_s}{2\,\bRQ^2}\,\bqt^2  \right) 	\frac{\JoV}{\bRQ} 
	 + \frac{2\,d\,(d-1)\,\Pbg}{\bRQ^2} \, \frac{2\,\bqt^2}{2+\BQT^2}\, \JoZ\\
&=
	  \frac{4\,(d-1)}{\nu_s}\, \Pbg\, \JoZ 
	-   \frac{d\,(d-1)}{2}\, \Pbg \,  \frac{d\, \nu_s}{2\,\bRQ^2} \,  \bqt^2
		 \left(1-\frac{d\,\nu_s}{2\,\bRQ^2}\,\bqt^2  \right) 	\frac{\JoV}{\bRQ} 
	 + \frac{2\,(d-1)\,\Pbg}{\nu_s} \, \BQT^2\, \JoZ\\
&=
	  \frac{d(d-1)}{2}\, \Pbg \left[\mathbf{W} 
	 - \left(1 + \frac{d\, \nu_s}{2\,\bRQ^2} \,  \bqt^2  -\left(\frac{d\,\nu_s}{2\,\bRQ^2}\,\bqt^2  \right)^2 \right)\frac{\JoV}{\bRQ}  + \frac{4}{d\,\nu_s}\, \BQT^2\, \JoZ \right]\\
&\approx
	  \frac{d(d-1)}{2}\, \Pbg \left[\mathbf{W} 
	 - \frac{2+\BQT^2}{2}\, \frac{\JoV}{\bRQ}  +  4\,\frac{\BQT^2}{d\,\nu_s}\, \JoZ\right] \\
&=
	 d\,(d-1)\, \Pbg \left[\frac{1}{2}\, \mathbf{W}  +\frac{\mathbf{C}}{r_+\, \bRQ} \right] .
\end{split}
\end{equation}
In the second line we used the definition of $\BQT$ from \eqref{eq:diagDefs} to simplify the ratio $\frac{\bqt^2}{2+\BQT^2}$, cf., \cref{fn:BQTprops}. The next step was to exploit
\eqref{eq:Wsource} to rewrite part of the $\JoZ$ source  in terms of $\mathbf{W}$. We then guessed that the contributions multiplying $\JoV$ resum into $1+\frac{1}{2}\, \BQT^2$.\footnote{
	This is the second instance after \eqref{eq:WGZero} where we are assuming that we can resum terms into a function of $\BQT$.  } 
Finally, we rewrote the combination of sources in terms of the deformed chemical potential using \eqref{eq:Csource} in the last line. One useful identity in deriving these expressions is 
\begin{equation}\label{eq:usefulidentity}
\frac{\sdc}{1+Q^2} = \frac{d(d-2)}{2\,\bRQ^2} = \frac{d\, (d-1)}{2}\, \frac{d\,\nu_s}{2\,\bRQ^2} \,.
\end{equation}	
Hence, we altogether claim
\begin{equation}
k^2\, \mathbf{G}
= 
		\frac{1}{d-2}  \left(\omega^2-\frac{k^2}{d-1}\right)  \MZ_{_\text{ren}} + 
	 d\,(d-1)\, \Pbg \left[\frac{1}{2}\, \mathbf{W}  +\frac{\mathbf{C}}{r_+\, \bRQ} \right] .
\end{equation}	
It is only in this derivation of $k^2\,\mathbf{G}$ that we have allowed ourselves to assemble terms in the momentum gradient expansion into functions of $\BQT$ -- these are the terms that combine into $\mathbf{C}$. 

One justification for this bold leap is the fact that the deformed chemical potential contribution is necessary to satisfy the conservation equation, since the Joule heating term in \eqref{eq:TJcons} produces a contribution proportional to the deformed chemical potential. Furthermore,  
with this rewriting we can simplify the pressure term $T_P$ considerably. Plugging \eqref{eq:k2Gsimplify} into  \eqref{eq:TPsimpA} we  find
\begin{equation}
\begin{split}
T_P
&=
	 \frac{k^2}{(d-1)^2}\, \MZ_\text{ren} -\frac{d-2}{2}\, \mathbf{W}  = 
	-\frac{T_2}{d-1}\,.
\end{split}
\end{equation}
This ensures that the stress tensor we obtain is manifestly trace-free.

\subsection{The on-shell action}
\label{sec:Zosbdy}

The on-shell action of the grSK solution can be evaluated starting from the final answer for $S[\Vd,\Zd]$ in \eqref{eq:ZVactionfinal}.  At quadratic order this is just a boundary term, which will give the generating function of boundary correlation functions with sources on the two boundaries of the grSK geometry. 

To evaluate the Wilsonian influence function, we Legendre transform  the resulting generating functional. The two steps can be concatenated by adding a suitable boundary term to implement the Legendre transform. To wit, consider the following:
\begin{equation}\label{eq:wifZA}
S[\Vd,\Zd] + 2\, \int_k \left(k^2 \,\Np{\Vd} \, \PiV\, \Vd + \frac{\Np{\Zd}}{\Lk} \, \PiZ \, \Zd \right) .
\end{equation}	
This additional boundary term cancels the $\delta \Vd$ and $\delta \Zd$ variational terms in \eqref{eq:deltaSZV}, effectively quantizing the fields with (renormalized) Dirichlet boundary conditions.  

The evaluation of the on-shell action produces two sets of  contributions. Terms of  the form 
$\PiV\, \Vd$ and $\PiZ\,\Zd$ combine from the bulk action, the variational boundary terms, and the Legendre transform, to produce the dynamical piece of the WIF, i.e., the terms in \eqref{eq:WIF}. One can write this down by inspection since with renormalized Dirichlet boundary conditions, the boundary sources are $\PiV$ and $\PiZ$ with the renormalized field values being the boundary operator expectation values \eqref{eq:ZVvevdefs}. We end up with
\begin{equation}\label{eq:wifB}
\begin{split}
 S_\text{WIF}[\PoV,\PoZ] 
&=
	 \int_k\, \frac{1}{2} \, \left[
	 	k^2\, \Np{\Vd}(\BQT)\, \left(\Vd^\dag_\text{ren} \, \PiV + \PiV^\dag\,\Vd_\text{ren}  \right)
	 	+
	 	\frac{\Np{\Zd}(\BQT)}{k^2}\, \left(\Zd^\dag_\text{ren} \, \PiZ + \PiZ^\dag\,\Zd_\text{ren}  \right) \right]
	 	 \bigg|_{r = r_c +i0}^{r=r_c  -i0} \\
&=
	-\int_k\, k^2 \left( \frac{r_+}{\sigma_\text{dc}}\, 
		\Np{\Vd}(\BQT)\, \PoV_d^\dag\,\Kc(\omega,\bk)\,  \left[\PoV_a + \left(\nB+\frac{1}{2}\right) \PoV_d\right] +\text{cc}\right) \\
&\qquad
		-\int_k \left( \frac{r_+^2}{2\,d\,(d-1)^2\,\Pbg}\, 
		\Np{\Zd}(\BQT)\, \PoZ_d^\dag\,\KS(\omega,\bk)\,  \left[\PoZ_a + \left(\nB+\frac{1}{2}\right) \PoZ_d\right] +\text{cc}\right) \,.
\end{split}
\end{equation}	
We introduced a large radius regulator at $r=r_c$ to aid the extraction of the on-shell action. 

The second contribution is a contact term, which originates from the $(\Dz_+\MW)^2$ boundary term in \eqref{eq:BdyDFinal}. It can be also evaluated directly  using \eqref{eq:JMWWbdy}, and the contribution factorizes on the grSK contour. We end up with  
\begin{equation}\label{eq:ScontactB}
S_\text{contact} [\Vd,\Zd] 
\supset \int_k\,  \frac{(d-1)\, (d-2) \,(d-6)}{8} \,\Pbg\, \left[\mathbf{W}_\skR^\dag\, \mathbf{W}_\skR - \mathbf{W}_\skL^\dag\, \mathbf{W}_\skL\right]  .
\end{equation}	
This is the quadratic piece of the contact term quoted in \eqref{eq:Scontact}. The additional pieces originate from the fact that the \RNAdS{} geometry has a non-vanishing free energy, leading to the constant and linear terms in  the first line of \eqref{eq:Scontact}.


\providecommand{\href}[2]{#2}\begingroup\raggedright\endgroup

\end{document}